\documentclass[11pt,a4paper]{article}
\pdfoutput=1
\usepackage{jheppub}
\usepackage{amsmath,amsfonts,amssymb,bbm}
\usepackage{hyperref, csquotes}
\usepackage{graphicx, color}
\usepackage{tikz,pgf}
\usepackage{physics,tensor}
\usepackage{mathtools, physics}
\usepackage{comment}
\usepackage[nottoc,numbib]{tocbibind}
\usepackage{accents}

\tikzstyle{wv} = [circle, inner sep=0.1pt, draw=black, minimum size=2mm]
\tikzstyle{bv} = [circle, inner sep=0.1pt, fill=black, minimum size=2mm]

\renewcommand{\eqref}[1]{Eq.\,(\ref{#1})}
\renewcommand\[{\begin{equation}}
\renewcommand\]{\end{equation}}
\def\e{\textrm e}
\def\i{\mathrm i}
\def\d{\mathrm{d}}

\renewcommand{\i}{\textrm{i}}

\newcommand{\cutoff}{L}
\newcommand{\hyper}{\text{H}}
\newcommand{\SU}{\text{SU$(2)$}}
\newcommand{\SL}{\text{SL$(2,\C)$}}

\newcommand{\qmarks}[1]{``#1''}

\newcommand{\R}{\mathbb R}
\newcommand{\C}{\mathbb C}
\newcommand*\diff{\mathop{}\!\mathrm{d}}

\newcommand{\ld}{{d_{\textrm{loc}}}}

\newcommand{\xiloc}{\xi_{\textrm{loc}}} %matter correlation length
\newcommand{\xinloc}{\xi_{\textrm{nloc}}}%geometry correlation length
\newcommand{\cas}{\textrm{Cas}}     % Casimir
\newcommand{\rk}{r}         		% rank of group field
\newcommand{\gf}{\Phi}       		% group field

\newcommand{\ff}{\phi}       		% frame field
\newcommand{\vff}{\boldsymbol{\phi}}       % frame field
\newcommand{\vg}{\boldsymbol{g}}

\newcommand{\nv}{n}

\definecolor{jred}{rgb}{0.8,0,0}
\definecolor{jgreen}{rgb}{0,0.7,0}
\definecolor{jblue}{rgb}{0,0,0.8}

\usetikzlibrary{arrows,patterns,shapes,positioning,fit}
\tikzstyle{c} =	[coordinate]
\tikzstyle{vc} = [circle, draw=black, line width=1pt, inner sep=0pt, minimum size=2mm]
\tikzstyle{v} =  [circle, draw=black, line width=.2pt, fill=black, inner sep=0pt, minimum size=1.5mm]
\tikzstyle{vb} =[circle, draw=black, line width=.1pt, fill=jred, inner sep=0pt, minimum size=1mm]
\tikzstyle{vh} =[circle, draw=black, line width=.1pt, fill=jblue, inner sep=0pt, minimum size=1.5mm]
\tikzstyle{vs} =[coordinate]%[circle, draw=black, line width=.1pt, fill=jgreen, inner sep=0pt, minimum size=1mm]
\tikzstyle{e} =	[draw=jred,line width=1pt]
\tikzstyle{eb}= [draw=jgreen,line width=1pt]
\tikzstyle{es}= [dashed, line width=1pt]
\tikzstyle{ev}= [draw=jgreen,line width=1pt]
\tikzstyle{f} = [line width=0.01pt,dotted,fill=blue, fill opacity=.1]
\tikzstyle{cs}= [ellipse, draw=black, line width=.1pt, fill=jred, inner sep=1pt, minimum size=2mm]%[draw=none,fill=jred, fill opacity=.7]

\newcommand{\cvf}{
\begin{tikzpicture}[scale=2,x=2ex,y=2ex,baseline={([yshift=-.59ex]current bounding box.center)}] 
\scriptsize{
\node [v]	(1)	at (0,0)	{};
\node [vb, label=above:]	(b1)	at (-1,1)	{};
\node [vb, label=below:]	(w1)	at (-1,-1)	{};
\node [vb, label=below:]	(b2)	at (1,-1)	{};
\node [vb, label=above:]	(w2)	at (1,1)	{};
\path
\foreach \i/\j in {1/2}{
   	(w\i) edge [ev]   (b\j)
	(b\i) edge [ev]  node [c] {} (w\j)
	}
\foreach \i in {1,2}{
	(w\i)   edge 	[ev]				(b\i)
	(w\i)	edge [ev,bend left=30]	(b\i)
	(w\i)	edge [ev,bend right=30]	(b\i)
	}
\foreach \i in {w1,b1,w2,b2}{
   	(\i) edge [e] (1)
	};
}
\end{tikzpicture}
}

\newcommand{\cvft}{
\begin{tikzpicture}[scale=2,x=2ex,y=2ex,baseline={([yshift=-.59ex]current bounding box.center)}] 
\node [v]	(v)	at (0,0) 	{};
\node [vb]	(1)	at (-1,1)	{};
\node [vb]	(2)	at (-1,-1)	{};
\node [vb]	(3)	at (1,-1)	{};
\node [vb]	(4)	at (1,1)	{};
\path
\foreach \i in {1,2,3,4}{
    (\i) edge [e] (v)
    }
\foreach \i/\j in {1/2,2/1,3/4,4/3}{
   (\i)	edge [ev,bend left=15]	(\j)
   (\i)	edge [ev,bend left=40]	(\j)
   };
\end{tikzpicture}
}

\newcommand{\cvfn}{
\begin{tikzpicture}[scale=2,x=2ex,y=2ex,baseline={([yshift=-.59ex]current bounding box.center)}] 
\scriptsize{
\node [v]	(1)	at (0,0)	{};
\node [vb, label=above:]	(b1)	at (-1,1)	{};
\node [vb, label=below:]	(w1)	at (-1,-1)	{};
\node [vb, label=below:]	(b2)	at (1,-1)	{};
\node [vb, label=above:]	(w2)	at (1,1)	{};
\path
\foreach \i/\j in {1/2,2/1}{
   	(w\i) edge [ev,bend left=30]  (b\j)
   	(w\i) edge [ev,bend right=30] (b\j)
	}
\foreach \i in {1,2}{
	(w\i)	edge [ev,bend left=30]  (b\i)
	(w\i)	edge [ev,bend right=30]	(b\i)
	}
\foreach \i in {w1,b1,w2,b2}{
   	(\i) edge [e] (1)
	};
}
\end{tikzpicture}
}

\newcommand{\cvfs}{
\begin{tikzpicture}[scale=2,x=2ex,y=2ex,baseline={([yshift=-.59ex]current bounding box.center)}] 
\scriptsize{
\node [v]	(0)	at (0,0)	{};
\foreach \i in {72,144,216,288,360}{
\begin{scope}[rotate=\i]
\node [vb]	(\i)	at (0,1.6)	{};
\end{scope}
}
\path
\foreach \i in {72,144,216,288,360}{
    \foreach \j in {72,144,216,288,360}{
   	    (\i) edge [ev]  (\j)
	    }
    }
\foreach \i in {72,144,216,288,360}{
  	(\i) edge [e] (0)
	};
}
\end{tikzpicture}
}

\begin{document}

\title{Phase transitions in TGFT: a Landau-Ginzburg analysis of Lorentzian quantum geometric models}

\author[a,b,c]{Luca Marchetti,}
\author[a,d]{Daniele Oriti,}
\author[a,d]{Andreas G. A. Pithis,}
\author[e,f]{Johannes Th\"urigen}

\emailAdd{luca.marchetti@phd.unipi.it}
\emailAdd{daniele.oriti@physik.lmu.de}
\emailAdd{andreas.pithis@physik.lmu.de}
\emailAdd{johannes.thuerigen@uni-muenster.de}

\affiliation[a]{Arnold Sommerfeld Center for Theoretical Physics,\\ Ludwig-Maximilians-Universit\"at München \\ Theresienstrasse 37, 80333 M\"unchen, Germany, EU}
\affiliation[b]{Università di Pisa,\\Lungarno Antonio Pacinotti 43, 56126 Pisa, Italy, EU}
\affiliation[c]{Istituto Nazionale di Fisica Nucleare sez. Pisa,\\Largo Bruno Pontecorvo 3, 56127 Pisa, Italy, EU}
\affiliation[d]{Munich Center for Quantum Science and Technology (MCQST), Schellingstr. 4, 80799 M\"unchen, Germany, EU}
\affiliation[e]{Mathematisches Institut der Westf\"alischen Wilhelms-Universit\"at M\"unster\\ Einsteinstr. 62, 48149 M\"unster, Germany, EU}
\affiliation[f]{Institut f\"ur Physik/Institut f\"ur Mathematik der Humboldt-Universit\"at zu Berlin\\
Unter den Linden 6, 10099 Berlin, Germany, EU
}
\date{today}

\begin{abstract}
 {
%Research question
In the tensorial group field theory (TGFT) approach to quantum gravity, the basic quanta of the theory correspond to discrete building blocks of geometry. It is expected that their collective dynamics gives rise to continuum spacetime at a coarse grained level, via a process involving a phase transition. 
%Method/model
In this work we show for the first time how phase transitions for realistic TGFT models can be realized using Landau-Ginzburg mean-field theory. 
More precisely, we consider models generating 4-dimensional Lorentzian triangulations formed by spacelike tetrahedra the quantum geometry of which is encoded in non-local degrees of freedom on the non-compact group $\mathrm{SL}(2,\mathbb{C})$ and subject to gauge and simplicity constraints. 
Further we include $\mathbb{R}$-valued variables which may be interpreted as discretized scalar fields typically employed as a matter reference frame. 
%Result
{We apply the Ginzburg criterion finding that fluctuations around the non-vanishing mean-field vacuum remain small at large correlation lengths regardless of the combinatorics of the non-local interaction validating the mean-field theory description of the phase transition.}
This work represents a first crucial step to understand phase transitions in compelling TGFT models for quantum gravity and paves the way for a more complete analysis via functional renormalization group techniques. 
Moreover, it supports the recent extraction of effective cosmological dynamics from TGFTs in the context of a mean-field approximation.
}
\end{abstract}

\setcounter{tocdepth}{2}

\maketitle

\newpage

\section{Introduction}

Coarse graining methods are instrumental for bridging the gap between microscopic and macroscopic scales
in models for quantum gravity based on discrete fundamental building blocks, by analogy with tenets from statistical physics~\cite{Goldenfeld:1992qy}. 
In particular, they allow to test if a smooth spacetime geometry can emerge from a discrete quantum geometric substrate and whether its dynamics is (approximately) captured by general relativity (GR) in an appropriate limit. 
It is generally expected that such a process is associated to some form of critical behaviour~\cite{Konopka:2008hp,Koslowski:2011vn,Oriti:2013jga}. 
Indeed, in an array of quantum gravity approaches, models display a rich phase structure and some of the phase transitions can potentially be related to relevant continuum limits~\cite{Oriti:2007qd,Gurau:2016cjo,Eichhorn:2018phj,Loll:2019rdj,Steinhaus:2020lgb}.
Approaches where this idea features prominently and coarse graining methods are employed are 
for instance causal sets~\cite{Surya:2019ndm}, 
loop gravity~\cite{Ashtekar:2004eh} 
and spin foam models~\cite{Perez:2003vx,Perez:2012wv,Rovelli:2011eq,Conrady:2010kc,Conrady:2010vx}, 
quantum Regge calculus~\cite{williams2009quantum}, 
dynamical triangulations~\cite{Ambjorn:2012jv}, tensor models~\cite{Gurau:2016cjo,gurau2017random,Gurau:2019qag}, 
and tensorial group field theories~\cite{Freidel:2005qe,Oriti:2011jm,Krajewski:2011zzu,Carrozza:2013oiy,Oriti:2014uga}, with the last two attempting to generalize the accomplishments of matrix models~\cite{DiFrancesco:1993cyw} for $2d$ gravity to higher dimensions. In fact, most of these formalisms exhibit a number of very strict structural connections, in particular tensorial group field theories can also be understood as providing a complete definition of spin foam models and a second quantized formulation of loop quantum gravity~\cite{Oriti:2014uga}.

Tensorial group field theories (TGFT) are combinatorially non-local field theories with tensor fields exciting geometric degrees of freedom. 
Such a rank-$r$ tensor field lives on $r$~copies of a Lie group and the quanta of the theory correspond to $(r-1)$-simplices, while their (perturbative) interaction processes correspond to $r$-dimensional simplicial complexes. The precise combinatorial properties as well as the expression of the transition amplitudes are dictated by the TGFT action, i.e. by the choice of specific TGFT model. 
For models which aim to describe physical quantum geometries, the Lie group relates to the local gauge group of gravity. 
%The TGFT action dictates how these building blocks are glued together to form $r$-dimen\-sional discrete lattice geometries perturbatively. 
To understand the phase structure of such theories better and in particular to determine under which conditions well-defined continuum spacetime geometries can emerge, one can apply coarse graining methods and functional renormalization group (FRG) techniques, as well as powerful approximation techniques such as mean-field theory, well known from the local field theory context and quantum many-body physics~\cite{sachs2006elements,Kopietz:2010zz,zinn2021quantum}. 
However, neither the application of these methods to TGFT nor the interpretation of their results is immediate, for two main reasons: first, because of the combinatorial non-locality of TGFT interactions, which requires to adapt standard RG techniques; second, because quantum gravity requires a manifestly background-independent form of coarse graining prescription~\cite{Pereira:2019dbn,Eichhorn:2021vid}, that, in particular, does not refer directly to spatiotemporal scales (distances, energies) and are necessarily of a more abstract nature. 
Despite these challenges, much progress has been achieved in recent years and these techniques have successfully been extended to the context of matrix, tensor and group field theory models~\cite{Eichhorn:2013isa,Eichhorn:2014xaa,Eichhorn:2017xhy,Eichhorn:2018phj,Eichhorn:2018ylk,Eichhorn:2019hsa,Castro:2020dzt,Eichhorn:2020sla,Benedetti:2015et,BenGeloun:2015ej,BenGeloun:2016kw,Benedetti:2016db,Carrozza:2016vsq,Carrozza:2016tih,Carrozza:2017vkz,BenGeloun:2018ekd,Pithis:2018bw,Pithis:2020sxm,Pithis:2020kio,Marchetti:2021xvf,Baloitcha:2020lha,Lahoche:2022gkz}.

Indeed, the RG analysis of various TGFT models has corroborated the conjecture of the existence of condensate phases, i.e. non-perturbative vacua. In turn, in the TGFT condensate cosmology program ~\cite{Gielen:2016dss,Oriti:2016acw,Pithis:2019tvp,Oriti:2021oux} the mean-field hydrodynamics of quantum geometric TGFTs has been mapped to an effective continuum cosmological dynamics, with a number of interesting features (e.g. a quantum bounce, a Friedmann regime, a late time acceleration, control over geometric fluctuations and the non-trivial dynamics of cosmological perturbations)~\cite{Gielen:2013kla,Gielen:2013naa,Oriti:2016qtz,deCesare:2016rsf, Marchetti:2020umh, Marchetti:2021xvf,Marchetti:2021gcv,Oriti:2021oux, Oriti:2016ueo, Jercher:2021bie}. 
Importantly, from experience with local theories~\cite{strocchi2013introduction,strocchi2005symmetry}, one expects that non-trivial vacua with non-vanishing expectation value of the field operator can only be obtained in the thermodynamic limit, i.e. for infinite system size. Hence, one expects that the domain of the TGFT field should be non-compact, and that otherwise quantum fluctuations would lead us back to the trivial vacuum in the \lq IR\rq ~\cite{Pithis:2020sxm,Pithis:2020kio,Pithis:2018bw,Marchetti:2021xvf}.

For a TGFT on a compact group, non-trivial vacua can be obtained either by taking the thermodynamic limit as a large-volume limit of a compact group domain~\cite{BenGeloun:2015ej,BenGeloun:2016kw,Pithis:2018bw,Pithis:2020sxm,Pithis:2020kio} or by extending the TGFT field domain by non-compact (local) directions~\cite{Marchetti:2021xvf,BenGeloun:2022xyz}, in addition to those corresponding to the gauge group of gravity.
The latter option is especially interesting from a physical perspective, since it is the result of matter coupling. For example, coupling scalar fields to geometric degrees of freedom adds flat local directions in the domain of the TGFT field~\cite{Oriti:2016qtz,Li:2017uao,Gielen:2018fqv, Oriti:2006jk}. Moreover, minimally coupled, massless and free scalar fields can be used as simple material reference frames, which allows to extract the dynamics of quantum geometry in relational terms; this is indeed the strategy pursued in TGFT cosmology ~\cite{Oriti:2016qtz,Gielen:2018fqv,Marchetti:2020umh,Marchetti:2021gcv}, as commonly done in classical and quantum gravity literature~\cite{Brown:1994py,Rovelli:2001bz,Dittrich:2005kc,Ashtekar:2011ni,Giesel:2012rb,Oriti:2016qtz,Gielen:2018fqv,Carrozza:2022xut,Goeller:2022rsx}. 
Indeed, recent works confirm the expectation that adding such degrees of freedom leads to a non-trivial and interesting phase structure for such hybrid models~\cite{Marchetti:2021xvf,BenGeloun:2022xyz}. 

For quantum geometric TGFT models with Lorentzian signature, on the other hand, the field domain is essentially given by (copies of) the Lorentz group, and it is therefore non-compact from the start. Thus, one would expect an interesting phase structure generically. However, so far the precise RG analysis as well as the simpler Landau-Ginzburg analysis of quantum geometric Lorentzian TGFT models has been uncharted territory due to towering technical challenges.
In particular, the analysis of such models requires command over infinite-dimensional group representations. 
Moreover, a regularization scheme has to be put in place since non-compactness together with non-locality of the interactions leads to infinite volume factors  when uniform field configurations are considered~\cite{Marchetti:2021xvf}.
Further complications come from so-called closure and simplicity constraints, the imposition of which is needed in order to ensure the geometric nature of the simplicial (and, in perspective, continuum) structures appearing in the models. 

\

% Method/model
The goal of our present work is to overcome these hurdles and to give a first glimpse at the phase properties of quantum geometric TGFT models, understanding better the general conditions under which critical behaviour occurs therein. 
To this end, we exploit the field-theoretic setting of TGFT and employ Landau-Ginzburg mean-field theory, which is known to capture the basic structure of the phase diagram of local field theories~\cite{sachs2006elements,Kopietz:2010zz,zinn2021quantum}. It has already been shown that this method is sufficient to scrutinize the basic phase properties also of TGFTs~\cite{Pithis:2018bw,Pithis:2019mlv,Marchetti:2021xvf}, at least for simpler models.

Applying Landau-Ginzburg theory to Lorentzian quantum geometries coupled to local scalar matter, we can build on our previous results for such simplified models on compact groups~\cite{Pithis:2018bw,Marchetti:2021xvf}.
The central challenge is to engage with the non-local and non-compact aspects of the quantum geometric degrees of freedom, thus in a context in which a full set of geometricity constraints is imposed, which also requires a careful regularization of the models. 
In particular, we apply this mean-field method to the Lorentzian Barrett-Crane (BC) model~\cite{Jercher:2021bie, Jercher:2022mky} which provides a quantization of Lorentzian Plebanski gravity (reducing to Palatini gravity in first-order formulation upon imposition of constraints), and related models with the same geometric building blocks but also including tensor-invariant interactions~\cite{Carrozza:2013oiy,Carrozza:2016vsq}. 
We restrict ourselves to the case where the TGFT field corresponds to spacelike tetrahedra only and postpone the inclusion of timelike and lightlike components to later work. 

%results
As a result, we find that mean-field theory is reliable for this Lorentzian model mainly due to the hyperbolic structure of the Lorentz group.
Like for the models on compact groups previously investigated~\cite{Marchetti:2021xvf}, 
non-locality of the interactions yields a contribution of so-called \enquote{zero-modes} specific to their combinatorics which modifies the expression of the dimension in the correlation function.
Also the number of local degrees of freedom adds to that dimension.
Further, the scaling of the mass term with correlation length is $\mu\sim1/\xi$ due to the hyperbolic part of the group $\mathrm{SL}(2,\mathbb{C})$, instead of $\mu\sim1/\xi^2$ on flat space.
But most importantly, due to this hyperbolicity there is a exponential factor which suppresses fluctuations at large correlation lengths, independent of the modified dimension.
The quantum geometric aspects (closure and simplicity constraints) do not alter this qualitative aspect of the result. Thus, the Ginzburg criterion for the reliability of the mean-field approximation is fulfilled independently of the rank of the non-local part of the field, of the number of additional local degrees of freedom and of the combinatorics of the interactions. 

%Meaning/Consequences
These results indicate that constitutive features of the Lorentz group lead to the generation of non-trivial phases also in quantum geometric models and allow for a valid description of the associated phase transition via mean-field theory. This is of direct relevance for the TGFT condensate cosmology program relying on such a mean-field approximation, and, more generally, for strengthening the evidence for the existence of a meaningful continuum gravitational limit in TGFT quantum gravity. 

The set up of this article is as follows. In Section~\ref{sec:LorentzianTGFTsandLandauGinzburg} we introduce tensorial group field theory for Lorentzian quantum geometries in the so-called extended formulation, include additional local non-compact degrees of freedom and expound on the relevant theory space. Then, we carry over Landau-Ginzburg mean-field theory to this context and discuss in detail the required regularization scheme which enables us to compute the correlation function of order parameter fluctuations. In Section~\ref{section:correlationlength} we discuss general features of the correlation function which allow us to calculate the correlation length via two complementary and mutually supporting methods well-known in statistical field theory. In Section~\ref{section:Ginzburg Q} we investigate the conditions under which Landau-Ginzburg theory is self-consistently applicable to the presented models via the Ginzburg criterion. 
Finally, we summarize our results in Section~\ref{sec:discussionandconclusion} and discuss shortcomings of our work and prospects for future investigations. 
We complement the main Sections in Appendix~\ref{appendix:harmonic analysis on SL2C} where we discuss details of the harmonic analysis on the Lorentz group, give an analogue presentation for $\text{Spin}(4)$ in Appendix~\ref{app:Spin4} and present explicit
calculations of the correlation functions in Appendix~\ref{app:integrals}.

\section{Landau-Ginzburg theory for Lorentzian TGFTs with local directions}\label{sec:LorentzianTGFTsandLandauGinzburg}

In this Section we apply Landau-Ginzburg mean-field theory to tensorial group field theories for quantum geometries with Lorentzian signature including local degrees of freedom.
The latter may be interpreted as massless and free scalar fields minimally coupled to the discrete geometry~\cite{Li:2017uao, Marchetti:2020umh}. We are thus dealing with hybrid field theories of both local and non-local degrees of freedom. 
The groundwork for this was laid in Ref.~\cite{Marchetti:2021xvf}, which focused on simplified models on Abelian groups. 
 
The models here are more realistic ones in that the non-local geometric degrees of freedom of the group field live on the (double covering of the) restricted Lorentz group $\mathrm{SL}(2,\mathbb{C})$ and are subject to gauge and simplicity constraints; this allows for a geometric interpretation of the discrete structures generated by the model in the sense of simplicial geometry. Working within the \enquote{extended formulation}~\cite{Baratin:2011tx,Jercher:2021bie,Jercher:2022mky}, wherein the domain of the group field is extended by a timelike normal vector such that the fields correspond to spacelike tetrahedra, these symmetries can be imposed in a covariant and commuting way. 

Landau-Ginzburg mean-field theory was originally developed, within statistical field theory, for local scalar fields with generic action functional $S[\Phi]$ in odd and/or even powers of the field and its gradient~\cite{Wipf:2021mns,zinn2021quantum}, also as a field description of lattice systems~\cite{Kopietz:2010zz}. 
Since a detailed evaluation of the partition function of such systems is extremely challenging in general, Landau-Ginzburg theory represents a key approximation scheme, aiming at providing at least a crude account of the phase diagram. 
In a mean-field setting one fundamentally assumes that the system exhibits a separation of scales which allows to average over the microscopic details~\cite{hohenberg2015introduction,zinn2021quantum}. 
This leads to a model which only involves scales which extend from the mesoscale to the macroscale. The field variable is an averaged quantity (the order parameter) which only reflects general features of the system such as symmetries and the dimensionality of the domain. 
The action functional is mostly restricted to the form of the classical action and further microscopic details are encoded by the values of couplings in the action. Clearly, under such coarse graining different microscopic theories lead to the same, i.e. universal, description on larger scales provided they share the same general features of the order parameter. 

For instance, ordinary mean-field theory studies the behaviour of (uniform) field con\-figurations $\Phi_0$ which minimize the action~\cite{sachs2006elements}. In other words, it corresponds to a saddle-point approximation of the partition function. This is rather unrefined since the impact of fluctuations $\delta\Phi$ over these background configurations are neglected. Landau-Ginzburg mean-field theory improves on this matter by retaining quadratic fluctuations around the saddle point. Their systematic treatment then requires to inject the background configuration together with the perturbations into the classical equations of motion while only terms to linear order in $\delta\Phi$ are kept~\cite{Kopietz:2010zz}. 
This allows to solve for the correlation function and the correlation length $\xi$ which extends from the macroscale to the mesoscale and diverges at criticality. 
It sets the scale beyond which correlations between order parameter fluctuations decay exponentially. 
For self-consistency one requires fluctuations of the order parameter up to the scale of %over a given volume determined by 
the correlation length to be much smaller than background configuration~$\Phi_0$. This is the so-called Ginzburg (or Levanyuk-Ginzburg~\cite{levanyuk1959contribution,ginzburg1961some}) criterion. 
Using it, one can e.g. extract for a local scalar field theory on $\mathbb{R}^d$ that Landau-Ginzburg theory is self-consistently applicable in dimensions $d$ larger than the critical dimension $d_{\textrm{crit}}=4$ while in lower dimensions results become inaccurate.%
\footnote{In fact, fluctuations on all scales and of higher order make non-negligible contributions in the critical region and have to be considered to account for the accurate quantitative critical behaviour. This can be done using the Wilsonian renormalization group formalism~\cite{wilson1983renormalization}. 
The key insight which led to its development is that at criticality there is no preferred scale, i.e., one has to look for a theory where the probability distribution exhibits scale-invariance. 
A particularly effective implementation of this setting is provided by the functional renormalization group methodology~\cite{Dupuis:2020fhh}.
}
For a study of this situation on a sphere and hyperboloid in $d$ dimensions we refer to~\cite{Benedetti1403}, which is also of direct technical relevance to this work.

\subsection{A Lorentzian TGFT model including local directions}\label{sec:modelsinlorentziansetting}

In current tensorial group field theories aiming to describe four-dimensional quantum geo\-metry with Lorentzian signature, the real- or complex-valued field%
\footnote{The remainder of this article is only concerned with real-valued fields. The main conclusions are not altered by this choice and could be easily carried over to the complex-valued case.
}  
$\gf$ lives on $r=4$ copies of the Lie group $G=\mathrm{SL}(2,\mathbb{C})$.
These degrees of freedom are subject to combinatorially non-local interactions. 
We extend the domain of the TGFT field to include the upper sheet of the $3$-hyperboloid $\hyper_{+}^3$ (see Fig.~\ref{fig:hypersurfaces} in Appendix~\ref{appendix:SL2Cgroupstructure}), 
according to the prescription of the extended Lorentzian Barrett Crane TGFT model in Ref.~\cite{Jercher:2021bie,Jercher:2022mky}. This version of the Riemannian and Lorentzian Barrett Crane model~\cite{Barrett:1999qw,Perez:2000ec,Perez:2000ep,DePietri:1999bx} was developed in  Refs.~\cite{Baratin:2011tx,Jercher:2021bie,Jercher:2022mky} to resolve the issues of non-covariant and non-commutative imposition of simplicity and closure constraints of the earlier formulations. 
Finally, we introduce frame coordinates $\vff=(\ff_1,...,\ff_{\ld})\in\mathbb{R}^{\ld}$ which are local in the sense of point-like interactions. Thus, altogether the TGFT fields 
\begin{equation}\label{eq:gftfield}
   \Phi(\boldsymbol{\phi},\boldsymbol{g},X)= \Phi(\phi_1,...,\phi_{\ld},g_1,...,g_4,X): \mathbb{R}^{\ld}\times\mathrm{SL}(2,\mathbb{C})^4\times \hyper_{+}^3\to \mathbb{R} %~ \text{or}~\mathbb{C}
\end{equation}
are defined as square-integrable functions $\Phi,\Phi'\in L^2(\mathbb{R}^{\ld}\times\mathrm{SL}(2,\mathbb{C})^4\times \hyper_+^3)$ with respect to the inner product
\begin{equation}
    \left( \Phi , \Phi'\right) = \int_{\mathbb{R}^{\ld}} \text{d}\boldsymbol{\phi}\int_{\mathrm{SL}(2,\mathbb{C})^4}\text{d}\boldsymbol{g}\int_{\hyper_+^3}\text{d}X~\Phi(\boldsymbol{\phi},\boldsymbol{g},X)  \Phi'(\boldsymbol{\phi},\boldsymbol{g},X),
\end{equation}
wherein $\d\vff$ is the Lebesgue measure and $\text{d}g$ the Haar measure on $\mathrm{SL}(2,\mathbb{C})$, see also~\ref{eq:HaarmeasureSL2C}.
However, the application of the Landau-Ginzburg method will necessitate to enlarge the space of functions to that of hyperfunctions at specific points~\cite{Ruehl1970}, as explained later. 
The interpretation of $X\in \hyper_+^3$ is that of a timelike vector normal to the tetrahedra described by the TGFT fields\footnote{As discussed in detail in ~\cite{Jercher:2022mky} , the choice between upper or lower parts of $\hyper^3_{\pm}$ is in fact irrelevant for the construction of the models. Here, we restrict our analysis to timelike normal vectors which lie in the upper $3$-hyperboloid.
}
and $\text{d}X$ denotes the respective integration over it.%
\footnote{Notice that the volumes of the Lorentz group and the homogeneous space $\hyper_+^3$ are infinite. Employing the Cartan decomposition of $\mathrm{SL}(2,\mathbb{C})$ it is clear that via their hyperbolic parts both Haar measures are equally divergent and their rotation parts contribute factors of one. For the purpose of the Landau-Ginzburg analysis with a uniform mean-field ansatz this will necessitate a regularization procedure in terms of a Wick rotation together with a compactification of the Lorentz group and the associated $3$-hyperboloid to bypass the occurrence of unphysical divergences, as treated in a moment in Section~\ref{section:wickrotation}.
} 
The field is subject to the following symmetries 
\begin{align}
    &\Phi(\boldsymbol{\phi}, g_1,g_2,g_3, g_4, X)=\Phi(\boldsymbol{\phi}, g_1 u_1,g_2 u_2,g_3 u_3, g_4 u_4, X),~~~\forall u_i \in \text{SU}(2)_X,\label{eq:simplicity}\\
    &\Phi(\boldsymbol{\phi}, g_1,g_2,g_3, g_4, X)=\Phi(\boldsymbol{\phi}, g_1 h^{-1},g_2 h^{-1},g_3 h^{-1}, g_4 h^{-1}, h\cdot X),~~~\forall h \in \text{SL}(2,\mathbb{C}),\label{eq:gaugeinvariance}
\end{align}
known as simplicity and right covariance. The latter implies that the flux variables dual to the group elements in~\eqref{eq:gftfield} close to form a $3$-simplex, i.e.~a tetrahedron, and that the Feynman amplitudes of the TGFT model assume the form of generalized lattice gauge theory amplitudes. Typically, the simplicity constraint converts the TGFT description of topological $BF$-theory in $4d$, the Ooguri model~\cite{Ooguri:1992eb}, into one describing gravitational degrees of freedom.%
\footnote{While in case of the BC model the simplicity constraints turn the Ooguri model into one for first-order Plebanski (then Plebanski, after constraints are imposed) gravity, in the case of the EPRL-like GFT model~\cite{Oriti:2016qtz} it is turned into one for Plebanski-Holst (then, Palatini-Holst) gravity.
}
Implementing the \textit{geometricity} constraints in terms of the normal~$X$ warrants that the constraints are covariantly imposed and commute with each other. 
However, since the field is not dynamic with respect to the normal $X$ in the sense that interactions are trivial in $X$, it does not appear in the perturbative expansion of the path integral, as explained below.

The geometric interpretation of the field configurations is most transparent in the flux representation~\cite{Baratin:2010wi,Guedes:2013vi,Oriti:2018bwr} which also expatiates the relation of quantum geometric TGFTs and simplicial path integrals~\cite{Oriti:2011jm,Finocchiaro:2018hks}. It is defined by the non-commutative Fourier transform of the field, i.e.
\begin{equation}\label{eq:ncftfield}
    \tilde{\Phi}(\boldsymbol{\phi}, \boldsymbol{B},X)= \tilde{\Phi}(\boldsymbol{\phi}, B_1,...,B_4,X)=\int \text{d}\boldsymbol{g}~ e_{g_1}(B_1)... e_{g_4}(B_4)\Phi(\boldsymbol{\phi},g_1,...,g_4, X),
\end{equation}
wherein $e_{g_i}$ denote non-commutative plane waves. Their product is non-commutative as indicated by the star product $e_{g}(B)\star e_{g'}(B)=e_{gg'}(B)$~\cite{Guedes:2013vi,Oriti:2018bwr}. The flux variables $B_i\in \mathfrak{sl}(2,\mathbb{C})$ are bivectors associated to triangles labelled with $i=1,2,3,4$ of a tetrahedron $\tau$ and their norm yields the area of the respective triangle. Bearing the vector space isomorphism $\mathfrak{sl}(2,\mathbb{C})\cong \bigwedge^2 \mathbb{R}^{1,3}$ in mind, the simplicity condition~(\ref{eq:simplicity}) enforces that bivectors are simple with respect to the timelike normal $X$, that is
\begin{equation}
    X_A \left(* B^{AB}\right)=0 \Rightarrow \exists u_i, v_i \in \mathbb{R}^{1,3}: B_i=u_i\wedge v_i,
\end{equation}
with Lorentz indices $A,B$. 
Moreover, due to the right-covariance condition~(\ref{eq:gaugeinvariance}), one finds that the bivectors $B_i$ at the tetrahedron $\tau$ close after integrating out the timelike normal, i.e.
\begin{equation}
    \sum_{i=1}^4 B_i = 0.
\end{equation}

To determine the correlation function from the linearized equation of motion later on, we will work in \enquote{Fourier} representation space. 
For this we give the expansion of the group field in terms of representations of the unitary principal series of $\mathrm{SL}(2,\mathbb{C})$ labelled by $\rho\in\R$ and $\nu\in\mathbb{Z}/2$.
In fact, due to the simplicity condition, the second $\mathrm{SL}(2,\mathbb{C})$-Casimir operator $\text{Cas}_2=B_i\cdot * B_i$ with eigenvalues $\rho_i \cdot \nu_i$ vanishes~\cite{Barrett:1999qw,Jercher:2022mky}. 
In this work, we focus on the solutions given by $\nu_i=0$ such that the fields expand as
\begin{equation}\label{eq:fieldfourier1}
    \Phi(\boldsymbol{\phi}, \boldsymbol{g},X)=\prod_{i=1}^4\left(\int\text{d}\rho_i \rho_i^2\sum_{j_i,m_i} D^{(\rho_i, 0)}_{j_i m_i 0 0}(g_i X)\right) \Phi_{j_1 m_1 j_2 m_2 j_3 m_3 j_4 m_4}^{\rho_1 \rho_2 \rho_3 \rho_4}(\boldsymbol{\phi}),
\end{equation}
where
$\Phi_{j_1 m_1 j_2 m_2 j_3 m_3 j_4 m_4}^{\rho_1 \rho_2 \rho_3 \rho_4}\equiv\Phi_{j_1 m_1 0 0 j_2 m_2 0 0 j_3 m_3 0 0 j_4 m_4 0 0}^{\rho_1 0 \rho_2 0 \rho_3 0 \rho_4 0}$ and $D^{(\rho, \nu)}_{j m l n}$ are the matrix coefficients  ($\mathrm{SL}(2,\mathbb{C})$-Wigner matrices) in the $\rho,\nu$ representation
(referring for further details on the representation theory to Appendix~\ref{appendix:SL2Charmonic}). Physically, the $\nu_i=0$ solutions cor\-respond to integrating out the rotational subgroup $\text{SU}(2)$ leading to the homogeneous space $\text{SL}(2,\mathbb{C})/\text{SU}(2)$. 
By plugging this solution into the first Casimir, one observes that the corresponding bivectors are spacelike.\footnote{We call a bivector spacelike, lightlike or timelike if $B\cdot B$ is positive, zero or negative, respectively~\cite{Barrett:1999qw,Perez:2000ep}.} Hence, the fields expressed by~\eqref{eq:fieldfourier1} form spacelike tetrahedra.%
\footnote{Notice that the second solution $\rho_i=0$ is realized for the field configurations by integrating out the subgroup $\text{SU}(1,1)$ which leads to the homogeneous space $\mathrm{SL}(2,\mathbb{C})/\text{SU}(1,1)$. Its normal $X$ is spacelike and its bivectors can be either timelike or spacelike, as explained in detail in Refs.~\cite{Jercher:2021bie,Jercher:2022mky}.} 
Consequently, the perturbative expansion of the models considered in this article yields cellular complexes only formed by spacelike components which represent a very special class of triangulations of Lorentzian manifolds.

Since the timelike normals solely assist as auxiliary variables containing extrinsic information about the embedding of the tetrahedra, they are integrated out and thus do not appear in the Feynman amplitudes of corresponding GFT models. Importantly, together with~\eqref{eq:gaugeinvariance} this leads to the closure of the Barrett-Crane tetrahedron~\cite{Jercher:2021bie}.\footnote{Alternatively, the timelike normal could also be fixed to some value $X_0$ like in the time-gauge $X_0=e$, inducing a preferred (and therefore undesirable) spatial foliation structure. Moreover, for the geometric interpretation of the field as a tetrahedron to be well-defined, an additional closure condition would have to be added by hand. Keeping the normal arbitrary and averaging over it, corresponds to a covariant treatment wherein the closure and thus the BC intertwiner shows up directly and all spatial foliations are treated on an equal footing, see also Ref.~\cite{Jercher:2021bie,Jercher:2022mky}.} The expansion of the fields is then given by
\begin{align}\label{eq:groupfieldSL2C}
    \Phi(\boldsymbol{\phi},\boldsymbol{g})&=\int_{\hyper_+^3}\text{d}X\Phi(\boldsymbol{\phi},\boldsymbol{g},X)\nonumber\\&=\prod_{i=1}^4\left(\int\text{d}\rho_i \rho_i^2\sum_{\substack{j_i,m_i;\\l_i,n_i}}D^{(\rho_i,0)}_{j_i m_i l_i n_i}(g_i)\right) B^{\rho_1 \rho_2 \rho_3 \rho_4}_{l_1 n_1 l_2 n_2 l_3 n_3 l_4 n_4} \Phi_{j_1 m_1 j_2 m_2 j_3 m_3 j_4 m_4}^{\rho_1 \rho_2 \rho_3 \rho_4}(\boldsymbol{\phi}),
\end{align}
wherein the so-called Barrett-Crane (BC) intertwiner~\cite{Barrett:1999qw,Oriti:2003wf} is defined by
\begin{equation}\label{eq:SL2CBCintertwiner}
    B^{\rho_1 \rho_2 \rho_3 \rho_4}_{j_1 m_1 j_2 m_2 j_3 m_3 j_4 m_4}\equiv\int_{\hyper_+^3}\text{d}X\prod_{i=1}^4 D^{(\rho_i,0)}_{j_i m_i 0 0}(X).
\end{equation}
The Fourier transform from $\vff\in\mathbb{R}^{\ld}$ to momentum space variables $\boldsymbol{k}\in\mathbb{R}^{\ld}$ 
is as usual
\begin{equation}
     \Phi_{j_1 m_1 j_2 m_2 j_3 m_3 j_4 m_4}^{\rho_1 \rho_2 \rho_3 \rho_4}(\boldsymbol{\phi})=\int_{\mathbb{R}^{\ld}}\frac{\text{d}\boldsymbol{k}}{(2\pi)^{\ld}} \Phi_{j_1 m_1 j_2 m_2 j_3 m_3 j_4 m_4}^{\rho_1 \rho_2 \rho_3 \rho_4}(\boldsymbol{k})\text{e}^{i\boldsymbol{\phi}\cdot \boldsymbol{k}},
\end{equation}
wherein $\boldsymbol{\phi}\cdot \boldsymbol{k}\equiv\sum_{i=1}^{\ld}k_i\phi_i$. In this article we only consider frame coordinates $\vff$ with Euclidean signature. 
The reason is that,
in quantum geometric TGFT models, these flat directions correspond to several scalar fields coupled to gravity which all
appear on equal footing in the fundamental dynamics and 
acquire specific properties (e.g. an interpretation as a Lorentzian reference frame) only through the use of special quantum states~\cite{Marchetti:2021gcv}.

\

\begin{figure}
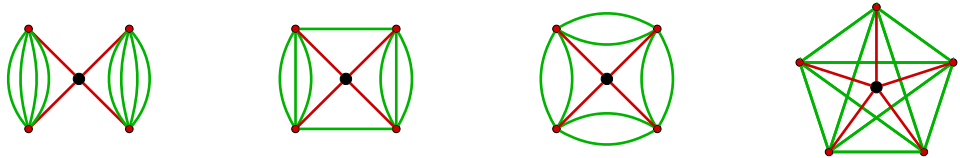

    \centering
        \cvft~~~~~~~~~
        \cvf~~~~~~~~~
        \cvfn~~~~~~~~~
        \cvfs
    \caption{From left to right: Diagrammatic representation of double-trace melonic, melonic, necklace and simplicial interaction vertex graphs for rank-$4$ fields. Green half-edges indicate the pairwise convolution of non-local group arguments $g_i^a$ while red ones represent local variables labelled with momenta $k_i$. Red vertices correspond to fields $\Phi(\boldsymbol{\phi},\boldsymbol{g}_i,X_i)$ while black ones represent point-like interactions with respect to $\boldsymbol{\phi}$.}
    \label{figure:interactions}
\end{figure}

The TGFT field interacts in a combinatorially non-local way with respect to the geo\-metric degrees of freedom $\boldsymbol{g}\in\mathrm{SL}(2,\mathbb{C})^4$ 
while the interactions are local from the point of view of the frame coordinates $\boldsymbol{\phi}\in\mathbb{R}^{\ld}$. 
Since the timelike normals $X\in\hyper_+^3$
play an ancillary role, they appear in the interactions without any coupling among fields and are just identified in the kinetic kernel. Hence, the TGFT action on $\mathbb{R}^{\ld}\times\mathrm{SL}(2,\mathbb{C})^4\times \hyper_+^3$ assumes the general form

\begin{align}\label{eq:TGFTaction}
    S[\Phi]&= %K+V=
    \left(\Phi,\mathcal{K}\Phi\right)+\sum_\gamma \frac{\lambda_\gamma}{\nv_\gamma} 
    \prod_{i=1}^{\nv_\gamma}\left(\int_{\hyper_+^3}\text{d}X_i\right)\int_{\mathbb{R}^{\ld}}\text{d}\boldsymbol{\phi}\Tr_\gamma\left(\Phi\right) \nonumber\\ 
&\equiv  \int_{\mathbb{R}^{\ld}}\text{d}\boldsymbol{\phi}\int_{\mathrm{SL}(2,\mathbb{C})^4}\text{d}\boldsymbol{g}\int_{{\hyper_+^3}}\text{d}X\Phi(\boldsymbol{\phi},\boldsymbol{g},X)\frac{1}{2}\left(-\sum_{i=1}^{\ld}\alpha_i\partial_{\phi_i}^2-\sum_{c=1}^{4}\Delta_{c}+\mu\right)\Phi(\boldsymbol{\phi},\boldsymbol{g},X)\nonumber\\
&\quad +\sum_{\gamma} \frac{\lambda_\gamma}{\nv_\gamma}
\int_{\mathbb{R}^{\ld}}\text{d}\boldsymbol{\phi}\int_{\mathrm{SL}(2,\mathbb{C})^{4\times \nv_\gamma}}\int_{\hyper_+^{3\times \nv_\gamma}}\prod_{i=1}^{\nv_\gamma}\text{d}\boldsymbol{g}_i\text{d}X_i \prod_{(i,a; j,b)}\delta\left(g_i^a (g_j^{b})^{-1}\right)\prod_{i=1}^{\nv_\gamma}\Phi(\boldsymbol{\phi},\boldsymbol{g}_i,X_i).
\end{align}
The kinetic operator 
\begin{equation}\label{eq:kineticoperator}
\mathcal{K} = \frac{1}{2}\left(-\sum_{i}\alpha_i\partial_{\phi_i}^2-\sum_{c}\Delta_{c}+\mu\right)
\end{equation}
contains second-order derivatives with respect to the local variables $\ff_i$ which in general are weighted by positive coefficients $\alpha_i=\alpha_i(\vg)$, %, though we consider them to be constant here, $\alpha_i\equiv \alpha$; 
$\Delta_c$ is the Laplacian on the group $\mathrm{SL}(2,\mathbb{C})$ and $\mu\in\R$ a mass parameter.
The interaction part is a sum over a set of $4$-regular vertex graphs $\gamma$  with $\nv_\gamma$ denoting the number of vertices therein (see Fig.~\ref{figure:interactions} for examples). 
The product of Dirac delta distributions runs over the edges of $\gamma$ which are labelled by $(i,a; j,b)$ with $i,j\in\{1,2,...,\nv_\gamma\}$ and $a,b\in\{1,2,3,4\}$. 
While in the local frame coordinates the interactions are point-like wherefore a single integration $\text{d}\vff$ appears,
the combinatorial non-locality with respect to the geometric degrees of freedom describes through its pairing pattern how different spacelike $3$-simplices are glued together across their faces to form the spacelike boundary of a $4d$ cellular complex. The timelike normals are not coupled between fields.
% For the sake of transparency, we briefly write out some interaction terms $S_{\text{IA}}$ at rank-$4$ of this type which are most relevant to this work, i.e.

The meaning of the terms in the action is the following.
The kinetic term specifies how to glue together two $3$-simplices across a shared $2$-simplex. The timelike normals are identified since they correspond to auxiliary variables. 
The Laplacian on the group manifold is suggested by radiative corrections generated under the renormalization group in the Boulatov and Ooguri GFT models ~\cite{BenGeloun:2011jnm,BenGeloun:2011rc,BenGeloun:2013mgx}. These are at the core general QFT arguments and we leave it to future research to substantiate the expectation that these also directly apply to models with simplicity constraints. 
The second derivatives with respect to the reference variables $\phi_i$ correspond to the lowest order of a series expansion of
second derivatives which arises from the discretization of the continuum action of such free massless and minimally coupled scalar fields over the simplicial geometry, see~\cite{Oriti:2016qtz,Li:2017uao} for details. Such a derivative truncation is expected to be justified from an effective perspective (which, as mentioned above, is the one adopted in this work), as suggested by~\cite{Marchetti:2020umh} in the context of TGFT condensate cosmology. Moreover, in line with basic tenets of Landau-Ginzburg theory~\cite{sachs2006elements,Kopietz:2010zz}, i.e. working in the Gaussian approximation, one only considers terms with the lowest number of derivatives in~\eqref{eq:kineticoperator}. 
The factors $\{\alpha_i\}$ % which are positive functions of the group variables that 
encode non-trivial features of the minimal coupling of the scalar fields to the discrete geometry~\cite{Oriti:2016qtz,Li:2017uao,Gielen:2018fqv} which is why they are functions of $\vg$. In Section~\ref{sec:asymptoticscorrelatorlocal}, we will impose some condition on (the zero modes of) these functions.
%Here we consider them to be equal and $\vg$-independent, i.e. $\alpha_i(\vg)\equiv \alpha$, so that all frame variables are equally weighted. 
The \enquote{mass term} $\mu$ can be motivated by the correspondence of TGFTs perturbative amplitudes with spin foam amplitudes, where it would correspond to spin foam edge weights~\cite{Perez:2012wv,Carrozza:2016vsq}. Here it serves as a control parameter allowing us to differentiate between a trivial and nontrivial vacuum state describable by mean-field theory. Notice that the non-trivial propagator allows to introduce a notion of scale%
\footnote{Correspondingly, the introduction of a scale allows to define a renormalization group flow~\cite{Rivasseau:2016zco,Gurau:2016cjo, Benedetti:2015et,Benedetti:2016db,BenGeloun:2016kw,Carrozza:2016tih,Carrozza:2017vkz,BenGeloun:2018ekd,Pithis:2020sxm,Pithis:2020kio}.} 
which in the context of Landau-Ginzburg theory gives rise to the notion of correlation length~\cite{Marchetti:2021xvf}, detailed for the present models in Section~\ref{section:correlationlength}.

In general, the non-local interactions perturbatively generate 2-complexes which for specific classes of vertex graphs have further structure of four-dimensional simplicial complexes. 
The most common interaction is the simplicial one
\begin{align}
\label{equation:intsimplex}
S_{\text{IA,simplex}}[\Phi]=&\frac{\lambda}{5!}\int_{\mathbb{R}^{\ld}}\text{d}\boldsymbol{\phi}\int_{\mathrm{SL}(2,\mathbb{C})^{10}}[\text{d}g]^{10}\int_{\hyper_+^{3\times 5}}[\text{d}X]^5\times\nonumber\\&\times \Phi_{1234}(\boldsymbol{\phi},X_1)\Phi_{4567}(\boldsymbol{\phi},X_2)\Phi_{7389}(\boldsymbol{\phi},X_3)\Phi_{9620}(\boldsymbol{\phi},X_4)\Phi_{0851}(\boldsymbol{\phi},X_5),
\end{align}
corresponding to the connected graph with $\rk+1=5$ vertices (Fig.~\ref{figure:interactions}) which generates diagrams that are gluings of 4-simplices.
The above geometricity conditions are specific to this 4-simplex interaction. 
However, it has the disadvantage that the generated gluings are not abstract simplicial complexes in general since highly pathological configurations corresponding to singular topologies occur~\cite{DePietri:2000ii,Gurau:2010nd,Gurau:2010mhz}. 
This issue can be resolved by considering an $\rk+1=5$-tuple of fields $\gf^{(c)}$, labelled by a ``colour'' index $c=0,1,2,...,\rk=4$, such that the simplicial interaction $\prod_c \gf^{(c)}$ with convolutions as~\eqref{equation:intsimplex}~\cite{Jercher:2022mky} generates coloured $r$-simplices. This reduces the combinatorial complexity of the Feynman diagrams which are then $\rk+1$ edge coloured graphs and these are bijective to $\rk$-dimensional simplicial pseudo-manifolds~\cite{Gurau:2009tw, Gurau:2011xp}. 

Another related example of interactions generating four-dimensional simplicial pseudo-manifolds are tensor invariants~\cite{Bonzom:2012hw}.
If a vertex graph $\gamma$ is $\rk$-valent edge colourable (e.g. the first three examples in Fig.~\ref{figure:interactions}), the corresponding interaction is invariant under tensorial symmetry, that is invariant under orthogonal transformations of the group field in each field argument. 
Interactions with such tensorial combinatorics can be obtained as effective interactions from coloured simplicial ones by integrating out all coloured fields $\gf^{(c)}$ but one, e.g. $\gf^{(0)}$. 
As a consequence the Feynman diagrams are still $\rk+1$ edge coloured graphs, but only the colour-0 edges describe propagation of $\gf^{(0)}$ 
while the connected components of the Feynman diagram upon 0-edge deletion are the vertex graphs of tensorial interactions. 

The relation of simplicial and tensorial interactions is more intricate for models with 
non-trivial propagators and geometricity constraints. As explained in~\cite{Carrozza:2016vsq}, there are two possibilities: 
$(1)$ All coloured fields have non-trivial propagators and geometricity constraints imposed. 
Then, the effective action gained upon integration of all but one coloured fields will have effective interactions of a 
highly complicated form. 
$(2)$ All but one coloured field are auxiliary fields with trivial propagator ($\delta$ convolution).
This leads to tensorial interactions with trivial convolutions $\delta(g_i^a (g_j^b)^{-1})$.
The models we consider here,~\eqref{eq:TGFTaction}, cover interactions of this second type.%
\footnote{As shown in Ref.~\cite{Carrozza:2016vsq}, the two strategies are closely related for the coloured Boulatov model which is a simplicial rank-$3$ GFT on $G=\text{SU}(2)$ with closure constraint providing a model for Euclidean quantum gravity in three dimensions. 
In this model the non-trivial propagator effectively generates at large $N$ tensorial interactions with derivatives.
This result should generalize to other models with only closure constraint such as the Euclidean Ooguri model, a rank-$4$ GFT model for $BF$-theory in $4d$. 
However, it is less clear how this point unfolds for Lorentzian models and models with simplicity constraints, like the recently formulated coloured complete BC GFT model~\cite{Jercher:2022mky}. Confronted with these challenges, in this article we impose closure and simplicity constraints onto the group fields in the spirit of the latter model and assume an effective field theory point of view in the sense that we work with ad hoc introduced uncolored tensor-invariant interactions and the plain simplicial interaction term. We leave it to future research to clarify their relation to the coloured BC GFT model but strongly suspect that these terms play an integral role in the definition of its complete theory space. 
Further motivation to study the critical behaviour of models with tensor-invariant interactions terms in Gaussian approximation also comes from the spin foam perspective. There, it is known that the most divergent radiative corrections correspond to spin foam amplitudes which can be reabsorbed into effective tensor-invariant coupling constants~\cite{Riello:2013bzw,Carrozza:2016vsq}.
} 
Note, however, that 
in Landau-Ginzburg theory, one makes an ansatz for the coarse-grained, effective interactions at meso- to macroscale. These are not necessarily interactions occurring in a stable regime of renormalization at the microscale.
Since there are no insights into the renormalization group flow of the model, we do not know which effective interactions are most relevant at the macroscale and thus consider the general class of \emph{any} interaction vertex graphs $\gamma$.

The type of interactions, in particular whether of even or odd power in the group field, has a direct impact on the type of phase transitions to be expected.
Landau-Ginzburg mean-field theory is most commonly employed to describe a second-order transition between a symmetric and a broken phase of a global symmetry of the action such as $\mathbb{Z}_2$, $\text{U}(1)$ or $\text{O}(N)$. 
Such symmetries are only possible for interactions of even powers which includes all tensorial interactions (since the edge colourable vertex graphs need to have an even number of vertices). On the contrary, the simplicial interaction at rank $\rk=4$  (Fig.~\ref{figure:interactions}) is quintic and thus does not accommodate this type of symmetry. Notice, however, that Landau-Ginzburg mean-field theory can also be applied to models including an odd-order term in the potential. 
These always force the transition to be of first-order and they do not entail a change in global symmetry~\cite{landau2013statistical,dmitriev1996reconstructive}.\footnote{We note that phase transitions of first-order are not uncommon in discrete quantum gravity approaches. For instance, it is well-known that they compete with second-order transitions in EDT~\cite{Catterall:1994pg,Bialas:1996wu,Coumbe:2014nea,Laiho:2016nlp} and CDT~\cite{Ambjorn:2012jv,Ambjorn:2022dvx}.} 

\subsection{Regularization of the models via compactification and Wick rotation}\label{section:wickrotation}

Considering the non-local geometric degrees of freedom to live on non-compact domain, we would encounter unphysical divergences due to empty integrals over $\mathrm{SL}(2,\mathbb{C})$ associated to the non-locality of the interactions and the closure constraint in combination with the uniform mean-field ansatz in the ensuing Landau-Ginzburg analysis.%
\footnote{We note here that by virtue of the Cartan decomposition of the Haar measure on $\mathrm{SL}(2,\mathbb{C})$, see Appendix~\ref{eq:Cartan decomposition of g}, one can clearly observe that the hyperbolic part of the measure is as divergent as the measure on $\hyper^{3}$, while the respective rotation parts contribute a factor of one. In this sense they contribute with the same degree of divergence in empty integrals over the group or the normal which is why we regulate them in just the same way hereafter.
} 
This necessitates to give a proper regularization of the Lorentzian formulation of the models introduced above.
From a formal field-theoretic point of view one may receive this as an IR regularization~\cite{BenGeloun:2015ej,BenGeloun:2016kw}. 
Removing the regularization consistently at the end of our analysis, will allow us to scrutinize the mean-field behaviour for the actual $\mathrm{SL}(2,\mathbb{C})$-valued domain.

The key ingredient for the regularization is a mapping from $\text{SL}(2,\mathbb{C})$ to $\text{Spin}(4)$, introduced in~\cite{Dona:2021ldn}, which effectively compactifies the domain of the non-local geometric degrees of freedom. This is accomplished by means of a simultaneous analytic continuation between the Lie algebras, Lie group elements and unitary irreducible representations of $\text{SL}(2,\mathbb{C})$ and $\text{Spin}(4)$. 
In close analogy with the well-known operation in field theory which shifts between Euclidean and Lorentzian signature for the underlying spacetime manifold this operation is referred to as \enquote{Wick rotation} in~\cite{Dona:2021ldn}. 
We emphasize that the operation presented there actually amounts to a regularization of $\mathrm{SL}(2,\mathbb{C})$
with a subsequent analytic continuation of (regularized) hyperbolic $3$-space to spherical $3$-space leading to $\text{Spin}(4)$, and thus involves also a compactification of the underlying manifold. While the mapping conveniently liberates us from all the aforementioned volume divergences, it also has other crucial technical advantages. 
Since the Lie group $\text{SL}(2,\mathbb{C})$ is non-compact, one deals with infinite-dimensional unitary representations (see App.~\ref{appendix:harmonic analysis on SL2C}) which are hard to manage. 
In contrast, $\text{Spin}(4)$ is compact and has finite-dimensional unitary irreducible representations (App.~\ref{app:Spin4}) which are more tractable.

In the following, we detail the regularization as given in~\cite{Dona:2021ldn}, adapt it to our needs and emphasize that in fact it involves two steps. 
Locally, the regularization corresponds to a map between the Lie algebras of $\text{SL}(2,\mathbb{C})$ and $\text{Spin}(4)$, %Schematically, one has
\begin{equation}
    \mathfrak{spin}(4)\cong\mathfrak{su}(2)\oplus\mathfrak{su}(2)\leftrightarrow \mathfrak{sl}(2,\mathbb{C})\cong\mathfrak{su}(2)\oplus i \mathfrak{su}(2),
\end{equation}
which rotates the generators of Euclidean and Lorentzian boosts (see App.~\ref{appendix:harmonic analysis on SL2C} and~\ref{app:Spin4}) into each other, i.e.
\begin{equation}
    (\vec{L},-i\vec{K}_E)\cong\mathfrak{sl}(2,\mathbb{C})~~~\text{and}~~~(\vec{L},i\vec{K})\cong\mathfrak{spin}(4),
\end{equation}
and thus amounts to an isomorphism of Lie algebras. At the global level, this permits the construction of a map between group elements of both Lie groups by virtue of the matrix exponential and their respective Cartan decompositions. For $\mathrm{SL}(2,\mathbb{C})$, this decomposition is given by (App.~\ref{appendix:harmonic analysis on SL2C})
\begin{align}
    \text{SU}(2)\times \text{A}^{+}\times \text{SU}(2)&\to \text{SL}(2,\mathbb{C})\nonumber\\
    (u, \e^{\frac{1}{2}\frac{\eta}{a}\sigma_3},v)&\mapsto u\,\text{e}^{ \frac{1}{2}\frac{\eta}{a}\sigma_3 }\,v^{-1}
\end{align}
with
\begin{equation}\label{eq:nctorussemigroup}
    \text{A}^{+}=\{\text{e}^{\frac{1}{2}\frac{\eta}{a}\sigma_3}|\eta\in \R_{+}\}
\end{equation} 
wherein $\eta$ denotes the rapidity parameter. In contrast, for $\text{Spin}(4)$ the Cartan decomposition, as given in~\cite{Dona:2021ldn} (see App.~\ref{app:Spin4} for further details), is 
\begin{align}
    \text{SU}(2)\times \text{T}^{+}\times \text{SU}(2)&\to \text{Spin}(4)\nonumber\\
    (u, \text{e}^{-i\frac{1}{2}\frac{t}{a}\sigma_3},v)&\mapsto (u\,\text{e}^{-i\frac{1}{2}\frac{t}{a}\sigma_3}\,\textbf{}v^{-1}, u\,\text{e}^{i\frac{1}{2}\frac{t}{a}\sigma_3}\,v^{-1})
\end{align}
where
\begin{equation}\label{eq:torussemigroup}
    \text{T}^{+}=\{\text{e}^{-i\frac{1}{2}\frac{t}{a}\sigma_3}|t\in[0,2\pi a[\,\},
\end{equation} 
see Appendix~\ref{app:Spin4} for further details.
In both cases we introduced a scale $a$ which corresponds to the radius of the corresponding homogeneous spaces 
and remains untouched by the analytic continuation. In the former case, $a$ is known as the skirt radius of the hyperboloid while in the latter it is simply the radius of the hypersphere. 
Sending $a$ to large values effectively flattens out the spaces. 
In this sense, this \enquote{curvature} scale $a$ is another important control parameter and will prove useful further below.

To relate the two Lie groups to each other, one has to give a mapping between the non-compact Cartan subgroup $\text{A}^{+}$ of $\mathrm{SL}(2,\mathbb{C})$ and the compact Cartan subgroup $\text{T}^{+}$ of $\text{Spin}(4)$. 
This is a two-step procedure. 
First, we have to regularize the infinite %appreciate that the 
volume of $\mathrm{SL}(2,\mathbb{C})$ introducing a cut-off $\cutoff$ in $\text{A}^{+}$ as
\begin{equation}\label{eq:regularizedA}
    \text{A}^{+}_{\cutoff}:=\{\text{e}^{\frac{1}{2}\frac{\eta}{a}\sigma_3}|\eta\in [0,\cutoff[\,\}.
\end{equation}
Using the Haar measure~\eqref{eq:HaarmeasureSL2C} this yields the regularized volume
\begin{equation}
    \text{vol}(\mathrm{SL}(2,\mathbb{C})_{\cutoff})=\text{vol}(\text{A}^{+}_{\cutoff})=\int_0^{\cutoff}\sinh^2{\left(\frac{\eta}{a}\right)}\frac{\text{d}\eta}{a},
\end{equation}
where we take the compact $\text{SU}(2)$ parts of the measure to be normalized to one. Then, the second step consists in the analytic continuation $\eta\rightarrow -i t$ to $\text{A}^{+}_{\cutoff}$ and identifying $\cutoff$ with $2\pi a$ which leads to $\text{T}^{+}$. 

A more geometric perspective on these points is obtained when realizing that the compactification together with the analytic continuation in fact maps the homogeneous spaces $\text{SL}(2,\mathbb{C})/\text{SU}(2)\cong \hyper^3$ and $\text{Spin}(4)/\text{SU}(2)\cong S^3$ into each other. To see this, consider the mapping between the respective metrics, i.e.
\begin{equation}\label{eq:metricshypsphere}
    \text{d}H_{\cutoff}^2=\left(\left(\frac{\text{d}\eta}{a}\right)^2+\sinh^2{\left(\frac{\eta}{a}\right)}\text{d}\Omega_2\right)
    \xleftrightarrow[\cutoff\leftrightarrow2\pi a]{\eta\leftrightarrow -i t}
    -\text{d}S^2
    =-\left(\left(\frac{\text{d}t}{a}\right)^2+\sin^2\left(\frac{t}{a}\right)\text{d}\Omega_2\right),
\end{equation}
wherein $\text{d}H_{\cutoff}^2$ denotes the metric element on the regulated $3$-hyperboloid, $\text{d}S^2$ that of the hypersphere and $\text{d}\Omega_2$ that of the two-sphere. 
Note that in line with the Haar measures, we dropped prefactors of $a^2$ on the right-hand sides of the respective metrics. 
From the regulated and then analytically continued metric we obtain
\begin{equation}\label{eqn:volumemap}
    \text{vol}(\mathrm{SL}(2,\mathbb{C})_\cutoff)
    =\text{vol}(\text{A}^{+}_{\cutoff})
    \to -\int_{0}^{2\pi a} \sin^2\left(\frac{t}{a}\right) \frac{\text{d}t}{a} 
    %= -\pi 
    =-\text{vol}(\text{T}^{+})=-\text{vol}(\text{Spin}(4)).
\end{equation}
We keep the minus sign as a book-keeping tool although it has no influence on our further arguments.

One can relate the unitary irreducible representations in the principal series of $\text{SL}(2,\mathbb{C})$ to those of $\text{Spin}(4)$
by virtue of the isomorphism of Lie algebras. To this end one maps the representation labels $(\rho,k)\leftrightarrow (-i p,k)$ where $p,k\in \mathbb{Z}/2$ and $\rho\in\mathbb{R}$. 
Thus, the first and second Casimir operators~\eqref{eq:sl2casimirs} and~\eqref{eq:spin4casimirs} transform as
\begin{align}\label{eqn:casimirmap}
    \frac{1}{a^2}\text{Cas}_{1,\text{SL}(2,\mathbb{C})}(\rho,k)&\xleftrightarrow[]{\rho\leftrightarrow -i p}- \frac{1}{a^2}\text{Cas}_{1,\text{Spin}(4)}(p,k)\nonumber\\\frac{1}{a^2}\text{Cas}_{2,\text{SL}(2,\mathbb{C})}(\rho,k)&\xleftrightarrow[]{\rho\leftrightarrow -i p}-\frac{i}{a^2} \text{Cas}_{2,\text{Spin}(4)}(p,k),
\end{align}
while the Plancherel measures are related by
\begin{equation}\label{eqn:plancherelmap}
    \int\sum_{k}\text{d}\rho\left(\rho^2+k^2\right)\xleftrightarrow[]{\rho\leftrightarrow -i p}\frac{1}{-\text{vol}(\text{T}^{+})}\sum_{p,k}-\left(p^2-k^2\right)=\frac{1}{\text{vol}(\text{T}^{+})}\sum_{p,k}\left(p^2-k^2\right).
\end{equation}
As demonstrated in Ref.~\cite{Dona:2021ldn}, this allows to analytically continue the matrix coefficients of the Wigner matrices of $\text{SL}(2,\mathbb{C})$ to those of $\text{Spin}(4)$ and vice versa by transforming the decomposition in terms of the reduced Wigner matrices given in~\eqref{eq:definition of SL2C reduced Wigner matrix} to that of~
\eqref{eq:definition of reduced Spin 4 Wigner matrix}, i.e.
\begin{equation}\label{eqn:repfuncmap}
    d^{(\rho,k)}_{jlm}\left(\frac{\eta}{a}\right)
    \xleftrightarrow[\eta\leftrightarrow -it]{\rho\leftrightarrow -i p}
    d^{(p,k)}_{jlm}\left(\frac{t}{a}\right).
\end{equation}
Employing this, one can easily transform functions and their expansion in representations on $\text{SL}(2,\mathbb{C})$ to those on  $\text{Spin}(4)$ and conversely.

Indeed, when applying this regularization prescription to the group field given in~\eqref{eq:groupfieldSL2C}, we yield
\begin{align}\label{eq:groupfieldregularized}
    \Phi(\boldsymbol{\phi},\boldsymbol{g})&=\int_{S^3}\text{d}X\Phi(\boldsymbol{\phi},\boldsymbol{g},X)\nonumber\\&=\prod_{i=1}^4\left(\sum_{p_i}\sum_{\substack{j_i,m_i;\\l_i,n_i}} \frac{p_i^2}{\text{vol}(\text{T}^{+})}D^{(p_i,0)}_{j_i m_i l_i n_i}(g_i)\right) B^{p_1 p_2 p_3 p_4}_{l_1 n_1 l_2 n_2 l_3 n_3 l_4 n_4} \Phi_{j_1 m_1 j_2 m_2 j_3 m_3 j_4 m_4}^{p_1 p_2 p_3 p_4}(\boldsymbol{\phi}),
\end{align}
with the corresponding regularized BC intertwiner
\begin{equation}\label{eq:BCintertwinerregularized}
    B^{p_1 p_2 p_3 p_4}_{j_1 m_1 j_2 m_2 j_3 m_3 j_4 m_4}\equiv\int_{S^3}\text{d}X\prod_{i=1}^4 D^{(p_i,0)}_{j_i m_i 0 0}(X)
\end{equation}
and note here that due to the Wick rotation of the Haar measure of $\mathrm{SL}(2,\mathbb{C})$ also the integration measure over the normal incorporates an additional minus sign, i.e.
\begin{equation}\label{eq:WRednormalmeasure}
    \text{d}X\equiv-\sin^2\left(\frac{t}{a}\right)\frac{\text{d}t}{a}\text{d}\Omega_2.
    \end{equation}
Up to this minus sign, the expansion of the regularized field~(\ref{eq:groupfieldregularized}) corresponds to the one obtained for the Riemannian model, see~\eqref{eq:fieldfourier1riemanniannormalintegrated}.

With this, we have for the regularized version of the general TGFT action~\eqref{eq:TGFTaction}
\begin{align}\label{eq:TGFTactionregularized}
    S[\Phi]&=\int_{\mathbb{R}^{\ld}}\text{d}\boldsymbol{\phi} \int_{\mathrm{Spin}(4)^4}\text{d}\boldsymbol{g} \int_{{S^3}}\text{d}X\, \Phi(\boldsymbol{\phi},\boldsymbol{g},X) \frac{1}{2} \bigg(-\sum_{i=1}^{\ld}\alpha_i\partial_{\phi_i}^2-\sum_{c=1}^{4}\Delta_{c}+\mu\bigg) \Phi(\boldsymbol{\phi},\boldsymbol{g},X) \nonumber\\
    & ~~+\sum_{\gamma} \frac{\lambda_\gamma}{\nv_\gamma}
    \int_{\mathbb{R}^{\ld}}\text{d}\boldsymbol{\phi}\int_{\mathrm{Spin}(4)^{4\times \nv_\gamma}}\int_{S^{3\times \nv_\gamma}}\prod_{i=1}^{\nv_\gamma}\text{d}\boldsymbol{g}_i\text{d}X_i \prod_{(i,a; j,b)}\delta(g_i^a (g_j^{b})^{-1})\prod_{i=1}^{\nv_\gamma}\Phi(\boldsymbol{\phi},\boldsymbol{g}_i,X_i).
\end{align}
Note that $\alpha_i$ denotes here the analytically continued version of the one given in the Lorentzian models. Since its precise form has no bearing on the ensuing analysis, we do not specify any differences between them.

To summarize, in this section we have introduced a regularization scale $\cutoff$, which will allow us in the following to characterize the asymptotic behavior of quantities of interest. The domain of the field-theoretic system is therefore characterized now by two scales, $\cutoff$~and the \enquote{curvature} scale $a$. Their interplay is crucial and, as we have seen above, defines different regimes of the theory:
\begin{enumerate}
    \item\label{item:noncompact} First, let us consider the limit $\cutoff/a\to\infty$ with finite $a$. In this case, $\eta\in[0,\cutoff]\to\mathbb{R}^+$, and thus we asymptotically reach the non-compact regime $\mathrm{SL}(2,\mathbb{C})_{\cutoff}\to\mathrm{SL}(2,\mathbb{C})$. 
\item\label{item:flat} Second, let us consider the limit $\cutoff\to\infty$ with $\cutoff/(2\pi a)=1$. It is still clearly a non-compact regime, but in this case the metric gets increasingly closer to a flat one, see e.g.~\eqref{eq:metricshypsphere}. We are thus reaching the flat, non-compact regime. 
\item\label{item:compact} Finally, we have the case in which $2\pi a=\cutoff=\text{finite}$. In this regularized case, via a Wick rotation we reach $\text{Spin}(4)$.
\end{enumerate}
While it will turn out to be much more convenient (for technical reasons) to perform most of the computations in the regime~\ref{item:compact} above, we remind the reader that eventually we will Wick rotate back and take a non-compact limit, to eventually reach the regime~\ref{item:noncompact} (and then also explore its flat regime~\ref{item:flat}).

\subsection{Gaussian approximation}\label{sec:gaussian}

In the following, we compute the $2$-point correlation function in the so-called Gaussian approximation, which quantifies correlations of fluctuations $\delta\Phi$ over the uniform background configuration $\Phi_0$.
 
To this aim, one starts off by computing the classical equations of motion from the generic action~\eqref{eq:TGFTaction} yielding
\begin{equation}\label{eq:eom}
    %\int_{S^3}\text{d}X
    \mathcal{K}\Phi+\sum_{\gamma} \frac{\lambda_\gamma}{\nv_\gamma}
    \prod_{i=1}^{\nv_\gamma-1}\left(\int_{S^3}\text{d}X_i\right)\sum_{v\in\mathcal{V}_{\gamma}}\Tr_{\gamma\backslash v}\left(\Phi\right)=0,
\end{equation}
where the variation of the field is done with respect to all variables, including the normal. The second summation exhausts all vertices in the vertex set $\mathcal{V}_\gamma$ of traces encoded by the graph $\gamma\backslash v$ obtained by erasing the vertex $v$ from the graph $\gamma$. Injecting uniform field configurations $\Phi(\boldsymbol{\phi},\boldsymbol{g},X)=\Phi_0$ therein, one yields
\begin{align}\label{eq:eomuniform}
    0&= \mu %\left((-1)a_{S^3}^3\text{vol}_{\text{T}^{+}}\right)
    \Phi_0+\sum_{\gamma} \lambda_\gamma%{\left(\nv_\gamma-1\right) !}
    \left(-%1a_{S^3}^3
    \text{vol}({\text{T}^{+}})\right)^{4\frac{\nv_\gamma-2}{2}+\nv_\gamma-1}\Phi_0^{\nv_\gamma-1} \nonumber\\
    &= \left(\mu+\sum_{\gamma} \lambda_\gamma%{\left(\nv_\gamma-1\right) !}
    \left(-%1a_{S^3}^3
    \text{vol}({\text{T}^{+}})\right)^{4\frac{\nv_\gamma-2}{2}+\nv_\gamma-1}\Phi_0^{\nv_\gamma-2}\right)%\left((-1)a_{S^3}^3\text{vol}_{\text{T}^{+}}\right)
    \Phi_0,
\end{align}
wherein the minus signs in front of the $\text{vol}({\text{T}^{+}})$-terms stem from the Wick rotated normal integration measure, see~\eqref{eq:WRednormalmeasure}. From~\eqref{eq:eomuniform} we obtain the minimizers of the classical action. These are given by the trivial solution and solutions to an algebraic equation of order two less than the interaction of highest-order. For instance, for vertex graphs $\gamma$ all with the same number of vertices $\nv_\gamma=\nv$, the minimizers are the $i=1,2,...,\nv-2$ roots 
\begin{equation}\label{eq:nonvvev}
    \left(-%1a_{S^3}^3
    \text{vol}({\text{T}^{+}})\right)^{\frac{4}{2}+\frac{\nv -1}{\nv -2}}\Phi_0=\zeta_i\left(-\frac{\mu}{\sum_\gamma {\lambda_\gamma}%{\left(\nv_\gamma-1\right) !}
    }\right)^{\frac{1}{\nv -2}},
\end{equation}
wherein $\zeta_i$ denotes the $i$'th root of unity,
though only real solutions of $\gf_0$ are of interest here.
Notice that compared to our previous work~\cite{Marchetti:2021xvf} this equation carries additional volume-factors due to the imposed geometricity conditions in the extended formalism. In particular, for a sum of quartic-order interactions, that is $\nv=4$, one obtains
\begin{equation}\label{eqn:minimumquartic}
    \Phi_0=0~\text{if}~\mu>0~~~\text{and}~~~\left(-%1a_{S^3}^3
    \text{vol}({\text{T}^{+}})\right)^{\frac{4}{2}+\frac{3}{2}}\Phi_0=\pm\sqrt{-\frac{\mu}{4\sum_\gamma{\lambda_\gamma}%{3!}
    }}~\text{if}~\mu<0.
\end{equation}
The latter corresponds to a non-vanishing mean order parameter which describes the phase of broken global $\mathbb{Z}_2$-symmetry.

In the Gaussian (or quasi-Gaussian) approximation one studies small fluctuations $\delta \Phi$ around this state, that is, one linearizes the equation of motion~\eqref{eq:eom} with the ansatz 
\begin{equation}
    \Phi(\boldsymbol{\phi},\boldsymbol{g},X)=\Phi_0+\delta\Phi(\boldsymbol{\phi},\boldsymbol{g},X)
\end{equation}
yielding 
\begin{equation}
  %\int_{S^3}\text{d}X
  \mathcal{K}\delta\Phi+\sum_{\gamma} {\lambda_\gamma}%{\nv_\gamma!}
  \prod_{i=1}^{\nv_\gamma-1}\left(\int_{S^3}\text{d}X_i\right)\sum_{v,v'\in\mathcal{V}_{\gamma}}\Tr_{\gamma\backslash v}\left(\Phi_0,\delta\Phi_{v'}\right)=0,  
\end{equation}
wherein $\delta\Phi_{v'}$ signifies the insertion of the field $\delta\Phi$ at $v'\in \gamma\backslash v$ while $\Phi_0$ is injected at all the other vertices. We can rewrite this expression in the following compact form
\begin{equation}\label{eq:eomvariationoffield}
    \left(
    %\int_{S^3}\text{d}X
    \mathcal{K}+F[\Phi_0]\right)\delta\Phi(\boldsymbol{\phi},\boldsymbol{g},X)=0
\end{equation}
with the Hessian of the interaction part
\begin{align}\label{eq:generalHessian}
    F[\Phi](\boldsymbol{\phi},\boldsymbol{g},X;\boldsymbol{\phi'},\boldsymbol{g'},X') 
    &:= \frac{\delta^2 S_{\text{IA}}[\Phi]}{\delta\Phi(\boldsymbol{\phi},\boldsymbol{g},X)\delta\Phi(\boldsymbol{\phi'},\boldsymbol{g'},X')} \nonumber\\ 
    &\;=\delta(\boldsymbol{\phi}-\boldsymbol{\phi'})\sum_{\gamma} \frac{\lambda_\gamma}{\nv_\gamma}
    \prod_{i=1}^{\nv_\gamma-2}\left(\int_{S^3}\text{d}X_i\right)\sum_{v,v'\in\mathcal{V}_{\gamma}}\Tr_{\gamma\backslash v\backslash v'}\left(\Phi\right),
\end{align}
which is computed at $\Phi(\boldsymbol{\phi},\boldsymbol{g},X)=\Phi_0$ and entails various combinations of Dirac distributions in the group variables. 

Let us further consider the special case of a sum over interaction terms $\gamma$ of the same order, that is the same number of vertices $\nv_\gamma=\nv$. As there is a growing number of graphs for a given number of vertices, this is usual in combinatorially non-local theories.
For example, already for the melonic quartic interaction,  Fig.~\ref{figure:interactions}, there are four different versions depending on which of the field arguments is attached to the single edges. 
Injecting~\eqref{eq:nonvvev} into~\eqref{eq:generalHessian}, we find in this case
\begin{align}
     F[\Phi_0](\boldsymbol{\phi},\boldsymbol{g},X;\boldsymbol{\phi'},\boldsymbol{g'},X') % \nonumber\\ 
     &= \!\left(-%1a_{S^3}^3
     \text{vol}(\text{T}^{+})\right)^{4\left(\frac{\nv}{2}-2\right)} \!\Phi_0^{\nv-2} \delta(\boldsymbol{\phi}-\boldsymbol{\phi'}) \!\prod_{i=1}^{\nv-2} \!\left(\int_{S^3} \!\!\!\text{d}X_i\right) \!\sum_{\gamma} \frac{\lambda_\gamma}{\nv} \mathcal{X}_{\gamma}(\boldsymbol{g},\boldsymbol{g'})\nonumber\\
     &=-\mu\delta(\boldsymbol{\phi}-\boldsymbol{\phi'})\frac{1}{\left(-%1a_{S^3}^3
     \text{vol}({\text{T}^{+}})\right)^{4+\nv-1}}\prod_{i=1}^{\nv-2}\left(\int_{S^3}\text{d}X_i\right)\sum_{\gamma}\tilde{\lambda}_{\gamma}\mathcal{X}_{\gamma}(\boldsymbol{g},\boldsymbol{g'})\nonumber\\&=-\mu\delta(\boldsymbol{\phi}-\boldsymbol{\phi'})\frac{1}{\left(-%1a_{S^3}^3
     \text{vol}({\text{T}^{+}})\right)^{4+1%-1
     }}\sum_{\gamma}\tilde{\lambda}_{\gamma}\mathcal{X}_{\gamma}(\boldsymbol{g},\boldsymbol{g'})
     ,
\end{align}
wherein
\begin{equation}
    \tilde{\lambda}_{\gamma}=\frac{\lambda_{\gamma}}{\sum_{\gamma'}\lambda_{\gamma'}},
\end{equation}
the integrations over $\text{d}X_i$ are empty since $F$ is evaluated on $\Phi_0$ and the operator $\mathcal{X}_{\gamma}(\boldsymbol{g},\boldsymbol{g'})$ corresponds to a sum of products of Dirac distributions the details of which depend on the combinatorics of the graph $\gamma$ (cf.~\cite{Marchetti:2021xvf} for examples). 

We are now in position to solve the equation of motion of the fluctuations~\eqref{eq:eomvariationoffield} using the Green's function method, from which we obtain the correlator in Fourier space. To this aim, we first integrate the above-given equation of motion over the normal so that we work with the constrained solution. This allows us to factorize the BC intertwiner from the kinetic and Hessian part. We also note that this additional integration just yields an additional volume-factor in the Hessian part of the equation which is easily cancelled by a corresponding factor in its denominator. Transferred to representation space, one then has for the Hessian
\begin{align}
    \hat{F}[\Phi_0]&(\boldsymbol{k},\boldsymbol{p},\boldsymbol{j},\boldsymbol{m};\boldsymbol{k'},\boldsymbol{p'},\boldsymbol{j'},\boldsymbol{m'})=\nonumber\\&-\mu\delta(\boldsymbol{k}+\boldsymbol{k'})\sum_{\gamma}\tilde{\lambda}_{\gamma}\hat{\mathcal{X}}_{\gamma}(\boldsymbol{p},\boldsymbol{j},\boldsymbol{m})\prod_{c=1}^4\delta_{p_c,p_c'}\mathbbm{1}_{p_c}\delta_{j_c,j_c'}\mathbbm{1}_{j_c}\delta_{m_c,m_c'}\mathbbm{1}_{m_c}
\end{align}
wherein
\begin{equation}\label{eq:chioperatorregularized}
   \hat{\mathcal{X}}_{\gamma}(\boldsymbol{p},\boldsymbol{j},\boldsymbol{m})=\sum_{p=0}^4\sum_{c_0,...,c_p}\mathcal{X}^{(\gamma)}_{c_0,...,c_p}\prod_{c=c_1}^{c_p}\frac{\delta_{p_c,1}}{p_c^2}\delta_{j_c,0}\delta_{m_c,0}
\end{equation}
with $\mathcal{X}^{(\gamma)}_{c_0,...,c_p}$ denoting combinatorial factors depending on the structure of the graph $\gamma$. For the aforementioned tensor-invariant interactions these are all non-trivial. Likewise, in representation space the kinetic operator reads
\begin{align}
    &\hat{\mathcal{K}}(\boldsymbol{k},\boldsymbol{p},\boldsymbol{j},\boldsymbol{m};\boldsymbol{k'},\boldsymbol{p'},\boldsymbol{j'},\boldsymbol{m'})=\nonumber\\&\left(\alpha_{\boldsymbol{p},\boldsymbol{j},\boldsymbol{m}}\sum_i k_i^2+\frac{1}{a^2}\sum_c \left(-\text{Cas}_{1,p_c}\right)+\mu\right)\delta(\boldsymbol{k}+\boldsymbol{k'})\prod_{c=1}^4\delta_{p_c,p_c'}\mathbbm{1}_{p_c}\delta_{j_c,j_c'}\mathbbm{1}_{j_c}\delta_{m_c,m_c'}\mathbbm{1}_{m_c}.
\end{align}

Bringing these points together, the $2$-point correlation function reads 
\begin{align}\label{eq:decompositioncorrelator}
    &C(\boldsymbol{\phi},\boldsymbol{g})=\int_{S^3}\text{d}X C(\boldsymbol{\phi},\boldsymbol{g},X)\nonumber\\&=
    \int_{\mathbb{R}^{\ld}}\frac{\text{d}\boldsymbol{k}}{(2\pi)^{\ld}}\text{e}^{i\boldsymbol{\phi}\cdot \boldsymbol{k}}\prod_{i=1}^4\left(\sum_{p_i} \frac{p_i^2}{\text{vol}(\text{T}^{+})}\sum_{\substack{j_i,m_i;\\l_i,n_i}}D^{(p_i,0)}_{j_i m_i l_i n_i}(g_i)\right) B^{p_1 p_2 p_3 p_4}_{l_1 n_1 l_2 n_2 l_3 n_3 l_4 n_4} \hat{C}_{j_1 m_1 j_2 m_2 j_3 m_3 j_4 m_4}^{p_1 p_2 p_3 p_4}(\boldsymbol{k}),
\end{align}
with Fourier coefficients
\begin{align}\label{eqn:fouriermodescorrfunc}
     \hat{C}_{j_1 m_1 j_2 m_2 j_3 m_3 j_4 m_4}^{p_1 p_2 p_3 p_4}(\boldsymbol{k})=\hat{C}(\boldsymbol{k},\boldsymbol{p},\boldsymbol{j},\boldsymbol{m})&=\nonumber\\&=\left(\hat{\mathcal{K}}+\hat{F}[\Phi_0]\right)^{-1}(\boldsymbol{k},\boldsymbol{p},\boldsymbol{j},\boldsymbol{m})\nonumber\\&=\frac{1}{\alpha_{\boldsymbol{p},\boldsymbol{j},\boldsymbol{m}}\sum_i k_i^2+\frac{1}{a^2}\sum_c \left(-\text{Cas}_{1,p_c}\right)+b_{\boldsymbol{p},\boldsymbol{j},\boldsymbol{m}}}
\end{align}
and the effective mass
\begin{equation}\label{eq:effectivemass}
    b_{\boldsymbol{p},\boldsymbol{j},\boldsymbol{m}}:=\mu\left(1-\sum_{\gamma}\tilde{\lambda}_{\gamma}\hat{\mathcal{X}}_\gamma(\boldsymbol{p},\boldsymbol{j},\boldsymbol{m})\right).
\end{equation}
Notice that in juxtapositon to local field theories where the effective mass is a constant, here it depends on the combinatorics of the non-local interactions.

In passing, we remark that, in unregularized form, a general $\mathrm{SL}(2,\mathbb{C})$-valued correlator would be mathematically ill-defined since its denominator via the unregularized version of~\eqref{eq:chioperatorregularized} would contain Dirac distributions each of which divided by infinite volume factors stemming from $\Phi_0$ and the non-compactness of $\mathrm{SL}(2,\mathbb{C})$. 

\section{Correlation function and correlation length}\label{section:correlationlength}

It is well-known from local field theory that the correlation length $\xi$ provides a characteristic length scale beyond which correlations between field fluctuations decay exponentially. As such, it plays a key role when computing the Ginzburg $Q$-parameter which, as we will see in Section~\ref{section:Ginzburg Q}, determines the domain of validity of Landau-Ginzburg mean-field theory by quantifying the strength of fluctuations with characteristic scale $\xi$.

For local field theories, the correlation length is most commonly defined either via the reciprocal value of the logarithm of the asymptotic correlation function in direct space or via the second moment of the correlation function, yielding an effective correlation radius~\cite{sachs2006elements,kostorz2001phase}. It was shown in~\cite{Pithis:2018bw,Marchetti:2021xvf} that these two strategies can be carried over to the context of TGFTs, the inherent non-locality of which requires however some extra care. The imposition of closure and simplicity together with the non-compactness and representation theoretic intricacies of $\mathrm{SL}(2,\mathbb{C})$ as done in this article add further technical challenges. 

To tackle these, in a preliminary step we discuss characteristics of the correlation function in Section~\ref{sec:correlationfunction} first with regard to the contributions stemming from the local variables (in Section~\ref{sec:correlationfunctionlocal}) and then in detail with respect to the non-local geometric degrees of freedom (in Section~\ref{sec:correlationfunctionnonlocal}). This allows us to compute the correlation length by studying the asymptotic form of the correlation function regarding both types of variables in Section~\ref{sec:asymptoticscorrelator}. This is then complemented by carrying over the alternative definition via the second-moment method to the present context in Section~\ref{subsection:secondmomentmethod}. 
Importantly, just as for local field theories, we find that the results for the correlation length obtained using the different methods agree, in spite of employing different simplifying assumptions. 

Before venturing forward, we would like to remark that given the absence of a spacetime interpretation of the domain of the TGFT field, the notion of correlation length discussed here is only understood as an internal scale and not as a distance in physical space. In this sense, our setup cannot address the question of how the critical behaviour manifests itself in terms of local degrees of freedom propagating on the effective spacetime geometry generated by the TGFT ones. To understand the latter issue, we need better control over such spacetime propagating degrees of freedom (see ~\cite{Marchetti:2021xvf} for a more detailed discussion).

\subsection{The correlation function}\label{sec:correlationfunction}

As we have just mentioned, the two main intricacies related to the computation of the correlation function in the case of a TGFT based on $\text{SL}(2,\mathbb{C})$ are the non-locality of the interactions and the complications related to the non-trivial group structure of $\text{SL}(2,\mathbb{C})$. Since both these issues are related to the group dependence of the TGFT field, it is very instructive to study the behaviour of the correlation function with respect the local and non-local degrees of freedom separately. In fact, this can be done since the local and non-local degrees of freedom enter differently into the dynamics of the models, as discussed in Section~\ref{sec:modelsinlorentziansetting}. To isolate the distinct contributions to the correlator and correlation length with respect to one type of variable, we follow the strategy of simply averaging over the other one, as discussed hereafter.

\subsubsection{The correlator with respect to the local variables}\label{sec:correlationfunctionlocal}

To study the dependence of the correlation function on local variables alone, one averages over all of the non-local variables in~\eqref{eq:decompositioncorrelator}, which yields
\begin{align}
    C(\boldsymbol{\phi})\equiv \int\text{d}\boldsymbol{g}\, C(\boldsymbol{\phi},\boldsymbol{g})&=
    \int_{\mathbb{R}^{\ld}}\frac{\text{d}\boldsymbol{k}}{(2\pi)^{\ld}}\text{e}^{i\boldsymbol{\phi}\cdot \boldsymbol{k}} B^{1111}_{00000000} \hat{C}_{00000000}^{1111}(\boldsymbol{k})\nonumber\\&= -\text{vol}(\text{T}^{+})\int_{\mathbb{R}^{\ld}}\frac{\text{d}\boldsymbol{k}}{(2\pi)^{\ld}}\text{e}^{i\boldsymbol{\phi}\cdot \boldsymbol{k}} \hat{C}_{00000000}^{1111}(\boldsymbol{k}),
\end{align}
wherein we used
\begin{equation}
    \int\diff g\, D^{(p,0)}_{jmln}(g)=\text{vol}(\text{T}^+)\delta_{p,1}\delta_{j,0}\delta_{m,0}\delta_{l,0}\delta_{n,0}\,.
\end{equation}
This correlation function takes a standard form known from local statistical field theories on $\mathbb{R}^{\ld}$. We will briefly come back to the discussion of its asymptotic behaviour in Section~\ref{sec:asymptoticscorrelatorlocal}.

\subsubsection{The correlator with respect to the non-local geometric variables}\label{sec:correlationfunctionnonlocal}

In contrast, if we want to scrutinize the correlator with respect to the non-local degrees of freedom we average over the local degrees of freedom in~\eqref{eq:decompositioncorrelator}, which leads us directly to
\begin{align}\label{eqn:corrfuncgspace}
    C(\boldsymbol{g})&=\prod_{i=1}^4\left(\sum_{p_i} \frac{p_i^2}{\text{vol}(\text{T}^{+})}\sum_{\substack{j_i,m_i;\\l_i,n_i}}D^{(p_i,0)}_{j_i m_i l_i n_i}(g_i)\right) B^{p_1 p_2 p_3 p_4}_{l_1 n_1 l_2 n_2 l_3 n_3 l_4 n_4} \hat{C}_{j_1 m_1 j_2 m_2 j_3 m_3 j_4 m_4}^{p_1 p_2 p_3 p_4}(\boldsymbol{0})\,.
\end{align}
The Fourier modes $\hat{C}_{j_1 m_1 j_2 m_2 j_3 m_3 j_4 m_4}^{p_1 p_2 p_3 p_4}(\boldsymbol{0})$ therein are given by~\eqref{eqn:fouriermodescorrfunc}. 

We discuss the non-locality issues in the following paragraph where we will also make use of the $\text{SL}(2,\mathbb{C})$-valued (i.e.\ Wick back-rotated and decompactified) residual correlation function which is the one from which the zero-mode contributions have been eliminated. This is a necessary step to compute the correlation length by means of the two aforementioned methods.

\paragraph{Zero-modes expansion of the correlation function.}\label{section:modebymodeexpansion}

As we have seen in Section~\ref{sec:gaussian}, the combination of non-local interactions and the projection onto uniform field configurations leads to an effective mass~\eqref{eq:effectivemass} which is a sum of products of Kronecker deltas fixing some representation labels to their trivial values (depending on the combinatorial pattern of the interactions). 
Only for these trivial or zero-modes the effective mass turns out to be different from $\mu$. For this reason, as argued in~\cite{Marchetti:2021xvf}, it is convenient to separate the contribution of zero-modes in the sums appearing in~\eqref{eqn:corrfuncgspace}. Introducing the multi-index $\vec{p}\equiv (p,j,m)$, we thus write
\begin{align}\label{eqn:cofg}
    C(\boldsymbol{g})&=\sum_{s=0}^4\sum_{(c_1,...,c_s)}\sum_{\substack{\vec{p}_{c_{s+1}},...,\vec{p}_{c_{r}}\neq(1,0,0) \\ \vec{p}_{c_{1}},...,\vec{p}_{c_{s}}=(1,0,0)}} \hat{C}_{j_1 m_1 j_2 m_2 j_3 m_3 j_4 m_4}^{p_1 p_2 p_3 p_4}(\boldsymbol{0})\nonumber%\\&\quad\times
    \int\diff X\,\prod_{i=1}^4\left(\frac{p_i^2}{\text{vol}(\text{T}^{+})}D^{(p_i,0)}_{j_i m_i 00}(g_iX)\right)\\
    &=\sum_{s=0}^4\sum_{(c_1,...,c_s)}\sum_{\substack{\vec{p}_{c_{s+1}},...,\vec{p}_{c_{r}}\neq(1,0,0) \\ \vec{p}_{c_{1}},...,\vec{p}_{c_{s}}=(1,0,0)}} \hat{C}_{j_1 m_1 j_2 m_2 j_3 m_3 j_4 m_4}^{p_1 p_2 p_3 p_4}(\boldsymbol{0})\times\nonumber\\
    &\quad\times\prod_{i=1}^4\left(\frac{p_i^2}{\text{vol}(\text{T}^{+})}\sum_{\substack{l_i,n_i}}D^{(p_i,0)}_{j_i m_i l_i n_i}(g_i)\right) B^{p_1 p_2 p_3 p_4}_{l_1 n_1 l_2 n_2 l_3 n_3 l_4 n_4}\,.
\end{align}
Alternatively, we can explicitly expand the sums, i.e.
\begin{align}\label{eqn:cofgv2}
C(\boldsymbol{g})    &=
    \prod_{i=1}^4\left(\sum_{\vec{p}_i\neq (1,0,0)}\frac{p_i^2}{\text{vol}(\text{T}^+)}\sum_{\substack{l_i,n_i}}D^{(p_i,0)}_{j_i m_i l_i n_i}(g_i)\right) B^{p_1 p_2 p_3 p_4}_{l_1 n_1 l_2 n_2 l_3 n_3 l_4 n_4}  \hat{C}_{j_1 m_1 j_2 m_2 j_3 m_3 j_4 m_4}^{p_1 p_2 p_3 p_4}(\boldsymbol{0})\nonumber
    \\&\quad+\sum_{c=1}^4\delta_{p_c,1}\delta_{j_c,0}\delta_{m_c,0}\prod_{i=1}^4\left(\sum_{\{\vec{p}_i\}\backslash \{\vec{p}_c\}\neq (1,0,0)}\frac{p_i^2}{\text{vol}(\text{T}^+)}\sum_{\substack{l_i,n_i}}D^{(p_i,0)}_{j_i m_i l_i n_i}(g_i)\right)\times\nonumber\\
    &\qquad\times B^{p_1 p_2 p_3 p_4}_{l_1 n_1 l_2 n_2 l_3 n_3 l_4 n_4}  \hat{C}_{j_1 m_1 j_2 m_2 j_3 m_3 j_4 m_4}^{p_1 p_2 p_3 p_4}(\boldsymbol{0})\nonumber\\
    &\quad+\dots\nonumber\\
    &\quad+\prod_{i=1}^4\left(\frac{1}{\text{vol}(\text{T}^+)}\sum_{\substack{l_i,n_i}}D^{(1,0)}_{00 l_i n_i}(g_i)\right)B^{1111}_{l_1 n_1 l_2 n_2 l_3 n_3 l_4 n_4}  \hat{C}_{00000000}^{1111}(\boldsymbol{0})\,.
\end{align}
Since $D^{(p_i,0)}_{j_i,m_i,00}(g_iX)\vert_{\vec{p}_i=(1,0,0)}=1$, the \textit{s-fold zero-mode} contribution in the above sum only depends on the leftover $4-s$ variables, i.e. the residual correlation function
\begin{align}
    C_s(g_{c_1},...,g_{c_{4-s}})&=\int\frac{\diff X}{\text{vol}(\text{T}^+)^{s}}\prod_{u=1}^{4-s}\left(\sum_{\substack{\vec{p}_{c_u}\neq (1,0,0) %\\ \forall u=1,\dots, r-s}
    }}\frac{p_{c_u}^2}{\text{vol}(\text{T}^+)}D^{(p_{c_u},0)}_{j_{c_u} m_{c_u} 0 0 }(g_{c_u} X)\right)\hat{C}_s(p_{c_1},\dots,p_{c_{4-s}})\nonumber\\
    &=\frac{1}{\text{vol}(\text{T}^+)^{s}}\prod_{u=1}^{4-s}\left(\sum_{\substack{\vec{p}_{c_u}\neq (1,0,0) %\\ \forall u=1,\dots, r-s}
    }}\frac{p_{c_u}^2}{\text{vol}(\text{T}^+)}\sum_{l_{c_u} n_{c_u}}D^{(p_{c_u},0)}_{j_{c_u} m_{c_u} l_{c_u} n_{c_u} }(g_{c_u})\right)\times\nonumber\\
    &\quad\times B^{p_{c_1}\dots p_{c_{4-s}}}_{l_{c_1}n_{c_1}\dots l_{c_{4-s}}n_{c_{4-s}}}\hat{C}_s(p_{c_1},\dots,p_{c_{4-s}})\,,
\end{align}
in which we have
\begin{equation}
    \hat{C}_{j_{c_1} m_{c_1}\dots j_{c_{4-s}} m_{c_{4-s}}}^{p_{c_1}\dots p_{c_{4-s}}}(\boldsymbol{0})=\frac{1}{\frac{1}{a^2}\sum_{u=1}^{4-s} \left(-\text{Cas}_{1,p_{c_u}}\right)+b_{c_1\dots c_s}}\equiv \hat{C}_s(p_{c_1},\dots,p_{c_{4-s}})
\end{equation}
and $B^{p_{c_1}\dots p_{c_{4-s}}}_{l_{c_1}n_{c_1}\dots l_{c_{4-s}}n_{c_{4-s}}}$ denotes the residual BC intertwiner contaminated by the $s$-fold zero-mode and $b_{c_1\dots c_s}$ corresponds to the evaluation of the effective mass of this mode. Each of these zero-mode contributions can then be Wick back-rotated and decompactified in order to study the behaviour of the correlation function on $\text{SL}(2,\mathbb{C})^{4-s}$. According to \eqref{eqn:volumemap}, \eqref{eqn:casimirmap}, \eqref{eqn:plancherelmap} and~\eqref{eqn:repfuncmap}, this procedure leads us to\footnote{Notice that in order to use the map in~\eqref{eqn:plancherelmap} we should add and subtract the contribution from $p_{c_u}=0$ to the sum. Then, we can map $\sum_{p_{c_u}}p_{c_u}^2/\text{vol}(\text{T}^+)\to \int\diff\rho_{c_u}\rho_{c_u}^2$; the remaining $p_{c_u}=0$-term of the sum (with a negative sign) is however suppressed by $\text{vol}(\text{T}^+)\to -\text{vol}(\text{A}^+_\cutoff)$ in the non-compact limit, and thus it can be safely neglected.}
\begin{align}\label{equation:wickbackrotatedcorrelator}
    C_s(g_{c_1},\dots,g_{c_{4-s}})
    &=\frac{1}{(-\text{vol}(\text{A}^+_\cutoff))^{s}}\prod_{u=1}^{4-s}\left(\sum_{\{j_{c_u},m_{c_u}\}\neq (0,0)} %\\ \forall u=1,\dots, r-s}
    \int\diff\rho_{c_u}\,\rho_{c_u}^2\sum_{l_{c_u} n_{c_u}}D^{(\rho_{c_u},0)}_{j_{c_u} m_{c_u} l_{c_u} n_{c_u} }(g_{c_u})\right)\nonumber\\
    &\quad\times B^{\rho_{c_1}\dots \rho_{c_{4-s}}}_{l_{c_1}n_{c_1}\dots l_{c_{4-s}}n_{c_{4-s}}}\hat{C}_s(\rho_{c_1},\dots,\rho_{c_{4-s}})\,,
\end{align}
where 
\begin{equation}
    \hat{C}_s(\rho_{c_1},\dots,\rho_{c_{4-s}})=\frac{1}{\frac{1}{a^2}\sum_{u=1}^{4-s} \left(\text{Cas}_{1,\rho_{c_u}}\right)+b_{c_1\dots c_s}}=\frac{1}{\frac{1}{a^2}\sum_{u=1}^{4-s} \left(\rho_{c_u}^2+1\right)+b_{c_1\dots c_s}}\,.
\end{equation}
Explicitly computing the integrals over $\rho_{c_u}$ in order to obtain the functional behaviour of the correlation function in group space is a Herculean task due to the presence of the residual BC intertwiner. However, it is possible to work around this issue when computing the correlation length with either of the two methods as we demonstrate hereafter.

\subsection{The correlation length via the asymptotic behaviour of the correlator}\label{sec:asymptoticscorrelator}

With this preparatory work in place, we are now able to discuss the asymptotic behaviour of the two-point function with respect to the two different types of variables.

\subsubsection{Asymptotic analysis with respect to the local variables}\label{sec:asymptoticscorrelatorlocal}

As explained in Section~\ref{sec:correlationfunctionlocal}, the averaging of the overall correlation function with regard to the non-local geometric degrees of freedom yields
\begin{align}
    C(\boldsymbol{\phi})=-\text{vol}(\text{T}^+)\int_{\mathbb{R}^{\ld}}\frac{\text{d}\boldsymbol{k}}{(2\pi)^{\ld}} \frac{\text{e}^{i\boldsymbol{\phi}\cdot \boldsymbol{k}}}{\alpha_{\boldsymbol{1},\boldsymbol{0},\boldsymbol{0}}\sum_i k_i^2+b_{\boldsymbol{1},\boldsymbol{0},\boldsymbol{0}}}\,.
\end{align}
From the results in Appendix~\ref{app:integrals} we infer that when $b_{\boldsymbol{1},\boldsymbol{0},\boldsymbol{0}}/\alpha_{\boldsymbol{1},\boldsymbol{0},\boldsymbol{0}}>0$%
\footnote{Since $b_{\boldsymbol{1},\boldsymbol{0},\boldsymbol{0}}>0$ always, this condition is satisfied as long as $\alpha_{\boldsymbol{1},\boldsymbol{0},\boldsymbol{0}}>0$ as well. From now on, therefore, we will restrict to this case only.
}, 
this function asymptotically exhibits an exponentially suppressed behaviour with the typical scale
\begin{equation}\label{eq:correlationlengthlocalmethod1}
    \xiloc^2=\frac{\alpha_{\boldsymbol{1},\boldsymbol{0},\boldsymbol{0}}}{b_{\boldsymbol{1},\boldsymbol{0},\boldsymbol{0}}}\propto \mu^{-1},
\end{equation}
which we identify as the correlation length associated to the local degrees of freedom. This behaviour is well-known from local statistical field theories. Hereafter, the factor $\alpha$, which controls the strength of the minimal coupling of the local and non-local degrees of freedom, will be absorbed into the definition of $\xiloc$.

\subsubsection{Asymptotic analysis with respect to the non-local geometric variables}\label{sec:asymptoticanalysisnonlocal}

Since we are interested only in the qualitative behaviour of the correlation function at large (group) \qmarks{distances} (i.e.\ for large values of $\eta/a$), the technical difficulties that one would encounter when trying to reconstruct the full correlation function in group space from its residual components can be easily bypassed, as we will see in a moment.

In order to study the asymptotic behaviour of $C_s(g_{c_1},\dots, g_{c_{4-s}})$ (and thus of the correlation function), it is convenient to use the Cartan decomposition on the representation functions given in~\eqref{eq:Cartan decomposition of SL-Wigner matrix}. This gives
\begin{align}
C_s(g_{c_1},...,g_{c_{4-s}})
    &=\frac{1}{(-\text{vol}(\text{A}^+_\cutoff))^{s}}\prod_{u=1}^{4-s}\Biggl(\sum_{\{j_{c_u},m_{c_u}\}\neq (0,0)} %\\ \forall u=1,\dots, r-s}
    \int\diff\rho_{c_u}\,\rho_{c_u}^2\sum_{l_{c_u} n_{c_u}}\times\nonumber\\
    &\quad\times \sum_{q_{c_u}=-\min(j_{c_u},l_{c_u})}^{\min(j_{c_u},l_{c_u})} D^{j_{c_u}}_{m_{c_u}q_{c_u}}(v_{c_u})d^{(\rho_{c_u},0)}_{j_{c_u}l_{c_u}q_{c_u}}\left(\frac{\eta_{c_u}}{a}\right)D^{l_{c_u}}_{q_{c_u}n_{c_u}}(v_{c_u}^{\prime {-1}})\Biggr)\times\nonumber\\
    &\quad\times B^{\rho_{c_1}\dots \rho_{c_{4-s}}}_{l_{c_1}n_{c_1}\dots l_{c_{4-s}}n_{c_{4-s}}}\hat{C}(\rho_{c_1},\dots,\rho_{c_{4-s}})\,,\nonumber\\
    &=\frac{1}{(-\text{vol}(\text{A}^+_\cutoff))^{s}}\prod_{u=1}^{4-s}\Biggl(\sum_{\{j_{c_u},m_{c_u}\}\neq (0,0)} \sum_{l_{c_u} n_{c_u}}\sum_{q_{c_u}=-\min(j_{c_u},l_{c_u})}^{\min(j_{c_u},l_{c_u})}\times\nonumber\\
    &\quad\times D^{j_{c_u}}_{m_{c_u}q_{c_u}}(v_{c_u})D^{l_{c_u}}_{q_{c_u}n_{c_u}}(v_{c_u}^{\prime {-1}})\Biggr) I^{j_{c_1}q_{c_1}\dots j_{c_{4-s}}q_{c_{4-s}}}_{l_{c_1}n_{c_1}\dots l_{c_{4-s}}n_{c_{4-s}}}\left(\frac{\eta_{c_1}}{a},\dots ,\frac{\eta_{c_{4-s}}}{a}\right)\,,
\end{align}
where
\begin{align}
    I^{j_{c_1}q_{c_1}\dots j_{c_{4-s}}q_{c_{4-s}}}_{l_{c_1}m_{c_1}\dots l_{c_{4-s}}}\left(\frac{\eta_{c_1}}{a},\dots ,\frac{\eta_{c_{4-s}}}{a}\right)&\equiv \prod_{u=1}^{4-s}\left(\int\diff\rho_{c_u}\,\rho_{c_u}^2d^{(\rho_{c_u},0)}_{j_{c_u}l_{c_u}q_{c_u}}\left(\frac{\eta_{c_u}}{a}\right)\right)\times\nonumber\\
    &\quad\times B^{\rho_{c_1},\dots, \rho_{c_{4-s}}}_{l_{c_1}n_{c_1}\dots l_{c_{4-s}}n_{c_{4-s}}}\hat{C}_s(\rho_{c_1},\dots,\rho_{c_{4-s}})\,.
\end{align}
Since we are interested in the behaviour of the $s$-fold zero-modes at large distances, it is exactly the last integral that we desire to evaluate. 
To this aim, we make a symmetry assumption according to which we will only restrict to isotropic configurations with $\eta \equiv \eta_{c_1}=...=\eta_{c_{4-s}}$. One expects them to already capture the qualitative behaviour of the correlation function at large distances, i.e. for ${\eta}/{a}\gg 1$.\footnote{Notice that these isotropic configurations are also those that are commonly considered in statistical field theory applications of the Landau-Ginzburg mean-field method, see~\cite{sachs2006elements,zinn2021quantum}} Since, from what we have seen in~\eqref{eqn:cofg}, the case in which $s=4$ is trivial (i.e.\ the contribution is constant), we will focus on $s<4$ from now on. 

To compute the integral, we recall from Appendix~\ref{app:repfunctionprop} that the representation functions $d^{(\rho_{c_u},0)}_{j_{c_u}l_{c_u}q_{c_u}}\left({\eta_{c_u}}/{a}\right)$ are \emph{entire} functions of $\rho_{c_u}$ (i.e.\ they are holomorphic on the whole complex plane). In addition, they are exponentially suppressed for large imaginary parts of $\rho_{c_u}$ and are even functions in $\rho_{c_u}$. The same properties of course hold for the (residual) BC intertwiner, which, after all, can be obtained as an integral of a product of the above representation functions, see~\eqref{eq:SL2CBCintertwiner}. As a result, the pole structure of the integrand in the above expression is determined by $\hat{C}_s$ alone.

Given these preliminary considerations, let us start the evaluation of $C_s(g_{c_1},...,g_{c_{4-s}})$ by performing one integration, say over $\rho_{c_1}$. We can use the residue theorem, closing the contour for instance on the upper half of the complex plane. The only pole encircled by the contour is then
\begin{equation}
\bar{\rho}_{c_1}=i\sqrt{\sum_{u=2}^{4-s}\rho_{c_u}^2+(4-s)+a^2b_{c_1,\dots,c_s}}
\end{equation}
Now, the asymptotic behaviour of the integral is of course determined by the asymptotic behaviour of the representation functions. As long as the values of $\rho_{c_u}\ne i\cdot n$ with $n$ an integer, this asymptotic behaviour is given by~\eqref{eqn:asymptoticrepfunctions}. On $\rho_{c_u}=i\cdot n$% those points
, the expansion of the Gauss hypergeometric function used in order to obtain~\eqref{eqn:asymptoticrepfunctions} %that result 
is not well defined. We assume that these points are avoided by appropriately deforming the contour of integration in the following computations; as we will see below, this assumption is a posteriori well motivated. Using the symmetries of the integral, we can again restrict any analytic continuation of the functions involved to the upper half of the complex plane and can thus approximate the reduced Wigner matrices by 
\begin{equation}
d^{(\rho_{c_u},0)}_{j_{c_u}l_{c_u}q_{c_u}}\left(\frac{\eta}{a}\right)\sim \text{e}^{-\frac{\eta}{a}}\text{e}^{i\rho_{c_u} \frac{\eta}{a}}c_{\rho_{c_u}}(j_{c_u},l_{c_u},q_{c_u}).
\end{equation}
The phase appearing in the integral $I^{j_{c_1}q_{c_1}\dots j_{c_{4-s}}q_{c_{4-s}}}_{l_{c_1}m_{c_1}\dots l_{c_{4-s}}}\left({\eta}/{a}\right)$, therefore, after being evaluated on $\bar{\rho}_1$ becomes
\begin{equation}
\phi=i\frac{\eta}{a}\sum_{u=1}^{4-s}\rho_{c_u} =\frac{\eta}{a}\left(-\sqrt{\sum_{u=2}^{4-s}\rho_{c_u}^2+(4-s)+a^2b_{c_1,\dots,c_s}}+i\sum_{u=2}^{4-s}\rho_{c_u}\right)\,.
\end{equation}
In the limit of large ${\eta}/{a}$, we can use the stationary phase method in order to evaluate the above integral. The stationary points of the phase are given by
\begin{equation}
\frac{\partial\phi}{\partial\rho_{c_u}}=-\frac{\rho_{c_u}}{\sqrt{\sum_{u=2}^{4-s}\rho_{c_u}^2+(4-s)+a^2b_{c_1,\dots,c_s}}}+i=0\,,
\end{equation}
which in turn implies that for any $u,u'=2,\dots,4-s$, we have that the stationary points $\bar{\rho}_{c_u}$ satisfy $\bar{\rho}_{c_u}=\bar{\rho}_{c_{u'}}$. We thus only need to solve the above equation for a single $\rho_{c_u}$, finding
\begin{equation}
\bar{\rho}_{c_u}=i\sqrt{1+\frac{a^2b_{c_1,\dots,c_s}}{4-s}}\,.
\end{equation}
Plugging this result into $\bar{\rho}_1$ we find $\bar{\rho}_{c_1}=\bar{\rho}_{c_u}$, $u=2,\dots,4-s$. Let us notice that exactly because of the presence of the effective mass, we are not reaching any singularity associated to the expansion of the hypergeometric functions we have used, thus confirming the self-consistency of our assumption. We can thus estimate the integral by evaluating the integrand on $\rho_{c_u}=\bar{\rho}_{c_u}$, $u=1,\dots,4-s$. There are three factors that contribute to the final result:
\begin{itemize}
\item First, $\hat{C}_s$. After the first evaluation with the residue theorem this amounts to
\begin{equation}
    2\bar{\rho}_1=2i\sqrt{1+\frac{a^2b_{c_1,\dots,c_s}}{4-s}}.
\end{equation}
Since $b\propto \mu$, in the limit of small enough $\mu$ this quantity never reaches zero.
\item Second, we have the BC intertwiner, evaluated on $\rho_{c_u}=\bar{\rho}_{c_u}$. Notice that when taking the $\mu\to 0$ limit, as we will in the following, this reduces to an evaluation of $\rho_u=i$ for each $u=1,\dots,4-s$. In this case the residual BC intertwiner is $B^{i...i}_{l_{c_1}n_{c_1}\dots l_{c_{4-s}}n_{c_{4-s}}}$.
\item Finally, we have the representation functions themselves. As we have seen, except for some unimportant factors, they behave collectively as
\begin{equation}
\exp\left(-(4-s)\frac{\eta}{a}\left(1+\sqrt{1+\frac{a^2b_{c_1,\dots,c_s}}{4-s}}\right)\right)\,.
\end{equation}
\end{itemize}
Thus, we conclude that
\begin{align}\label{eqn:ggeomasympt}
    &C_s(g_{c_1},\dots,g_{c_{4-s}})\bigg\vert_{\eta_{c_1}=...=\eta_{c_{4-s}}=\eta}
    \sim\frac{\exp\left(-(4-s)\frac{\eta}{a}\left(1+\sqrt{1+\frac{a^2b_{c_1,\dots,c_s}}{4-s}}\right)\right)}{(-\text{vol}(\text{A}^+_\cutoff))^{s}}\\
    &\quad\times\prod_{u=1}^{4-s} \Biggl(\sum_{\substack{(j_{c_u},m_{c_u}) \\ \neq (0,0)}} \sum_{l_{c_u} n_{c_u}}\sum_{q_{c_u}=-\min(j_{c_u},l_{c_u})}^{\min(j_{c_u},l_{c_u})}D^{j_{c_u}}_{m_{c_u}q_{c_u}}(v_{c_u})D^{l_{c_u}}_{q_{c_u}n_{c_u}}(v_{c_u}^{\prime {-1}})\Biggr) \tilde{I}^{j_{c_1}q_{c_1}\dots j_{c_{4-s}}q_{c_{4-s}}}_{l_{c_1}n_{c_1}\dots l_{c_{4-s}}n_{c_{4-s}}}\nonumber\,,
\end{align}
where $\tilde{I}^{j_{c_1}q_{c_1}\dots j_{c_{4-s}}q_{c_{4-s}}}_{l_{c_1}n_{c_1}\dots l_{c_{4-s}}n_{c_{4-s}}}$ are coefficients independent of $\eta$. 
From the above equation we see that independently of the sign of the (assumed to be small) effective mass of the zero-mode $b_{c_1,\dots c_{s}}$, the correlation function always exhibits an exponentially decaying behaviour at large ${\eta}/{a}$. 
The reason for this particular feature can be traced back to the hyperbolic properties of $\text{SL}(2,\mathbb{C})$. Indeed, this should be contrasted to what happens in the flat Abelian case where, instead, the correlation function has an oscillating (and polynomially suppressed) behaviour when the effective mass is for instance negative~\cite{Marchetti:2021xvf}. 

From the form of the above (isotropized) residual correlation function, we see that the curved nature of the $\mathrm{SL}(2,\mathbb{C})$ has a profound impact on the behavior of correlations. Indeed, even at criticality ($\mu\to 0$), the scale at which correlations decay exponentially remains finite, contrarily to what happens in the flat case. In order to isolate the scale of correlations associated to the transition $\mu\to 0$, we consider the \qmarks{weighted} residual correlation function
\begin{equation}
\tilde{C}_s\equiv\left\vert\det\left(J\left(\frac{\eta}{a}\right)\right)\right\vert^{4-s}C_s=\sinh^{2(4-s)}\left(\frac{\eta}{a}\right)C_s\,.\end{equation}
In the region where the phase transition is expected to take place, i.e. around $\mu=0$,  the asymptotic behaviour of this combined expression is given by
\begin{equation}\label{eqn:weightezeromodes}
    \tilde{C}_s(g_{c_1},...,g_{c_{4-s}})\vert_{\eta_{c_1}=\dots=\eta_{c_{4-s}}=\eta}\sim \frac{\e^{-\frac{1}{2}a\, b_{c_1,\dots,c_s}\eta}}{(-\text{vol}(\text{A}^+_\cutoff))^{s}}\,,
\end{equation}
where we neglected unimportant proportionality factors. This quantity would be exponentially diverging for negative effective mass $b_{c_1,\dots c_{s}}$ and large ${\eta}/{a}$. From the discussion in Section~\ref{sec:gaussian}, we see that  $b_{c_1,\dots c_{s}}$ can be negative only from
$s$-fold zero-modes with $s < s_0$, where $s_0$ is the minimum number of delta functions on the multi-index $\vec{p}$ appearing in the interactions (compare to~\cite{Marchetti:2021xvf}).

\paragraph{Correlation length.}
As explained above, one way to define the correlation length in statistical field theory is by means of the length after which correlations are exponentially suppressed. Given the form of~\eqref{eqn:weightezeromodes}, this characteristic length is immediately read off to be 
\begin{equation}
\xi_s=\frac{2}{a\,b_{c_1,\dots,c_s}}\propto (a\mu)^{-1}\,.
\end{equation}
The largest of the $\xi_s$ of all contributing zero modes $s\ge s_0$ thus determines the correlation length of the system, $\xinloc=\max \xi_s$. Here, one must exclude contributions characterized by $b_{c_1,\dots,c_s}\le 0$, in which case, as we have seen above, there is no effective exponential suppression. The correlation length, therefore, diverges as $\mu^{-1}$ as we approach the critical point\footnote{We recall that, by construction, $\xinloc\le \cutoff\to\infty$. In particular, notice that the order of the limits $\cutoff\to\infty$ and $\mu\to 0$ is important. The correlation length can only diverge if we send $\cutoff\to\infty$ first and $\mu\to 0$ second. Approaching the critical point while $\cutoff$ is still finite, instead, would lead to a finite correlation length, in agreement with the result that phase transitions are absent if the field domain is compact~\cite{Marchetti:2021xvf}.}. Notice that if we had defined $\xi_s$ (and thus $\xi$) by not taking into account the measure factor, the correlation function would remain finite even when $\mu\to 0$, and it would not provide a characterization of the phase transition itself. 
We also remark that our results are in accordance with those obtained for local scalar field theories on the $d$-dimensional hyberboloid obtained in~\cite{Benedetti1403}.

\paragraph{Flat limit.}
As we have mentioned in Section~\ref{section:wickrotation}, by taking the limit $\cutoff=a\to\infty$, we can reduce to the flat Abelian limit. In particular, by taking this limit before sending $\mu\to 0$, we can compare the results obtained here with those in our previous work~\cite{Marchetti:2021xvf}. In this limit, the Jacobian determinant becomes 
\begin{equation}
    \left\vert\det\left(J\left(\frac{\eta}{a}\right)\right)\right\vert^{4-s}\sim \left(\frac{\eta}{a}\right)^{2(4-s)}
\end{equation}
and thus it does not affect the exponential decay of the correlation function (and in turn the definition of the correlation length). To see this, we only have to consider the prefactor in~\eqref{eqn:ggeomasympt}, which for large $a$ and then small effective mass yields
\begin{equation}
\exp\left(-(4-s)\frac{\eta}{a}\left(1+\sqrt{1+\frac{a^2b_{c_1,\dots,c_s}}{4-s}}\right)\right)\sim \exp\left(-\eta\sqrt{(4-s)b_{c_1,\dots,c_s}}\right)\,.
\end{equation} 
From the last expression we easily infer the correlation length
\begin{equation}\label{eq:correlationlengthnonlocalflatmethod1}
\xi_{s,\text{flat}}=\frac{1}{\sqrt{(4-s)b_{c_1,\dots,c_s}}}\propto \mu^{-\frac{1}{2}}\,,
\end{equation}
which is exactly the same behaviour as found in~\cite{Marchetti:2021xvf}. Importantly, notice that $\xi_s\propto \xi^2_{s,\text{flat}}$ 
%(and thus $\xinloc\propto \xinloc^2_{,\text{flat}}$) 
which is another intriguing consequence of the non-flatness of the $\mathrm{SL}(2,\mathbb{C})$-valued group domain. Moreover, we remark that the correlation length of the non-local geometric degrees of freedom in the flat limit ~\eqref{eq:correlationlengthnonlocalflatmethod1} agrees in form with the result obtained for the local variables in~\eqref{eq:correlationlengthlocalmethod1}, as expected from~\cite{Marchetti:2021xvf}.

In the following, we will compute these correlation lengths also by means of another method which makes use of different simplifying assumptions and thus complements and supports the arguments just given. 

\subsection{The correlation length via the second moment of the correlator}\label{subsection:secondmomentmethod}

\subsubsection{General setup of the method}\label{section:secondmomentsetup}
In local statistical field theory, when studying long-range correlations, another way to define the correlation length is to express it via the second moment of the correlation function which is derived by expanding the Fourier representation of the correlation function up to second order in the momenta~\cite{doi:10.1002/9783527603978.mst0387,kostorz2001phase}. As was shown in~\cite{Marchetti:2021xvf}, this strategy can be carried over to TGFTs including also local degrees of freedom. In the following, we will adapt this method to the present case and will spell out all imposed assumptions. Notice that due the non-locality of the geometric degrees of freedom together with the gauge-invariance condition one is required to introduce a regularization scheme when working with uniform field configurations, as discussed above. For this reason, we formulate this method in regularized form and undo it where possible towards the end of Section~\ref{section:nonlocalcontributioncorrelationlength}.

To start off, we consider the Fourier transform of the (regularized) group field on the configuration space $D\cong\mathbb{R}^{\ld}\times \mathrm{Spin}(4)^4$ which encompasses the non-commutative Fourier transform with respect to the non-local geometric degrees of freedom~\eqref{eq:ncftfield} (adapted to $\mathrm{Spin}(4)$) \textit{and} the Fourier transform with respect to the local degrees of freedom, i.e. 
\begin{align}
    \hat{\Phi}(\boldsymbol{k},\boldsymbol{B},X)
    &=\int_D\text{d}^4 g\text{d}\boldsymbol{\phi}\, \text{e}^{i\boldsymbol{\phi}\cdot\boldsymbol{k}}\left(\prod_{\ell=1}^4 \text{e}_{g_{\ell}}(B_{\ell})\right)\Phi(\boldsymbol{\phi},\boldsymbol{g},X).
\end{align}
In the next step, we use this to give the Fourier expansion of the correlator while expanding the combined plane waves on the product domain to second order in the momenta $(\boldsymbol{k},\boldsymbol{B})$. While the expansion with respect to $\boldsymbol{k}$ goes through in a standard fashion, with regard to the fluxes it does not so easily, due to the star-product. In fact, the non-commutative plane waves expand as
\begin{equation}
    \text{e}_{g}(B)=\text{e}^{i k(g)\cdot B}=\sum_{n=0}^{\infty}\frac{i^n}{n!}k(g)^{i_1}\cdots k(g)^{i_n} B_{i_1}\star\cdots\star B_{i_n},
\end{equation}
wherein $k(g)=-i\ln(g)\in\mathfrak{spin}(4)$ denote canonical coordinates obtained through the logarithm map. In the following, we make the crucial assumption that the correlation function displays rotational symmetry with respect to the local and non-local degrees of freedom (i.e. left- and right-invariance with respect to $K=\text{SU}(2)$), that is $C(\boldsymbol{\phi},\boldsymbol{g},X)\equiv C(||\boldsymbol{\phi}||,t_1,...,t_4,X)$ which effectively Abelianizes the product domain. In particular, leaving the local degrees of freedom briefly aside, with respect to the non-local degrees of freedom the domain reduces to the direct product of the non-compact Abelian subgroup~$\text{T}^{+}$. This simplifies the non-commutative star product to the standard point-wise one~\cite{Baratin:2010wi,Guedes:2013vi,Oriti:2018bwr}. Using this, the expansion of the correlator in momentum space up to second order yields
\begin{align}\label{equation:correlatorsecondmoment1}
    \hat{C}(\boldsymbol{k},\boldsymbol{B},X)&\approx \int\text{d}^4 g\text{d}\boldsymbol{\phi}
    \left(1-\frac{1}{2}\left((\boldsymbol{\phi}\cdot \boldsymbol{k})^2+\sum_{\ell}\tr\left((k(g_{\ell}) B_{\ell}\right) \tr\left(k(g_{\ell}) B_{\ell}\right)\right)\right)
    C(\boldsymbol{\phi},\boldsymbol{g},X)\nonumber\\&=
    \int\text{d}^4 g\text{d}\boldsymbol{\phi}
    \left(1-\frac{1}{2}\left(\frac{\boldsymbol{k}^2\boldsymbol{\phi}^2}{ \ld}+\frac{1}{d_{\text{nloc}}}\biggl(\sum_{\ell}b_\ell^2\biggr)\biggl(\sum_{\ell}t_\ell^2\biggr)\right)\right)
    C(\boldsymbol{\phi},\boldsymbol{g},X),
\end{align}
wherein $d_{\text{nloc}}\equiv \frac{4}{2}\dim{(\mathrm{Spin}(4))}=4\cdot 3$ (the factor of $4$ corresponds to the rank of the group field while that of $3$ accounts for the number of dynamically relevant $\mathrm{Spin}(4)$-variables) and we note that due to the symmetry assumptions the first order terms vanish and the second order terms greatly simplify.\footnote{To see this for the case of the trace-terms, let us write  $g\in \mathrm{Spin}(4)$ as~\cite{grouptheoryphysicists} $g=\text{e}^{i(\vec{\rho}\cdot\vec{L}+\vec{t}\cdot\vec{K}_E)}$, where $\vec{L}\equiv\frac{\vec{\sigma}}{2}$ and $\vec{K}_{E}\equiv \frac{\vec{\sigma}}{2}$ are the generators of rotations and Euclidean boosts respectively, see Appendix~\ref{app:Spin4}, and $\vec{\sigma}$ denotes the Pauli vector built from the Pauli matrices~\cite{Ruehl1970,Oriti:2018bwr}. With this, one finds $\tr\left(k(g)B\right)=\frac{1}{2}\tr\left((\vec{\rho}\cdot\vec{\sigma}+\vec{t}\cdot\vec{\sigma})(\vec{B}\cdot\vec{\sigma})\right)=(\vec{\rho}+\vec{t})\cdot\vec{B}$. Under the given symmetry assumptions, we neglect the contributions from the $\rho_i$ and require the $t_3$-axis to be along $\vec{B}$ with angle $\theta$, then $\tr\left(k(g)B\right)=t_3 ||\vec{B}||=t_3 b=t\cos(\theta) b$. 
Adapting the Haar measure $\text{d}g=\text{d}u\text{d}v\text{d}\mu(t)$ on $\text{Spin}(4)$ to these symmetry assumptions and coordinates, gives $\text{d}g=\sin^2(\theta)\text{d}\theta\text{d}\mu(t)$. Consequently, upon integration over the remaining rotational contributions an integrand comprising of the correlator multiplied by linear terms in the traces vanishes while one with quadratic terms still contributes. In fact, this is completely analogous to what happens to the corresponding terms for the local degrees of freedom.
} 
With this, we can equivalently write
\begin{align}
    \hat{C}(\boldsymbol{k},\boldsymbol{B},X)\approx&\biggl(\int \text{d}^4 g \, \text{d}\boldsymbol{\phi} \, C(\boldsymbol{\phi},\boldsymbol{g},X)\biggr)\times\nonumber\\&~~~\times\Biggl(1-\frac{1}{2}\frac{\int \text{d}^4 g \, \text{d}\boldsymbol{\phi}\left(\frac{\boldsymbol{k}^2\boldsymbol{\phi}^2}{ \ld}+\frac{1}{d_{\text{nloc}}}\biggl(\sum_{\ell}b_\ell^2\biggr)\biggl(\sum_{\ell}t_\ell^2\biggr)\right)C(\boldsymbol{\phi},\boldsymbol{g},X)}{\int \text{d}^4 g \, \text{d}\boldsymbol{\phi} \,  C(\boldsymbol{\phi},\boldsymbol{g},X)}\Biggr).
\end{align}
and from the respective second moments we may define the corresponding correlation lengths
\begin{equation}\label{equation:correlationlengthlocal}
    \xiloc^2\equiv\frac{1}{2\ld}\frac{1}{\int\text{d}\boldsymbol{g}\int\text{d}X\int\text{d}\boldsymbol{\phi}C(\boldsymbol{\phi},\boldsymbol{g},X)}\int\text{d}\boldsymbol{g}\int\text{d}X\int\text{d}\boldsymbol{\phi}||\boldsymbol{\phi}||^2 C(\boldsymbol{\phi},\boldsymbol{g},X)
\end{equation}
and 
\begin{equation}\label{equation:correlationlengthnonlocal}
    \xinloc^2\equiv\frac{1}{2d_\text{nloc}}\frac{1}{\int\text{d}\boldsymbol{g}\int\text{d}X\int\text{d}\boldsymbol{\phi}C(\boldsymbol{\phi},\boldsymbol{g},X)}\int\text{d}\boldsymbol{g}\int\text{d}X\int\text{d}\boldsymbol{\phi}\left(-||\boldsymbol{g}||^2\right)C(\boldsymbol{\phi},\boldsymbol{g},X).
\end{equation}
Note that we also included the integration over the normal $X$ to eliminate all redundant information about the embedding and $||\boldsymbol{g}||^2$ corresponds to $\sum_\ell t_\ell^2$.%
\footnote{
We remark that the quadratic terms of the parameters $t_\ell$ appearing in~\eqref{equation:correlatorsecondmoment1} can also be understood as the squared geodesic distance on the group $\mathrm{Spin}(4)$ assuming individual invariance with respect to the left and right $\text{SU}$-subgroups. 
To see this, we remind the reader that a (bi-invariant) Riemannian metric on $\mathrm{Spin}(4)$ is naturally induced by the inner product $\langle X,Y\rangle= \Re\tr(XY^{\dagger})$, with $X,Y\in\mathfrak{spin}(4)$. In this way, for each $g=\text{e}^{\lambda_a X^a}=\text{e}^{X}\in \mathrm{Spin}(4)$ with $\lambda_a\in\mathbb{C}$ and  $X^a\in\mathfrak{spin}(4)$ the quantity $||g||^2\equiv \langle X,X\rangle= \Re\tr(XX^{\dagger})$ corresponds to the geodesic distance between $g$ and the identity $e$~\cite{gallier2020differential,alexandrino2009introduction}. 
(In fact, the geodesic distance is Lipschitz equivalent to any matrix norm on $\mathrm{Spin}(4)$~\cite{einsiedler2013ergodic}.) Working then for instance in the Cartan decomposition of $\mathrm{Spin}(4)$ and assuming a large parameter $t$ compared to the other angles therein, one finds for the Riemannian distance $d(e,g)=||g||^2\sim t^2$ and $||\boldsymbol{g}||^2\sim \sum_{\ell} t_{\ell}^2$ on the product domain $\mathrm{Spin}(4)^4$. 
Note that a large $t$ naturally relates to a large boost parameter $\eta$ upon Wick rotation such that $||g||^2\sim \eta^2$ for $\text{SL}(2,\mathbb{C})$.
} 
This splitting of the correlation length into both contributions is meaningful since both types of degrees of freedom behave differently at the dynamical level. Note that the minus sign in front of $||g||^2$ in the definition~\ref{equation:correlationlengthnonlocal} is required since we work at this stage with the regularized group field.

In the next step, following the same arguments as in Sections~\ref{sec:correlationfunctionlocal} and~\ref{sec:correlationfunctionnonlocal}, the denominators in the definitions of the two correlation lengths are readily computed since the integrals over the local and non-local degrees of freedom extend over the entire domain~$D$. 
This yields
\begin{equation}
    \int\text{d}\boldsymbol{g}\int\text{d}X\int\text{d}\boldsymbol{\phi}\,C(\boldsymbol{\phi},\boldsymbol{g},X)=%(a_{S^3}^3)^4a_{S^3}^3
    \frac{-\text{vol}(\text{T}^{+})}{b_{\boldsymbol{1},\boldsymbol{0},\boldsymbol{0}}}.
\end{equation}
In the subsequent sections we will also need the analytically continued version of the previous result which is just given by $\text{vol}(\text{A}^{+}_{\cutoff})/b_{\boldsymbol{1},\boldsymbol{0},\boldsymbol{0}}$.

\subsubsection{Contribution to the correlation length of the local variables}\label{section:localcontributioncorrelationlength}

To evaluate the contributions to the correlation length stemming from the local degrees of freedom, we can simply average over the data of the non-local variables in~\eqref{equation:correlationlengthlocal}, yielding
\begin{align}
        \xiloc^2&=\frac{b_{\boldsymbol{1},\boldsymbol{0},\boldsymbol{0}}}{2\ld}\int\text{d}\boldsymbol{\phi}\int_{\mathbb{R}^{\ld}}\frac{\text{d}\boldsymbol{k}}{(2\pi)^{\ld}}\frac{||\boldsymbol{\phi}||^2\text{e}^{i\boldsymbol{\phi}\cdot \boldsymbol{k}}}{\alpha_{\boldsymbol{1},\boldsymbol{0},\boldsymbol{0}}\sum_i k_i^2+b_{\boldsymbol{1},\boldsymbol{0},\boldsymbol{0}}}\nonumber\\&=\frac{\alpha_{\boldsymbol{1},\boldsymbol{0},\boldsymbol{0}}}{b_{\boldsymbol{1},\boldsymbol{0},\boldsymbol{0}}}\propto\mu^{-1},
\end{align}
which is conveniently obtained when first integrating over the momentum variables. As expected, this result agrees with the one obtained via asymptotic arguments in Section~\ref{sec:asymptoticscorrelatorlocal}. 

\subsubsection{Contribution to the correlation length of the non-local variables}\label{section:nonlocalcontributioncorrelationlength}

The evaluation of this part of the correlation length is more subtle, as suggested by the discussion of features of the correlation function with respect to the non-local geometric degrees of freedom in Section~\ref{sec:correlationfunctionnonlocal}.

To start off, one takes the mean over the local degrees of freedom in~\eqref{equation:correlationlengthnonlocal} which gives
\begin{align}
    \xinloc^2&
    =\frac{1}{2d_\text{nloc}}\frac{b_{\boldsymbol{1},\boldsymbol{0},\boldsymbol{0}}}{(-\text{vol}(\text{T}^{+}))}\int\text{d}\boldsymbol{g}\int\text{d}X \left(-||\boldsymbol{g}||^2\right) C(\boldsymbol{g},X).
\end{align}    
Injecting now the decomposition of the correlator subject to the discussed symmetry assumptions, we have 
\begin{align}
    \xinloc^2&=\frac{1}{2d_\text{nloc}}\frac{b_{\boldsymbol{1},\boldsymbol{0},\boldsymbol{0}}}{ (-\text{vol}(\text{T}^{+}))}\int\text{d}\boldsymbol{g}\left(-||\boldsymbol{g}||^2\right)\times\nonumber\\&\quad\times\prod_{i=1}^4\left(\sum_{p_i}\frac{p_i^2}{\text{vol}(\text{T}^{+})}\sum_{\substack{j_i,m_i;\\l_i,n_i}}D^{(p_i,0)}_{j_i m_i l_i n_i}(g_i)\right)  \frac{B^{p_1 p_2 p_3 p_4}_{l_1 n_1 l_2 n_2 l_3 n_3 l_4 n_4}
   }{\frac{1}{a^2}\sum_c \left(-\text{Cas}_{1,p_c}\right)+b_{\boldsymbol{p},\boldsymbol{j},\boldsymbol{m}}}\nonumber\\&=\frac{1}{2d_\text{nloc}}\frac{b_{\boldsymbol{1},\boldsymbol{0},\boldsymbol{0}}}{(-\text{vol}(\text{T}^{+}))}\prod_{i=1}^4\left(\int\sin^2{\left(\frac{t_i}{a}\right)}\frac{\text{d}t_i}{a}
   \sum_{p_i}\frac{p_i^2}{\text{vol}(\text{T}^{+})} \frac{1}{p_i}\frac{\sin{\left(\frac{p_i t_i}{a}\right)}}{\sin{\left(\frac{t_i}{a}\right)}}\right)\times\nonumber\\&\quad\times\frac{\left(-(t_1^2+...+t_4^2)\right)B^{p_1 p_2 p_3 p_4}_{00000000}
   }{\frac{1}{a^2}\sum_c \left(-\text{Cas}_{1,p_c}\right)+b_{\boldsymbol{p},\boldsymbol{0},\boldsymbol{0}}}.
\end{align}
To better understand, how different s-fold zero-modes contribute to this correlation length, it is expedient to employ the explicit decomposition of the correlator given in~\eqref{eqn:cofg} and~\eqref{eqn:cofgv2}, subject to the given symmetry assumptions, that is
\begin{align}\label{eq:secondmomentcorrelator1}
 \xinloc^2&=\frac{1}{2d_\text{nloc}}\frac{b_{\boldsymbol{1},\boldsymbol{0},\boldsymbol{0}}}{(-\text{vol}(\text{T}^{+}))}\left(\prod_{i=1}^4\int\sin^2{\left(\frac{t_i}{a}\right)}\frac{\text{d}t_i}{a}\right)\left(-\sum_{i=1}^{4} t_i^2\right)\times\nonumber\\&\quad\times\sum_{s=0}^4\sum_{(c_1,...,c_s)}\sum_{\substack{\vec{p}_{c_{s+1}},...,\vec{p}_{c_{r}}\neq(1,0,0) \\ \vec{p}_{c_{1}},...,\vec{p}_{c_{s}}=(1,0,0)}} \left(\prod_{i=1}^4\frac{p_i^2}{\text{vol}(\text{T}^{+})}\frac{1}{p_i}\frac{\sin{\left(\frac{p_i t_i}{a}\right)}}{\sin{\left(\frac{t_i}{a}\right)}}\right) \frac{B^{p_1 p_2 p_3 p_4}_{00000000}
   }{\frac{1}{a^2}\sum_c \left(-\text{Cas}_{1,p_c}\right)+b_{\boldsymbol{p},\boldsymbol{0},\boldsymbol{0}}}\,.
\end{align}
Now, we rearrange the sum $\boldsymbol{t}^2\equiv\sum_{i=1}^{4} t_i^2$ into two parts 
\begin{equation}
    \boldsymbol{t}^2=\boldsymbol{t}^2_s+\boldsymbol{t}^2_{4-s},
\end{equation}
which are associated to the modes $\vec{p}_{c_1},\dots,\vec{p}_{c_s}$ and $\vec{p}_{c_{s+1}},\dots,\vec{p}_{c_4}$, respectively. At first, we notice that the second line in~\eqref{eq:secondmomentcorrelator1} only depends on the latter $4-s$ variables, i.e. $t_{s+1},\dots t_{4}$, since the zero-modes were injected. Consequently, the contribution proportional to
\begin{align}
    \frac{1}{\text{vol}(\text{T}^{+})^s}\left(\prod_{i=1}^s\int\sin^2{\left(\frac{t_i}{a}\right)}\frac{\text{d}t_i}{a}\right)(\boldsymbol{t}^2_s)\left(\prod_{i=s+1}^4\int\sin^2{\left(\frac{t_i}{a}\right)}\frac{\text{d}t_i}{a}\frac{p_i^2}{\text{vol}(\text{T}^{+})}\frac{1}{p_i}\frac{\sin{\left(\frac{p_i t_i}{a}\right)}}{\sin{\left(\frac{t_i}{a}\right)}}\right)
\end{align}
therein gives $0$ for each $s<4$. This is due to the fact that while the integration over the $4-s$ variables yields Kronecker deltas $\delta_{p_i,1}$, by construction these vanish since the corresponding $p_i$ are always different from $1$. Secondly, we observe that when $s=4$, one obtains a contribution proportional to $a^2$ and in the same way as in~\cite{Marchetti:2021xvf}. We argue that it is in fact unphysical and should be subtracted to get a physical correlation length.\footnote{Likewise, if we work with the Wick back-rotated expression for this contributions, one yields a result proportional to $\cutoff^2 \text{vol}(\text{A}^{+}_{\cutoff})$.} Finally, we can scrutinize the contributions which concern the leftover $4-s$ variables. Since for these the effective mass is liberated from any Kronecker delta, the residual correlator effectively becomes that of a local theory on the remaining $4-s$-dimensional space. In this case, we can safely Wick back-rotate and decompactify the latter, so that with regard to the correlation length~\eqref{eq:secondmomentcorrelator1} we have to evaluate the contribution proportional to
\begin{align}
    \left(\prod_{u=s+1}^{4}\int\text{d}\rho_u\rho_u^2\right)&\frac{B^{\rho_{c_{s+1}}\dots \rho_{c_{4}}}_{00000000}
   }{\frac{1}{a^2}\sum_{v=s+1}^4 \text{Cas}_{1,\rho_v}+b_{c_1,...,c_s}}\times\nonumber\\&\times\left(\prod_{u=s+1}^{4}\int\sinh^2{\left(\frac{\eta_u}{a}\right)}\frac{\text{d}\eta_u}{a}\frac{1}{\rho_u}\frac{\sin{\left(\frac{\rho_u \eta_u}{a}\right)}}{\sinh{\left(\frac{\eta_u}{a}\right)}}\right)(\boldsymbol{\eta}^2_{4-s}).
\end{align}
To accomplish this, one first expresses $\boldsymbol{\eta}^2_{4-s}$ therein in terms of second derivatives with respect to the corresponding $\rho_{c_{s+1}},\dots,\rho_{c_4}$, then integrates out contributions over $\eta_{c_{s+1}},\dots,\eta_{c_4}$ which yields delta distributions (see~\eqref{eq:SL2Ctrick}) and finally changes the integration order, leading to
\begin{align}
    &\left(\prod_{u=s+1}^{4}\int\text{d}\rho_i\frac{1}{\rho_u}\delta(\rho_u-i)\right)\left(-a^2\sum_{u=s+1}^4 \partial^2_{\rho_u}\right)\frac{B^{\rho_{c_{s+1}}\dots \rho_{c_{4}}}_{00000000}\rho_{c_{s+1}}\cdot... \cdot\rho_{c_{4}}
   }{\frac{1}{a^2}\sum_{v=s+1}^4 \text{Cas}_{1,\rho_v}+b_{c_1,...,c_s}}\nonumber\\&\qquad\qquad=(4-s)\left(\frac{8 B^{iiii}}{a^2b_{c_1,...,c_s}^3}+\frac{6 B^{iiii}}{b_{c_1,...,c_s}^2}+\frac{4i B'^{iiii}}{b_{c_1,...,c_s}^2}+\frac{2ia^2 B'^{iiii}}{b_{c_1,...,c_s}}-\frac{a^2 B''^{iiii}}{b_{c_1,...,c_s}}\right).
\end{align}
Note that derivatives $\partial_{\rho_u}$ with respect to the residual intertwiner lead for each $u$ to the same result upon evaluation on $(\rho_{s+1},\dots,\rho_4)=(i,\dots,i)$ yielding the prefactor $(4-s)$. 
We also remark that the integration domain of the intertwiners extends here up to the cut-off~$\cutoff$. 
Bringing this together with the prefactor  $\frac{1}{2d_\text{nloc}} \frac{b_{\boldsymbol{1},\boldsymbol{0},\boldsymbol{0}}}{\text{vol}(\text{A}_{\cutoff}^{+})}$, gives for the correlation length two contributions which are independent of the regulator and two which do depend on it. 
In the same way as above for the case of the $s=4$ fold zero-mode, we argue that these are unphysical and play no role for the definition of the correlation length.%
\footnote{We remark that analogous arguments would hold in the simpler Abelian case with $G=\text{U}(1)$ or $\mathbb{R}$. There, the closure constraint in momentum space would simply yield a Kronecker or Dirac delta over the momenta which would lead to a similar unphysical contribution, see also~\cite{Marchetti:2021xvf}.
}. 
Hence, we have
\begin{equation}
    \xinloc^2=\sum_{s=s_0}^4\frac{(4-s_0)}{d_{\text{nloc}}}\sum_{(c_1,\dots,c_s)}b_{\boldsymbol{1},\boldsymbol{0},\boldsymbol{0}}\left(\frac{4}{a^2 b_{c_1,...,c_s}^3}+\frac{1}{ b_{c_1,...,c_s}^2}\right),
\end{equation}
and note that the first contribution in the bracket arises due to the hyperbolicity of the domain leading to the specific zero-mode structure discussed here. Clearly, in the limit $\mu\to 0$ the correlation length asymptotically behaves as
\begin{equation}\label{eq:correlationlengthnonlocalfinitea}
    \xinloc\propto(a\mu)^{-1},
\end{equation}
whereas in the flat limit, attained when first approaching the large $a$ regime, its asymptotics for $\mu\to 0$, are given by
\begin{equation}
    \xinloc\propto\mu^{-\frac{1}{2}}.
\end{equation}
Notably, these results agree with those obtained via studying the asymptotic behaviour of the correlation function in Section~\ref{sec:asymptoticanalysisnonlocal}, in spite of making different simplifying assumptions. 

\section{Ginzburg criterion}\label{section:Ginzburg Q}

If fluctuations of the order parameter $\Phi$ averaged over an appropriate region $\Omega$ are small compared to the order parameter $\Phi_0$ itself, averaged over that region, i.e. 
\begin{equation}\label{eq:Ginzburgmotivation}
    \left\langle(\delta\Phi)^2\right\rangle_\Omega\ll \left\langle\Phi_0^2\right\rangle_{\Omega}\,,
\end{equation}
then mean-field theory is self-consistently applicable. 
This is also known as the \textit{Ginzburg criterion}. 
Since the two-point function of the fluctuations is encoded by the correlation function~\eqref{eq:decompositioncorrelator}, the condition~\eqref{eq:Ginzburgmotivation}, as applied to the present context, is rewritten as 
\begin{equation}
    Q\equiv  \frac{\int_{\Omega} \text{d}^{4}g\, \text{d}X\, \text{d}^\ld \ff~C(\pmb{\ff},\pmb{g},X)}
    {\int_{\Omega}\text{d}^{4}g \, \text{d}X \, \text{d}^\ld\ff~ \gf_0^2}
\end{equation}
with $\vert Q\vert\ll 1$. One refers to the quantity $Q$ as the Ginzburg parameter.

It is important to specify the region $\Omega$ to be averaged over.
Since correlations are statistically relevant only up to distances of the order of the correlation length $\xi$, this region is parametrized by $\xi$, i.e. $\Omega\equiv\Omega_\xi$. 
Given that the local and non-local degrees of freedom enter differently into the dynamics of the models, it is sensible to distinguish two a priori independent parameters $\xiloc$ and $\xinloc$ and consider $\Omega_\xi$ as
\begin{equation}\label{eqn:integrationregion}
    \Omega_\xi = \Omega_{\xiloc}\times \Omega_{\xinloc} 
    \sim 
    [-\xiloc,\xiloc]^{\ld} \times \left(\SU \times A^+_{\xinloc} \times \SU\right)^4\times \hyper^3_\cutoff\,,
\end{equation}
meaning that the integration over the $\text{SL}(2,\mathbb{C})$-variables is performed over the whole compact components of $\text{SL}(2,\mathbb{C})$ characterizing its Cartan decomposition and with $\eta\le \xinloc$ and $\hyper^3_\cutoff$ is the regularized hyperboloid, parametrized on the non-compact direction by $0\le \eta\le 2\pi\cutoff$. 

The behaviour of $Q$ is of primary interest at the phase transition. 
A critical point is reached when $\mu\to 0$. 
Should $Q\gg 1$ at criticality, fluctuations are large. This would lead to an invalidation of Landau-Ginzburg mean-field theory. In contrast, if the Ginzburg parameter remains small, mean-field theory gives a trustworthy account of the system in the critical region.

The plan of this section is as follows. We start in Section~\ref{section:Ginzburgnonlocal} with the computation of $Q$ with respect to the non-local geometric variables only. The impact of the local degrees of freedom is then included in Section~\ref{section:Ginzburglocalnonlocal}.

\subsection{Ginzburg criterion for non-local variables}\label{section:Ginzburgnonlocal}

As we have seen in Section~\ref{sec:correlationfunctionnonlocal}, the correlation function for non-local geometric variables can be split in different contributions, characterized by the number $s$ of their zero-modes. 
We have also seen that, 
while in general their contribution to the correlation function are always exponentially suppressed regardless of the sign of the effective mass of each zero-mode, 
their weighted contributions are not. 
In particular we have seen (Sections~\ref{sec:asymptoticscorrelatorlocal}) that zero-modes with negative masses (i.e.\ with $s<s_0$) produce asymptotically diverging contributions. Consequently, these are long-range correlations that appear regardless of the physics of the phase transition. Following~\cite{Marchetti:2021xvf}, we thus choose to exclude them from the computation of the $Q$-integral.

In the following, we separately compute the denominator and the numerator of the $Q$-integral working with $\cutoff$-regularized $\text{SL}(2,\mathbb{C})$ data but in the large-$\cutoff$ limit which removes the regularization. 
We first limit ourselves to the case of a sum over quartic interactions, that is encoded by graphs $\gamma$ with $\nv_\gamma=4$, and generalize thereafter. 

Using~\eqref{eqn:minimumquartic} and considering $\xi\equiv \xinloc$, the denominator evaluates to
\begin{align}
%Q_D&\equiv
\int_{\Omega_{\xinloc}}\diff X \diff ^4 g\,\Phi_0^2 
&= \frac{-\mu}{\sum_\gamma {\lambda_{\gamma}}%{3!}
}\frac{\text{Vol}(\text{A}^+_\xi)^{4}}{\text{Vol}(\text{A}^+_\cutoff)^{6}}
%\textcolor{green}{\left(\frac{\text{Vol}(\text{A}^+_\xi)}{\text{Vol}(\text{A}^+_\cutoff)}\right)^1}
=\frac{-\mu}{\sum_\gamma {\lambda_{\gamma}}%{3!}
}\text{Vol}(\text{A}^+_\cutoff)^{-2}\left(\frac{\text{Vol}(\text{A}^+_\xi)}{\text{Vol}(\text{A}^+_\cutoff)}\right)^4 %\textcolor{green}{\left(\frac{\text{Vol}(\text{A}^+_\xi)}{\text{Vol}(\text{A}^+_\cutoff)}\right)^1}
,
\end{align}
where the integrations over the compact directions contribute with factors of unity. 
Let us now come to the numerator. In this case, we compute 
\begin{align}
%Q_N&\equiv
\int_{\Omega_{\xinloc}}\diff X\text{d}\boldsymbol{g}\,C(g_{1},\dots,g_{4},X) %\nonumber\\
&=\sum_{s=s_0}^4 \left(\frac{\text{Vol}(\text{A}^+_\xi)}{-\text{Vol}(\text{A}^+_\cutoff)}\right)^s\sum_{(c_1,\dots,c_s)}\int\diff^{4-s} g\,C_s(g_{c_1},\dots g_{c_{4-s}})
\end{align}
where we used~\eqref{equation:wickbackrotatedcorrelator} and that for any $s>s_0$ the correlation function is asymptotically exponentially decaying, so that we can extend the domain of integration over the whole $\text{SL}(2,\mathbb{C})^{4-s}$. The remaining integrals over the group variables can be computed in two ways. 
One can either use that, in the sense of hyperfunctions~\cite{Ruehl1970}, one has (see also~\eqref{eq:SL2Ctrick})
\begin{equation}
    \int_{\text{SL}(2,\mathbb{C})}\diff g D^{(\rho,0)}_{jmln}(g)=-\delta(\rho-i)\delta_{j,0}\delta_{m,0}\delta_{l,0}\delta_{n,0}\, .
\end{equation}
Alternatively, one may employ~\eqref{eqn:ggeomasympt} in the limit of small~$\mu$. Both lead to
\begin{equation}\label{eq:integrationoverresidualcorrelator}
\int\diff^{4-s} g\,C_s(g_{c_1},\dots g_{c_{4-s}})\sim \frac{\text{vol}(\text{A}^+_\cutoff)}{b_{c_1,\dots,c_s}}\,,
\end{equation}
wherein the $\text{vol}(\text{A}^+_\cutoff)$-factor stems from the BC intertwiner evaluated on $4$ zero-modes, i.e. $B^{iiii}_{00000000}=\text{vol}(\text{A}^+_\cutoff)$. We will further discuss the impact of the closure condition via the BC intertwiner in a moment. 
As a result, we have
\begin{align}\label{eqn:qintegralnonlocalstep1}
Q&\sim -\frac{\sum_{\gamma} {\lambda_{\gamma}}%{3!}
}{\mu}\text{vol}(\text{A}^+_\cutoff)^{3}\sum_{s=s_0}^4 \text{e}^{2(4-s)\frac{\cutoff-\xi}{a}}\sum_{(c_1,\dots, c_s)}\frac{1}{b_{c_1,\dots,c_s}}.
\end{align}
Introducing the functions
\begin{equation}
    f_s:= -\mu\sum_{(c_1,\dots, c_s)}\frac{1}{b_{c_1,\dots,c_s}},
\end{equation}
which is independent of $\mu$ and wherein the sum runs over all $s$-fold zero-modes $(c_1, ...,c_s)$, we yield
\begin{equation}
   Q\sim \frac{\sum_{\gamma} {\lambda_{\gamma}}%{3!}
   }{\mu^2}\text{vol}(\text{A}^+_\cutoff)^{3}\sum_{s=s_0}^4 f_{s}~\text{e}^{2(4-s)\frac{\cutoff-\xi}{a}}.
\end{equation}
As we are in the large-$\cutoff$ limit, the sum is dominated by the term involving $\text{e}^{-2(s-4)\cutoff/a}$ at $s=s_0$. That term precisely corresponds to the one which is least dominant in $\xi$ via the factor $\text{e}^{2(s_0-4)\xi/a}$. 
For the non-local geometric variables we found (at finite skirt radius $a$) for the correlation length $\xi\sim {1}/(a\mu)$, see for instance~\eqref{eq:correlationlengthnonlocalfinitea}.
Using that at large $\cutoff$ we have $\text{vol}(\text{A}^+_\cutoff) \sim \e^{2\cutoff/a}$ we then find
\begin{equation}
    Q\sim \sum_\gamma \lambda_{\gamma} \,\text{e}^{3\cdot 2 \cutoff/a} (\xi a)^2 f_{s_0} \text{e}^{(4-s_0)2\frac{\cutoff-\xi}{a}}
    = \xi^2\text{e}^{-(4-s_0)2\frac\xi{a}} \sum_\gamma \bar{\lambda}_\gamma
    %=\hat{\lambda}_{\gamma}
\end{equation}
with the rescaling in $a$ of the couplings $\bar{\lambda}_\gamma\equiv a^2 \text{e}^{(3+4-s_0)2\cutoff/a}\lambda_\gamma$. We see that the behaviour in $\xi$ is exponentially suppressed and dominated by the lowest physical zero-mode. 
In particular, the exponential suppression is always guaranteed since the case $s_0=4$ leads to unphysical contributions to the correlation length which we excluded, as discussed in Section~\ref{section:correlationlength}. This implies that the mean-field theory of such quartic models always yields a self-consistent description of the phase transition.

Our formalism permits the computation of $Q$ for interactions of any order, not merely of quartic order. 
Consider a single interaction of a graph $\gamma$ with $\nv_\gamma$ vertices which yields at least an $s_0$-fold zero-mode.
This gives rise to the non-trivial vacuum solution $\Phi_0$,~\eqref{eq:nonvvev}, such that 
\begin{align}\label{eq:GinzburgQnonlocal}
    Q&\sim \lambda_\gamma^{\frac{2}{\nv_\gamma-2}}\text{vol}(\text{A}^+_\cutoff)
    ^{2\left(\frac{\nv_\gamma-1}{\nv_\gamma-2}\right) + 4 -s_0 }
    %^{4+\frac{\nv_\gamma}{\nv_\gamma-2}-(s_0-1)}
   \text{vol}(\text{A}^+_{\xi})^{-(4-s_0)} \left(-\mu\right)^{-\frac{\nv_\gamma}{\nv_\gamma-2}}\nonumber\\&%\textcolor{red}{=\lambda^{\frac{2}{\nv_\gamma-2}}\text{vol}(\text{A}^+_\cutoff)^{2\left(\frac{\nv_\gamma-1}{\nv_\gamma-2}\right)}\left(\frac{\text{vol}(\text{A}^+_{\xi})}{\text{vol}(\text{A}^+_{\cutoff})}\right)^{s_0-4}\left(a\xi\right)^{\frac{\nv_\gamma}{\nv_\gamma-2}}}\nonumber\\&
    \propto \bar{\lambda}_\gamma^{\frac{2}{\nv_\gamma-2}}\xi^{\frac{\nv_\gamma}{\nv_\gamma-2}}\text{e}^{-2(4-s_0)\frac\xi{a}}
    =\hat{\lambda}_{\gamma}^{\frac{2}{\nv_\gamma-2}}
\end{align}
with the coupling rescaled in $a$
\begin{equation}\label{eq:rescalingcoupling}
    \bar{\lambda}_\gamma\equiv %a^{\frac{\nv_\gamma}{2}}\text{vol}(\text{A}^{+}_{\cutoff})^{\nv_\gamma-1+\frac{4-s_0}{2}(\nv_\gamma-2)}\lambda_\gamma\sim 
    a^{\frac{\nv_\gamma}{2}}\text{e}^{\left(\nv_\gamma-1 + (4-s_0)\frac{\nv_\gamma-2}{2}\right)\frac{2\cutoff}{a}}\lambda_\gamma.\footnote{Clearly, different regularization schemes could lead to results different to those obtained here. Note, however, that we absorbed the regulator-dependence of the result in terms of the cut-off $\cutoff$ into the coupling so that we yield a regulator-independent form of $Q$. Such a procedure is also applied in the local statistical field theory context, see~\cite{Kopietz:2010zz}. Hence, we do not expect our results to vary when a different regularization scheme is employed. We leave it to future investigations to scrutinize this expectation.}
\end{equation}

We remind the reader that in Landau-Ginzburg theory one typically focuses on selected interaction terms while one omits others. 
The possibly most dominant contributions have $s_0 = 0$ which corresponds to disconnected (multi-trace) interactions.
These are just disconnected fundamental melons. 
In that case, the result on the asymptotics of $Q$ is in accordance with what is obtained for a local quartic scalar field theory on the $3$-hyperboloid~\cite{Benedetti1403}. 
This is due to the fact that these multi-trace interactions can be reinterpreted as interactions of a local vector theory %, as explained in~
\cite{Rivasseau:2015ova}. 
One can also simply study the behaviour of $Q$ for subdominant interactions like the melonic ($s_0=1$) and that of necklace-type ($s_0=2$). However, to scrutinize the stability of an action containing such interactions along all scales, one has to resort to renormalization group analyses. 
Another relevant example is the simplicial interaction which has $\nv_\gamma=5$ and $s_0=3$. 

\paragraph{Flat limit.}
For infinite skirt radius $a$ the $3$-hyperboloid is turned into $\mathbb{R}^3$.%
\footnote{
Notice that this group corresponds to that of rotationless Galilean transformations (Galileian boosts). Interestingly, the non-relativistic limit of gravity theories in first-order formulation yields Newtonian gravity among others modified by torsion degrees of freedom~\cite{Banerjee:2016laq,Bergshoeff:2017dqq,Guerrieri:2020vhp}. We leave it to future research to clarify if the models considered here reduce to a quantum version of Newtonian models in the same flat limit.
} 
Taking this limit before computing the integrals in $Q$ reduces the exponential factors to simple powers. 
For instance, one finds $\text{vol}(\text{A}^+_\xi)\sim \frac4 3 (\xi/a)^3$ in this limit where the power of $3$ corresponds to the dimensionality of flat $3$-space. 
Using this and the relation $\xi_{\text{flat}}%\equiv\xi
\sim{\mu^{-1/2}}$ %, for the Ginzburg parameter, 
one has for a single interaction $\gamma$
\begin{equation}
Q\sim \lambda_{\gamma}^{\frac{2}{\nv_\gamma -2}}\frac{\xi^{\frac{2 \nv_\gamma }{\nv_\gamma -2 }-3(4-s_0)}}{\cutoff^{3(s_0-4)}
}.
\end{equation}
%As expected, t
This result is fully consistent with %what was obtained in 
our previous work for $G=\mathbb{R}$, see~\cite{Marchetti:2021xvf}, only that the cut-off $\cutoff$ used here plays the role of $a$ used in~\cite{Marchetti:2021xvf}. 
As before, there is a critical dimension $d_{\textrm{crit}}=\frac{2\nv_\gamma}{\nv_\gamma -2}$ 
but it is the effective dimension $3(4-s_0)$ of the model which needs to be greater than $d_{\textrm{crit}}$ for fluctuations to be small and mean-field theory to be self consistent.
For quartic interactions, $d_{\textrm{crit}}=4$ such that interactions with $s_0=0,1,2$ meet this criterion while those with $s_0=3$ do not. 
In particular,  a model with only quintic simplicial interaction has $s_0=3$ and thus effective dimension $3(4-3)=3$ which is still below the critical dimension $d_{\textrm{crit}}=2\cdot 5/(5-2)=10/3$ of order-five interactions; thus mean-field theory does not apply in the large-$a$ limit of such model.

\paragraph{On the imposition of the closure constraint.}

As commented on after~\eqref{eq:integrationoverresidualcorrelator}, the effect of the BC intertwiner and thus of the closure constraint onto the numerator of the Ginzburg parameter leads to a factor $\text{vol}(\text{A}_{\cutoff}^{+})$. It arises since the remaining $4-s$ integrations over $\mathrm{SL}(2,\mathbb{C})$ lock in the corresponding representation parameters to the complex unit in the intertwiner. Since from the denominator we obtain a complementary factor due to the integration over the normal $X$, these volume factors cancel each other. Alternatively, we could have also proceeded with $4-s-1$ group integrations, which would have yielded for the intertwiner $B^{iii\rho}_{00000000}$, that is a delta distribution on the last representation parameter shifted by the complex unit. Integration over the respective parameter would then simply lead to a group integration with the result $\text{vol}(\text{A}_{\xi}^{+})$ and thus one additional zero-mode. This would be in line with our results that for models on $G=\text{U}(1)$ or $\mathbb{R}$ where the imposition of the closure constraint leads to a shift $s_0\to s_0 + 1$ or alternatively for the rank $r\to r-1$ in the final result for the $Q$-parameter~\cite{Marchetti:2021xvf}. However, in our models, where the domain of the TGFT field is explicitly extended by one additional slot to account for the timelike normal $X$ over which we have to average, a complementary factor arises also in the denominator leading to its overall cancellation. Consequently,~\eqref{eq:integrationoverresidualcorrelator} already fully accounts for the impact of the closure condition via the BC intertwiner, the explicit representation of which as a function of $\rho_1,...,\rho_4$ (discussed in detail in~\cite{Barrett:1999qw}) is not required for this argument, and thus no shift in $s_0$ or the rank occurs. We stress that this is simply a feature of the extended formalism used to impose the geometricity constraints, and it does not imply a lack of imposition of the closure constraint.

In contrast, in models with no imposition of simplicity and no extension with a normal, an example of which would be the Ooguri model amended by a non-trivial kinetic term as specified in~\eqref{eq:TGFTaction}, a shift $s_0\to s_0+1$ in~\eqref{eq:GinzburgQnonlocal} and~\eqref{eq:rescalingcoupling} does in fact occur due to the closure condition. This is due to the fact that the denominator cannot produce a corresponding factor since there is simply no integration over a normal present. In addition, the absence of such integrations would also affect the rescaling of the coupling with respect to factors in $\text{vol}(\text{A}_{\cutoff}^{+})$ as compared to~\eqref{eq:rescalingcoupling}, yielding 
\begin{equation}
    \bar{\lambda}_\gamma\equiv %a^{\frac{\nv_\gamma}{2}}\text{vol}(\text{A}^{+}_{\cutoff})^{\nv_\gamma-1+\frac{4-s_0}{2}(\nv_\gamma-2)}\lambda_\gamma\sim 
    a^{\frac{\nv_\gamma}{2}}\text{e}^{2(\frac{\nv_\gamma-2}{2}+\frac{4-(s_0+1)}{2}(\nv_\gamma-2))\frac\cutoff{a}}\lambda_\gamma.
\end{equation}
Moreover, these arguments would also apply to the case of the older versions of the BC model which do not possess an extended domain and which differ in the way simplicity is imposed together with the closure condition, see~\cite{rovelli2004quantum} for a discussion. The computation of the Ginzburg parameter then leads to the same result as compared to those models where no simplicity is imposed at all, thus showing a degree of universality.

\paragraph{Note on rescaling of couplings and consequences for renormalization group analyses.}

In local statistical field theory it is possible to identify %determine 
from the scaling of $Q$ with correlation length $\xi$ the canonical dimension of couplings which for local interactions agrees with their scaling dimension with respect to scaling with the momentum scale $k\sim 1/ \xi$ around the Gaussian fixed point. 
In our previous work~\cite{Marchetti:2021xvf}, we found that the result of the $Q$ calculation delivers the correct scaling dimension as known from perturbative renormalization~\cite{BenGeloun:2014kpa}.
In the present context, the rescaling of the coupling takes a more unusual form with an exponential factor due to the hyperbolicity of the Lorentz group, that is rewriting~\eqref{eq:rescalingcoupling} in $k\sim 1/\xi$ as
\begin{equation}
    \hat{\lambda}_{\gamma}= k^{-\frac{\nv_\gamma}{2}}\text{e}^{-(\nv_\gamma-2)\frac{4-s_0}{k a}} \bar{\lambda}_\gamma
    % =\lambda_\gamma a^{\frac{\nv_\gamma}{2}}\text{e}^{2(\nv_\gamma-1+\frac{4-s_0}{2}(\nv_\gamma-2))\cutoff/a}k^{-\frac{\nv_\gamma}{2}}\text{e}^{-(\nv_\gamma-2)(4-s_0)/(k a)}.
\end{equation}
From this rescaling we deduce that the scaling dimension for the present models is effectively infinite. This scaling relation will be crucial in upcoming functional renormalization group studies of the models treated here.

However, given these results, we can already put forward the conjecture that the Gaussian fixed point is the only one in the phase diagram of TGFTs on hyperbolic spaces.
In~\cite{Benedetti1403} this was shown for a local quartic scalar field theory on the $3$-hyperboloid
by means of an FRG analysis in the so-called local potential approximation (LPA).
We expect this to be the case also for our TGFT models the qualitative behaviour of the  Ginzburg parameter $Q$ is predominantly determined by the exponential decay (per $r=4>s_0$)
and the specific effect of combinatorial non-locality (that is the occurrence of the $s_0$, the number of zero modes) only has a subordinate impact.
Previous results in the cyclic-melonic LPA$'$~\cite{Pithis:2020sxm,Pithis:2020kio} on $G=\mathbb{R}$ suggest that there may exist non-Gaussian fixed points even for finite dimensions beyond the critical dimension, due to the specific flow of the anomalous dimension in TGFT.
However, in the hyperbolic case here, the (scaling) dimension becomes effectively infinite in the IR such that any such possible fixed point vanishes in the IR limit. 
We leave the clarification of these matters to future investigations.

\subsection{Ginzburg criterion for local and non-local variables}\label{section:Ginzburglocalnonlocal}

The inclusion of the local degrees of freedom is straightforward at this point. 

We reinstate the differentiation between $\xiloc$ and $\xinloc$. For general interactions with graph~$\gamma$ and valency $\nv_\gamma$ we have for the denominator of the $Q$-integral
\[
%Q_D&=
\int_{\Omega_{\xiloc}}\d\boldsymbol{\phi}\int_{\Omega_{\xinloc}}\d X \diff ^4 g\,\Phi_0^2 %\nonumber\\
~\sim~ \text{vol}(\text{A}_{\cutoff}^+)^{1-2\left(\frac{\nv_\gamma-1}{\nv_\gamma-2}\right)}\left(\frac{\text{vol}(\text{A}^+_{\xinloc})}{\text{vol}(\text{A}^+_\cutoff)}\right)^{4}\left(\frac{-\mu}{\lambda_{\gamma}}\right)^{\frac{2}{\nv_\gamma-2}}
\]
For the numerator, instead, the integral over the local degrees of freedom can be easily extended over the whole domain since the correlation function shows exponentially decaying properties for scales larger than $\xiloc$, as demonstrated in Section~\ref{section:correlationlength}. Hence, the integration over the local degrees of freedom $\phi_i$ leads to delta distributions $\delta(k_i)$ over the corresponding momenta. As a result, we obtain exactly $C(\boldsymbol{g)}$ as if it had been independent of the local degrees of freedom from the outset. Consequently, the integration over the group variables in $Q_N$ goes then through as in Section~\ref{section:Ginzburgnonlocal}. In turn, this gives for the Ginzburg parameter 
\begin{align}
    Q&\sim \bar{\lambda}_{\gamma}^{\frac{2}{\nv_\gamma-2}}\xiloc^{-\ld}\xinloc^{\frac{\nv_\gamma}{\nv_\gamma-2}}\text{e}^{-2(4-s_0)\xinloc/a}\sim
    \bar{\bar{{\lambda}}}_{\gamma}^{\frac{2}{\nv_\gamma-2}}\xinloc^{\frac{\nv_\gamma}{\nv_\gamma-2}-\frac{\ld}{2}}\text{e}^{-2(4-s_0)\xinloc/a} %=\bar{\lambda}_{\gamma}^{\frac{2}{\nv_\gamma-2}},
\end{align}
where we used in the second step that $\xiloc^2\sim a\xinloc$ and applied the rescaled coupling $\bar{\lambda}_\gamma$ as in~\eqref{eq:rescalingcoupling} with $\bar{\bar{\lambda}}_\gamma\equiv a^{-\frac{\ld}{2}}\bar{\lambda}_\gamma$. Compared to the evaluation of $Q$ solely for the non-local geometric degrees of freedom in~\eqref{eq:GinzburgQnonlocal}, we observe that the local degrees of freedom lead to a mild suppression of the linear term in $\xi$. However, this is still rather irrelevant compared to the dominant exponential suppression stemming from the non-local degrees of freedom.

\paragraph{Flat limit.}
In the large $a$ limit, we note that $\xiloc=\xinloc\equiv\xi$ and we retrieve for the Ginzburg parameter
\begin{equation}
    Q\sim \lambda_{\gamma}^{\frac{2}{\nv_\gamma -2}}\frac{\xi^{\frac{2 \nv_\gamma }{\nv_\gamma -2 }-\ld-3(4-s_0)}}{\cutoff^{3(s_0-4)}}.
\end{equation}
Again, this agrees with the result obtained for $G=\mathbb{R}$ in~\cite{Marchetti:2021xvf} where the cut-off $\cutoff$ used here corresponds to $a$ therein.

\section{Discussion and conclusions}\label{sec:discussionandconclusion}

The central objective of this article was to investigate the phase structure and realization of phase transitions of Lorentzian TGFT models for quantum geometry minimally coupled with massless and free scalar fields, using Landau-Ginzburg mean-field theory.

We have transferred over the Landau-Ginzburg method from the local statistical field theory context, building on our recent work~\cite{Marchetti:2021xvf}, and geared it towards rank-$4$ TGFTs which generate Lorentzian quantum geometries. We have mostly worked in the context of the Lorentzian BC model and related models with tensor-invariant interactions, based on spacelike tetrahedra only. Due to the Lorentzian signature, the combinatorial non-locality of the interactions and the geometricity constraints required a thorough regularization scheme. This was built on an analytic continuation and compactification of the Lorentz group to $\text{Spin}(4)$, as used also in~\cite{Dona:2021ldn}. 

We have computed the correlation function and correlation length for such models and then evaluated, using them, the Ginzburg parameter.
The main result is that, due to the non-compactness and hyperbolic nature of the Lorentz group, we can always find a transition towards a phase with non-vanishing expectation value of the field (operator) and that this phase transition is always self-consistently described in terms of mean-field theory. 
Since such configurations are highly populated by GFT quanta, this is evidence for the existence of an interesting continuum geometric approximation to be studied in mean-field language in such TGFTs. %\footnote{} 
Such phases had so far only been conjectured to exist for Lorentzian GFT models and had been used as a working hypothesis for the TGFT condensate cosmology program~\cite{Gielen:2016dss,Oriti:2016acw,Pithis:2019tvp,Oriti:2021oux}, where cosmological dynamics is also extracted from the TGFT mean-field hydrodynamics, albeit around non-uniform field configurations. The latter approach receives therefore an important, if indirect, support from our results~\cite{Marchetti:2022nrf}.

In particular, we note that the expectation value of the number operator becomes infinite for the non-vanishing ground state expectation value of the TGFT field (operator) used here. 
Consequently, the occupation number is not a meaningful observable anymore. In fact, the system is then described by means of a non-Fock representation (of the canonical commutation relations)~\cite{Pithis:2016wzf, Pithis:2019mlv,Kegeles:2017ems,Kegeles:2018tyo} in close analogy to the local QFT context~\cite{strocchi2005symmetry}. Such representations are required to describe many particle systems in the thermodynamic limit. Only then inequivalent irreducible representations of the canonical commutation relations become available which correspond to different equilibrium states, i.e. phases of the system. Hence, such representations are necessary to consistently describe phase transitions in terms of quantum field theory. Moreover, in quantum optics it is well-known that states with an infinite number of photons, so-called non-Fock (coherent) states, give rise to the description of the system in terms of a classical radiation field~\cite{strocchi2005symmetry,honegger2015photons}. For these reasons, we deem the occurrence of such a representation an important prerequisite to capture continuum macroscopic information and to classicalize TGFT models.

Beyond these points, our analysis allows to slightly generalize our findings. If one keeps the rank~$r$ unspecified, one finds that the critical rank is in principle infinite %zero.
, essentially due to the exponential suppression factor caused by the hyperbolicity of the Lorentz group. This means that mean-field theory gives an accurate account of the phase properties of such models for any rank and any valency of interaction (the combinatorics of which are encoded by the minimum number of zero-modes $s_0$). Moreover, in agreement with the literature~\cite{Carrozza:2013oiy,Carrozza:2016vsq,Bonzom:2011zz,Gurau:2011tj,Gurau:2013cbh} our results demonstrate that melonic interactions propel the critical behaviour of models defined by different types of interactions (as they are the most divergent ones in terms of the cut-off $\cutoff$). Lastly, given the exponential suppression factor, the impact of the local degrees of freedom is not such as to affect the critical behaviour of the overall system, at finite skirt radius $a$. 

In the flat limit, where $a\to \infty$, on the other hand, we find that the mean-field critical behaviour is effectively the one of a local scalar field theory on $\mathbb{R}^{\ld+3(4-s_0)}$, consistently with our previous work~\cite{Marchetti:2021xvf}. Hence, the presence of the local degrees of freedom, enhances the validity of mean-field theory at criticality, in that limit. 

We emphasize that despite the simplifying assumptions of the mean-field setting (in particular the projection onto uniform field configurations), our results are in agreement with those obtained with more involved FRG studies of TGFTs~\cite{Pithis:2020kio}. 
This confirms the usefulness of this method, which can to some degree 
(in the Gaussian regime) 
effectively bypass the much more involved non-perturbative RG treatments of TGFTs.

In the end, we believe that our work constitutes an important step towards understanding the continuum phase properties of realistic interacting quantum matter-quantum geometry systems from the TGFT perspective.

\

In the following, we conclude by discussing shortcomings and possible future extensions of this work. In our analysis we restricted attention to 4d quantum geometric models with spacelike tetrahedra only. In recent work~\cite{Jercher:2022mky}, configurations with timelike and lightlike tetrahedra were included in a general formulation of the BC model. The analysis of the phase structure of such more general models (potentially including additional local causality conditions in their interactions, see also~\cite{Jordan:2013awa,Loll:2015yaa,Asante:2021phx}) should also be accessible via the Landau-Ginzburg method. In particular, we expect that the timelike configurations lead to a similar exponential suppression factor in the computation of the Ginzburg parameter $Q$. This is due to the fact that for these configurations the quantum geometry is encoded via the homogeneous space $\mathrm{SL}(2,\mathbb{C})/\text{SU}(1,1)\cong \hyper^{2,1}$ (the one-sheeted $3$-hyperboloid) the integration measure of which is proportional to $\cosh^2{(\eta)}\text{d}\eta$. 

In contrast, the EPRL TGFT model with spacelike tetrahedra~\cite{Oriti:2016qtz} can be written using the compact Lie group $\mathrm{SU}(2)$ in the domain (relegating the non-compact Lorentz data in the expression for the kinetic and interaction kernels). Hence, one would expect that, in order to obtain an interesting phase structure, non-compact directions should be added to the model (unless the non-compact data in the kernels modify substantially the analysis). This could be realized using the local $\mathbb{R}$-valued directions considered also here or generalizing the model to include timelike configurations (using the Conrady-Hnybida extension~\cite{Conrady:2010kc,Conrady:2010vx} of the EPRL model~\cite{Rovelli:2011eq,Perez:2012wv}), the geometric data of which require to include also the non-compact Lie group $\text{SU}(1,1)$ in the TGFT field domain. For a TGFT-toy model on the latter space, it has already been demonstrated that mean-field theory is sufficient to describe the phase structure~\cite{Pithis:2018bw,Pithis:2019mlv}, again, because an exponential suppression factor appears due to the hyperbolicity of the domain. Note, however, that the needed regularization scheme affects the results of the mean-field analysis. Whether this implies that the BC and EPRL-like GFT model lie in the same universality class from the point of view of continuum gravitational physics, as conjectured in~\cite{Dittrich:2021kzs,Jercher:2021bie}, is left to be clarified by future investigations. 

In passing, we note that the representations with $\rho=\pm i$ (defined in the sense of hyperfunctions~\cite{Ruehl1970}) were important to determine the phase structure and in particular of the critical behavior of the model considered here. The circumstance that they seem to correspond to lightlike bivectors is tantalizing and calls to understand the physics of the critical regime better.

It will certainly be important to supplement the approximation method we employed by non-perturbative renormalization group techniques (e.g. functional renormalization group methods~\cite{Berges:2002ga,Dupuis:2020fhh}) so that the effect of quantum fluctuations on all scales onto the critical properties of the models can be taken into account. 
There are several indications from FRG analyses of TGFTs~\cite{Benedetti:2015et,BenGeloun:2015ej,BenGeloun:2016kw,BenGeloun:2018ekd,Pithis:2020kio} that the mean-field account gets refined in interesting ways when fluctuations on length scales smaller than $\xi$ get involved. 
These are characterized by the wave function renormalization yielding the anomalous dimension. In particular, this is because of a tensor specific flow of this quantity (see also~\cite{Eichhorn:2017xhy,Eichhorn:2018ylk} for pure tensor models), which can lead to drastic consequences for the phase structure already in the flat case where $G=\mathbb{R}$~\cite{Pithis:2020kio}. 

Another important extension of our analysis would be to go beyond the projection onto uniform field configurations and to study quadratic fluctuations around non-uniform background configurations (which still minimize the action). The extraction of such configurations is arduous but has been explored for related GFT models, see for example~\cite{Fairbairn:2007sv,Livine:2011yb,BenGeloun:2018eoe}. Such groundwork could thus be used to check the robustness of and to enhance our results at the mean-field level. Notice that the majority of works on non-perturbative renormalization of TGFT models has in fact been conducted without making use of the projection onto uniform field configurations, see~\cite{Benedetti:2015et,Benedetti:2016db,BenGeloun:2016kw,Carrozza:2016tih,Carrozza:2017vkz,BenGeloun:2018ekd}. The latter washes out combinatorial subtleties between interaction terms of the same valency (while it retains other essential non-local information) with the trade-off of the obvious greater simplicity. Going beyond this limitation, by appreciating the difference between all interactions with the same valency at increasing order, is going to be technically challenging for full-grown TGFT models for 4d Lorentzian quantum gravity. Crossing this barrier, however, will ultimately be necessary. 
Concerning the matter coupling to the non-local geometric content of our models, we assumed it to be minimal, as encoded by the factor $\alpha$ in the kinetic operator which we then additionally set to a constant. These assumptions may be in fact the reason why the effect of the matter content is not more pronounced in the expression for the Ginzburg parameter. We expect this to change via the stepwise relaxation of these assumptions and in particular when the impact of more realistic (like for instance massive, self-interacting and non-minimally coupled) matter degrees of freedom is considered.

\subsection*{Acknowledgements}
The authors thank D. Benedetti, P. Dona and A. Jercher for discussions and an anonymous referee for comments which led to an improvement of this manuscript.

D. Oriti and A. Pithis acknowledge funding from DFG research grants OR432/3-1 and OR432/4-1.
The work of A. Pithis leading to this publication was also supported by the PRIME programme of the German Academic Exchange Service (DAAD) with funds from the German Federal Ministry of Education and Research (BMBF). A. Pithis also thanks for the generous financial support by the MCQST via the seed funding Aost 862983-4.
The work of J. Th\"{u}rigen was funded by the Deutsche Forschungsgemeinschaft (DFG, German Research Foundation) in two ways,
primarily under the author's project number 418838388 and
furthermore under Germany's Excellence Strategy EXC 2044--390685587, Mathematics M\"unster: Dynamics–Geometry–Structure. The work of L. Marchetti was funded by Fondazione Angelo Della Riccia. L. Marchetti also thanks the Ludwig Maximilians-Universit\"at (LMU) Munich for the hospitality.

\appendix

\section{Harmonic analysis on $\mathrm{SL}(2,\mathbb{C})$}\label{appendix:harmonic analysis on SL2C}

In this appendix we summarize important facts on the group structure, representation theory and harmonic analysis of $\mathrm{SL}(2,\mathbb{C})$, the special linear group of degree $2$ over the complex numbers. It is a double cover of the restricted Lorentz group $\mathrm{SO}^{+}(1,3)$ which is an isometry group of Minkowski spacetime.
The latter comprises of those Lorentz transformations which preserve the orientation of space, time and spacetime. The following presentation is based on~\cite{gel1950unitary,gel1963representations,naimark1964linear,duc1967theory,gelfand1968representation,Ruehl1970,knapp2016representation,gel2016generalized1,gel2016generalized2,turski1998harmonic} needed for the main sections of this article. A nice panorama on these matters can also be found in~\cite{Martin-Dussaud:2019ypf}.

\subsection{Basic group structure of $\mathrm{SL}(2,\mathbb{C})$}\label{appendix:SL2Cgroupstructure}

The non-compact, simple and simply connected six-dimensional Lie
group $G=\mathrm{SL}(2,\mathbb{C})$ is the group of $2\times 2$ complex matrices of determinant $1$, i.e.,
\begin{equation}
\mathrm{SL}(2,\mathbb{C})=\left\{g = \mqty(a & b\\ c & d)\; \Bigg{\vert} \; \{a,b,c,d\}\in\C,\; \det(g) = 1\right\},
\end{equation} 
wherein the group multiplication corresponds to matrix multiplication. The corresponding Lie algebra $\mathfrak{sl}(2,\mathbb{C})$ is the vector space of traceless $2\times 2$ complex matrices together with the Lie bracket being the commutator. Its basis is formed by the generators of boosts $K_i \equiv \frac{i}{2}\sigma_i$ and of rotations $L_i \equiv \frac{1}{2}\sigma_i$, where $\sigma_i$ with $i=1,2,3$ denote the Pauli matrices. These satisfy the commutation relations
\begin{equation}\label{eq:commutationrelationsSL}
[K_i,K_j] = -i\varepsilon_{ijk}L_k,\quad [L_i,K_j] = i\varepsilon_{ijk}K_k,\quad [L_i,L_j] = i \varepsilon_{ijk}L_k.
\end{equation}

\begin{figure}[htp]
\centering
\includegraphics[width=.45\textwidth]{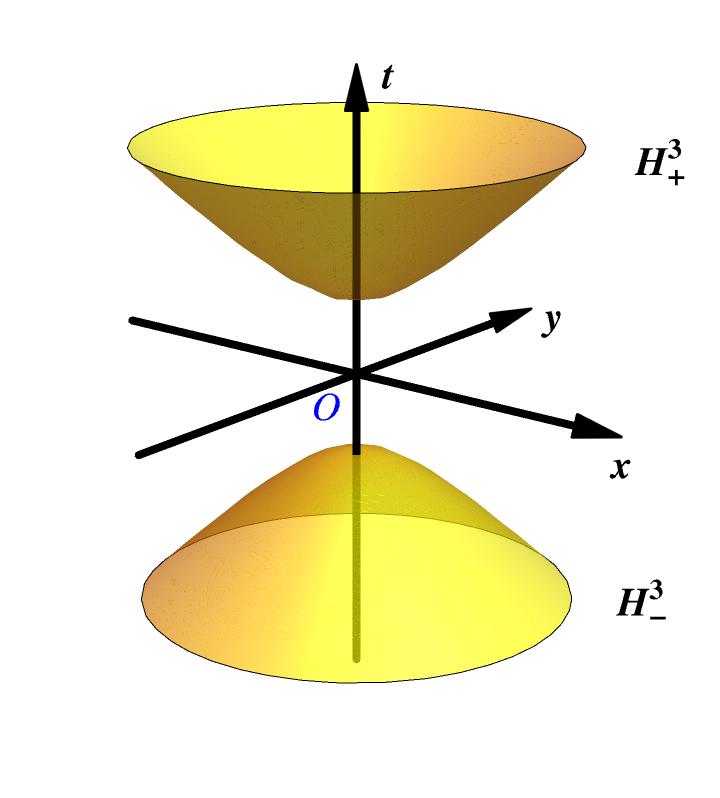}\hfill
\includegraphics[width=.45\textwidth]{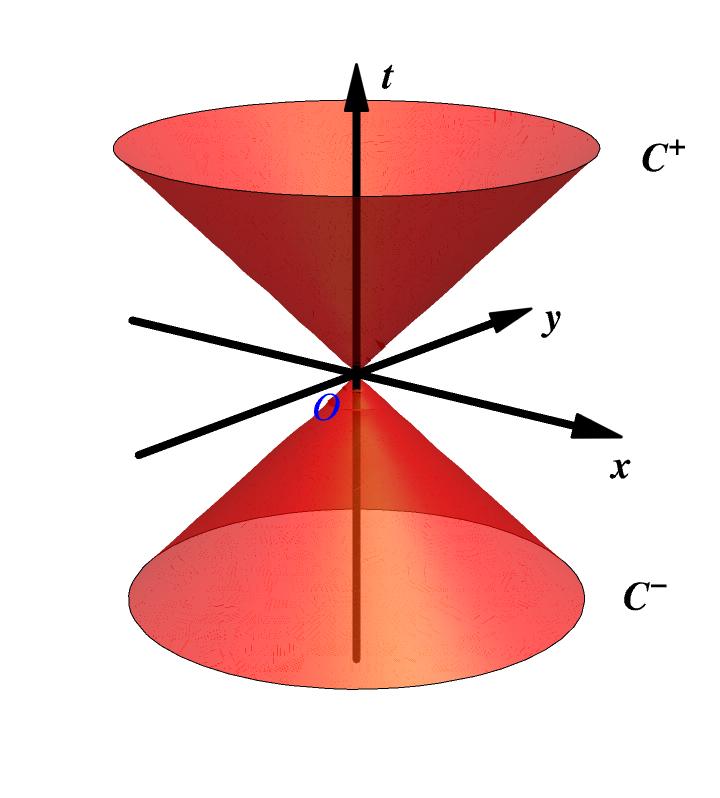}\hfill
\caption{Two distinguished hypersurfaces in Minkowski space from the perspective of a chosen observer at $O$. Left panel: The two-sheeted hyperboloid defined by $x^\mu x_\mu=1$ with $x\in \mathbb{R}^{1,3}$. Right panel: The lightcone defined via $x^{\mu}x_{\mu}=0$. The different hyperbolic shells $\hyper^{3}_{\pm}$ and corresponding sectors of the light cone can be selected by choosing either $x_0>0$ or $x_0<0$, respectively. We remark that the $z$-axis is suppressed to yield $3d$ pictures and that the so-called skirt radius $a$ of the hyperboloid is set to unity therein.}
\label{fig:hypersurfaces}
\end{figure}

$\SL$ acts transitively on the homogeneous space $\SL/\SU\cong \hyper^{3}$ which is the $2$-sheeted hyperboloid of unit-norm timelike vectors in Minkowski space (with signature of the metric $(+,-,-,-)$), depicted in Fig.~\ref{fig:hypersurfaces} hereafter. 
Let $X%\equiv [b]
\in\hyper^{3}$. %for $b\in\SL$ a  representative 
With this, the stabilizer subgroup $\SU_X\subset\SL$ is defined as
\begin{equation}\label{eq:invariance under seperate SU2}
\SU_X
=
\left\{ h\in\SL \; \vert \; h\cdot X = X\right\}.
\end{equation}

The Cartan decomposition of $\mathrm{SL}(2,\mathbb{C})$ is given by
\begin{equation}\label{eq:Cartan decomposition of g}
\forall g\in \mathrm{SL}(2,\mathbb{C})\;\exists\; u,v\in\mathrm{SU}(2),\eta\in \R_{+}: \quad
g = u \,\e^{\frac{\eta}{2}\sigma_3}\,v^{-1}
\end{equation}
and $\eta$ is known as the boost parameter. This is also conveniently written es $G=K \text{A}^{+} K$ with $K=\mathrm{SU}(2)$ and $\text{A}^{+}=\{\e^{\frac{\eta}{2}\sigma_3}: \eta\in \R_{+}\}$. The Haar measure $\text{d}g$ on $G$ upon Cartan decomposition is 
\begin{equation}\label{eq:HaarmeasureSL2C}
    \d g=\sinh^2(\eta)\d\eta \,\d u\, \d v,
\end{equation}
wherein $\text{d}u$ and $\text{d}v$ are Haar measures on $K=\mathrm{SU}(2)$. 
In these coordinates the round metric on the %above-introduced 
hyperboloid $\hyper^{3}\cong \SL/\SU$ is
\begin{equation}
    \text{d} H^2=\text{d}\eta^2+\sinh^2(\eta)\text{d}\Omega_2,
\end{equation}
wherein $\text{d}\Omega_2$ corresponds to the metric element on the $2$-sphere leading to the measure
\begin{equation}
    \d X=\sinh^2(\eta) \d\eta \,\d\Omega_2.
\end{equation}
Notice that $\text{d}g$ (and thus $\text{d}X$) is oftentimes normalized by $4\pi$ in the literature. 
In this work, we consider the compact $\mathrm{SU}(2)$ contributions to be normalized to $1$ 
and note that due to the infiniteness of the $\mathrm{SL}(2,\mathbb{C})$ group volume one can in principle normalize the overall Haar measure at will. 
As a minimal choice, we set it to $1$. 
We also remind the reader that in the main article we instate the skirt radius of the hyperboloid $a$ (such that $x^{\mu}x_{\mu}=a^2$) which essentially leads to a multiplication of the Haar measure with $a^3$ and to replacing $\eta$ with $\frac{\eta}{a}$ therein. 
In this appendix we refrain from the introduction of this scale and keep the exposition as close as possible to the literature where the radius is usually set to unity.

Together with the Haar measure %on $G$ 
we obtain the space of square-integrable functions on $\mathrm{SL}(2,\mathbb{C})$, denoted by $L^2(\mathrm{SL}(2,\mathbb{C}))$. 
We briefly remark here that to fully mount the Landau-Ginzburg analysis on the main part of this article, one has to extend the space of function to that of hyperfunctions~\cite{Ruehl1970,hormander2015analysis}, see also the next subsection~\ref{appendix:SL2Charmonic}. This is in complete analogy with what is done in the same type of analysis for local field theories on $d$-dimensional Euclidean space where one extends the space of functions to distributions.

\subsection{Harmonic analysis on $\mathrm{SL}(2,\mathbb{C})$}\label{appendix:SL2Charmonic}

In the following, we devote our attention to a synopsis on the unitary irreducible representations of $\SL$. 
For a probability interpretation we are interested in unitary representations. 
Apart from the trivial representation, this leaves us with the infinite-dimensional irreducible representations of $\SL$ which constitute the principal and the complementary %/supplementary 
series~\cite{gelfand1968representation}.

The infinite-dimensional representations can be realized over the space of homogeneous functions $f(z_0,z_1)$ with $z_0,z_1\in \mathbb{C}$ of degree $(\lambda,\mu)\in\mathbb{C}^2$ such that $\forall \alpha\in\mathbb{C}$ one has
\begin{equation}
    f(\alpha z_0, \alpha z_1)=\alpha^{\lambda}\bar{\alpha}^{\mu}f(z_0,z_1)
\end{equation}
and $\mu-\lambda\in\mathbb{Z}$ is required such that $\alpha^{\lambda}\bar{\alpha}^{\mu}$ is a single-valued function of $\alpha$. 
%Let $\mathcal{D}^{(\lambda,\mu)}[z_0,z_1]$ 
Consider the subspace of homogeneous functions of degree $(\lambda,\mu)$ which are infinitely differentiable over $\mathbb{C}^2 \backslash \{0\}$ in $z_0,z_1, \bar{z}_0$ and $\bar{z}_1$. 
A continuous representation of $g\in \mathrm{SL}(2,\mathbb{C})$ over this space is then defined by the action
\begin{equation}
    \left(\boldsymbol{D}^{(\lambda,\mu)}(g)f\right)(z_0,z_1)=f(g^T (z_0,z_1))=f(a z_0 + c z_1, b z_0+d z_1).
\end{equation}
A simple way to (projectively) realize these is for functions $\phi(z)=f(z,1)$ such that 
\begin{equation}
    \left(\boldsymbol{D}^{(\lambda,\mu)}(g)\phi\right)(z)= (b z + d)^{\lambda}(\overline{b z + d})^{\mu}\phi\left(\frac{a z +c}{b z +d}\right).
\end{equation}
Subsequently, we use the change of parameters $(\lambda,\mu)=(i\rho+\nu-1,i\rho-\nu-1)$.

%Notice that T
The only unitary irreducible representations of $\SL$~\cite{knapp2016representation} are,
up to unitary equivalence, the trivial representation, the unitary principal series and the complementary series. 
Elements of the complementary series are labelled by imaginary values $i\rho\in \,]{-1,0}[ %\cup ]0,1[
$ and $\nu=0$. In contrast, the unitary irreducible representation spaces in the principal series, which are the main focus in this work, are defined by pairs $(\rho,\nu)\in\R\times\mathbb{Z}/2$ in the space $L^2(\mathbb{C})$ with scalar product
\begin{equation}
    (\phi,\psi)=\int \d z \,  \bar{\phi}(z)\psi(z),
\end{equation}
wherein the measure is given by $\text{d}z=\text{d}\Re{(z)}\text{d}\Im{(z)}$. These representations are unitarily equivalent for $(\rho,\nu)$ and $(-\rho,-\nu)$. Note that the non-unitary principal series encompasses all the irreducible finite-dimensional representations $(\lambda,\mu)$ of $\SL$ as subrepresentations. The polynomials are then no more homogeneous and are  of degree $\leq \lambda$ in $z$ and $\leq \mu$ in $\bar{z}$~\cite{knapp2016representation}.

%Importantly, 
The limit $\rho\to i$ in the complementary series is special since it tends to (the semi-direct sum of) the trivial representation and the representation $(0,-1)$ of the unitary principal series. This is due to the fact that the space of all irreducible representations of $\mathrm{SL}(2,\mathbb{C})$ is not Hausdorff~\cite{naimark1964linear}. Interestingly, as we explain further below, when extending the scope to hyperfunctions~\cite{Ruehl1970}, it is possible to approach the trivial representation also from the point of view of the unitary principal series for the values $(\rho=\pm i,\nu=0)$ therein.

Focussing now on the unitary principal series, the operator %Wigner matrices %for representations 
$\boldsymbol{D}^{(i\rho+\nu-1,i\rho-\nu-1)}$ in the canonical %$\mathrm{SU}(2)$ 
basis %with basis vectors 
$\ket{\rho,\nu;j\,m}$
has matrix elements %$D^{(\rho,\nu)}_{jmln}(g)$ defined by
(see %for details on the construction of this basis we refer to
~\cite{Ruehl1970,Martin-Dussaud:2019ypf} for details)
\begin{equation}\label{eq:matrixelements}
    D^{(\rho,\nu)}_{jmln}(g)=\bra{\rho,\nu;j\,m} g\ket{\rho,\nu;l\,n},
\end{equation}
where
\begin{equation}\label{eq:rangesoflabels}
j,l\in\{\abs{\nu},\abs{\nu}+1,...\}, \quad m\in\{-j,...,j\},\quad n\in\{-l,...,l\}.
\end{equation}
As group homomorphisms the representations multiply for any two $g_1,g_2\in\SL$ as
%The representation of a product of two group elements $g_1,g_2\in\SL$ has an  expression as the product of two representation matrices, i.e.
\begin{equation}\label{eq:SL2C-Wigner matrix evaluated on product}
D^{(\rho,\nu)}_{jmln}(g_1g_2) 
=
\sum_{j'=\abs{\nu}}^\infty \sum_{m' = -j'}^{j'}D^{(\rho,\nu)}_{jmj'm'}(g_1)D^{(\rho,\nu)}_{j'm'ln}(g_2) .
\end{equation}
Furthermore, the representation matrices satisfy the orthogonality relation~\cite{Ruehl1970,Martin-Dussaud:2019ypf}
\begin{equation}\label{eq:orthogonality relation of SL2C wigner matrices}
\int\limits_{\SL}\dd{g}\overline{D^{(\rho_1,\nu_1)}_{j_1 m_1 l_1 n_1}(g)}D^{(\rho_2,\nu_2)}_{j_2 m_2 l_2 n_2}(g)
=
\frac{\delta(\rho_1-\rho_2)\delta_{\nu_1, \nu_2}}{\rho_1^2+\nu_1^2}
\delta_{j_1, j_2}\delta_{l_1, l_2}\delta_{m_1, m_2}\delta_{n_1, n_2},
\end{equation}
as well as the complex conjugation property
\begin{equation}\label{eq:complex conjugate of Wigner matrix}
\overline{D^{(\rho,\nu)}_{jmln}(g)}
=
(-1)^{j-l+m-n}D^{(\rho,\nu)}_{j\,-m\,l\,-n}(g),
\end{equation}
see for instance~\cite{Speziale:2016axj}. Notice that oftentimes the right-hand side of~\eqref{eq:orthogonality relation of SL2C wigner matrices} is multiplied by a factor of $\frac{1}{4}$ which we absorb into the Haar measure on the left-hand side. 
%This permits us to declot our formulas and smoothen the overall presentation.

As demonstrated in Ref.~\cite{Ruehl1970}, it is possible to relate the trivial representation to the unitary principal series. This can be accomplished using the Fourier transform of general distributions, also called analytic functionals or hyperfunctions~\cite{gel2016generalized1,hormander2015analysis}, which allow to define an analytic delta functional, see in particular Chapter $4$ in~\cite{Ruehl1970}. Employing this, one can safely write
\begin{align}\label{eq:SL2Ctrick}
    \int\text{d}g D^{(\rho,\nu)}_{j m l n}(g)&= \int\text{d}g D^{(\rho,\nu)}_{j m l n}(g) \mathbbm{1}\nonumber\\&=\int\text{d}g D^{(\rho,\nu)}_{j m l n}(g) \frac{1}{2}\left(D^{(i,0)}_{0 0 0 0}(g)+D^{(-i,0)}_{0 0 0 0}(g)\right)\nonumber\\&=\frac{1}{2}\left(\delta(\rho-i)+\delta(\rho+i)\right)\frac{\delta_{\nu,0}\delta_{j,0}\delta_{l,0}\delta_{m,0}\delta_{n,0}}{\rho^2+\nu^2}\nonumber\\&=-\delta(\rho-i)\delta_{\nu,0}\delta_{j,0}\delta_{l,0}\delta_{m,0}\delta_{n,0},
\end{align}
where in the last step we only retained the \enquote{zero-modes} at $\rho=i$ due to unitary equivalence. 
We use this trick repeatedly in the main text when working with the Fourier expansions of the correlation functions. 
The term \enquote{zero-mode} refers to the fact that the first Casimir, defined shortly hereafter, vanishes for such configurations. Note that the corresponding bivectors are lightlike.

%Using~\eqref{eq:Evaluating SL-Wigner matrix on SU elements} together with
The Cartan decomposition %of $\SL$
\eqref{eq:Cartan decomposition of g}
%, we can decompose the Wigner 
in terms of the representation 
matrices is
\begin{equation}\label{eq:Cartan decomposition of SL-Wigner matrix}
D^{(\rho,\nu)}_{jmln}(g) 
=
\sum_{q = -\min(j,l)}^{\min(j,l)}D^j_{mq}(u) \, d^{(\rho,\nu)}_{jlq}(\eta) \, D^l_{qn}(v^{-1}),
\end{equation}
with the \textit{reduced} $\SL$-Wigner matrix given by
\begin{equation}\label{eq:definition of SL2C reduced Wigner matrix}
d^{(\rho,\nu)}_{jlq}(\eta) \equiv D^{(\rho,\nu)}_{jqlq}\left(\e^{\frac{\eta}{2}\sigma_3}\right).
\end{equation}
and using that
%Working in this basis, evaluating 
the $\SL$-Wigner matrices relate to $\SU$-Wigner matrices $D^j_{mn}$ when evaluated on $u\in\SU$ as
\begin{equation}\label{eq:Evaluating SL-Wigner matrix on SU elements}
D^{(\rho,\nu)}_{jmln}(u) = \delta_{j,l}D^j_{mn}(u).
\end{equation}
Furthermore, integration over $\SU$ leads to
\begin{equation}
\int\limits_{\SU}\dd{u}D^{(\rho,\nu)}_{jmln}(u)
=
\delta_{j,l}\delta_{j,0}\delta_{m,0}\delta_{n,0}\delta_{\nu,0}.
\end{equation}
The factor $\delta_{\nu,0}$ appears since $j$ is restricted to zero by virtue of~\eqref{eq:rangesoflabels}. 
This is to be highlighted since when imposing simplicity with respect to a timelike normal one restricts to representations with labels $(\rho,0)$.

The spectrum of the unitary irreducible presentations $(\rho,\nu)$ is the following.
The two Casimir operators of $\SL$ are constructed from the generators of $\SL$  given in~\eqref{eq:commutationrelationsSL} and read
\begin{equation}\label{eq:sl2casimirs}
%\begin{aligned}
\cas_1  = \vb{K}^2-\vb{L}^2 
\,, \quad %\\[7pt]
\cas_2  = \vb{K}\cdot\vb{L}.
%\end{aligned}
\end{equation}
They act on states $\ket{\rho,\nu;j,m}$ %\in\mathcal{D}^{(\rho,\nu)}$ 
in the canonical basis as
\begin{align}%\label{eq:diagonalizedSL2CCasimirs}
\cas_1\ket{\rho,\nu;j,m} & =  \rho^2-\nu^2+1\ket{\rho,\nu;j,m},\label{eq:definition of cas1}\\%[7pt]
\cas_2\ket{\rho,\nu;j,m} & = \rho\nu\ket{\rho,\nu;j,m}\label{eq:definition of cas2}.
\end{align}
When imposing the simplicity constraint the second Casimir vanishes such that %. By virtue of~\eqref{eq:definition of cas2} one sees that then 
either $\rho$ or $\nu$ has to be zero. 
In this work, we focus on %representations 
the imposition of simplicity by means of a timelike normal vector ($\cas_1\ge 0$) so that we restrict our attention to representations labelled by $(\rho,0)$~\cite{Barrett:1999qw,Jercher:2021bie,Jercher:2022mky}.

The expansion (Plancherel decomposition) of a function $f\in L^2\left(\SL^4\right)$ in terms of unitary irreducible representations is
\begin{equation}\label{eq:General expansion of function on SL2C^4}
\begin{aligned}
f(g_1 ,g_2 ,g_3 ,g_4)
& =
\left[\prod_{i=1}^4 \int\displaylimits_{\R}\dd{\rho_i}\sum_{\nu_i\in\frac{\mathbb{Z}}{2}}\left(\rho_i^2+\nu_i^2\right)\sum_{j_i = \abs{\nu_i}}^{\infty}\sum_{l_i = \abs{\nu_i}}^{\infty}\sum_{m_i = - j_i}^{j_i}\sum_{n_i = -l_i}^{l_i}\right]\times\\[7pt]
&~~~\times f^{\rho_1 \rho_2 \rho_3 \rho_4 \nu_1 \nu_2 \nu_3 \nu_4}_{j_1 m_1 j_2 m_2 j_3 m_3 j_4 m_4 l_1 n_1 l_2 n_2 l_3 n_3 l_4 n_4}\prod_{i=1}^4 D^{(\rho_i,\nu_i)}_{j_i m_i l_i n_i}(g_i).
\end{aligned}
\end{equation}
Setting the Fourier coefficients therein to unity, one obtains the expansion of the corresponding $\delta$-distribution. Note that we dropped the factor of $4$ from the Plancherel measure as mentioned already above to simplify the presentation. To complete the picture, we refer to the detailed explanations given in Section~\ref{sec:modelsinlorentziansetting}, for how these expansions are changed when extending the domain of the function by a timelike normal $X$ and imposing geometricity constraints.

\subsection{Asymptotic behaviour of reduced Wigner functions}\label{app:repfunctionprop}

The dependence of the representation functions on $\eta$ is encoded entirely in the reduced Wigner functions $d^{(\rho,\nu)}_{jlq}$ introduced in~\eqref{eq:Cartan decomposition of SL-Wigner matrix}. These have the explicit form (our notation is identical to~\cite{Speziale:2016axj}, and similar to~\cite{Dona:2021ldn})
\begin{align}
d^{(\rho,\nu)}_{jlq}(\eta) =& \, (-1)^{\frac{j-l}{2}}\text{e}^{i\Phi^\rho_j}\text{e}^{-i\Phi^\rho_l}\frac{\sqrt{d_j}\sqrt{d_l}}{(j+l+1)!}
\sqrt{\prod_{+,-} (j\pm\nu)!(j\pm q)!(l\pm\nu)!(l\pm q)!} \,
% \nonumber\\&\times \left[(j+\nu)!(j-\nu)!(j+q)!(j-q)!(l+\nu)!(l-\nu)!(l+q)!(l-p)!\right]^{1/2}
\text{e}^{-(\nu-i\rho+q+1)\eta} \nonumber\\
& \times \sum_{s,t} \frac{
(-1)^{s+t}\text{e}^{-2t\eta} (\nu+q+s+t)! (j+l-\nu-q-s-t)!}
{s!(j-\nu-s)!(j-q-s)!(\nu+q+s)!t!(l-\nu-t)!(l-q-t)!(\nu+q+t)!} \nonumber\\
& \quad \quad \quad \times _{2}\!F_1\left(l+1-i\rho,\nu+q+1+s+t,j+l+2; 1-\text{e}^{-2\eta}\right),
\end{align}
where the ranges of $s$ and $t$ are determined by the conditions for the existence of the factorials (see~\cite{rashid}, but also~\cite{Ruehl1970}) and where
\begin{equation}\label{eqn:phasereduced}
\text{e}^{i\Phi^\rho_j}\equiv \frac{\Gamma(j+i\rho+1)}{\vert\Gamma(j+i\rho+1)\vert}\,,\qquad d_j\equiv (2j+1)\,.
\end{equation}
The reduced representation function satisfies the properties
\begin{equation}\label{eqn:parityproperty}
d^{(\rho,\nu)}_{jlq}(\eta)=d^{(-\rho,-\nu)}_{jlq}(\eta)\,,\qquad d^{(\rho,\nu)}_{jlq}(\eta)=d^{(\rho,-\nu)}_{jl-q}(\eta)\,.
\end{equation}
Moreover, as a function of $\rho$, $d^{(\rho,\nu)}_{jlq}(\eta)$ is entire: This follows from the fact that the hypergeometric function is an entire function of its first argument~\cite{hyperentire,hyperentire2} and from the fact that the phases~\eqref{eqn:phasereduced} are constructed such that the poles of the $\Gamma$-function are canceled. 

The dominating contributions to the above equations are determined depending on the behaviour of the hypergeometric function with respect to $\eta$. The expansion of the hypergeometric function $_{2}\!F_1[a,b,c;z]$ around $z=1$, for $c-a-b\neq\mathbb{Z}$, is given by~\cite{hypergeo}
\begin{align}
_{2}\!F_1[a,b,c;z]&\propto \frac{\Gamma(c)\Gamma(a+b-c)}{\Gamma(a)\Gamma(b)}(1-z)^{c-a-b}\left(1+\mathcal{O}(z-1)\right)\nonumber\\
&\quad+\frac{\Gamma(c)\Gamma(c-a-b)}{\Gamma(c-a)\Gamma(c-b)}\left(1+\mathcal{O}(z-1)\right).
\end{align}
In our case, the crucial combination $c-a-b$ is given by $j-q-s-t+i\rho$, so that the asymptotic behaviour of $d^{(\rho,\nu)}_{jlq}(\eta)$ is given by, for fixed $s$ and $t$
\begin{align}
d^{(\rho,0)}_{jlq}(\eta)&\sim \sum_{s,t}\left[c_1^\rho(j,l,q,s,t)\text{e}^{-\eta(q+1-i\rho-2t)}+c_2^\rho(j,l,q,s,t)\text{e}^{-\eta(q+1-i\rho+2t)-2\eta(j-q-s-t+i\rho)}\right]\nonumber\\
&\sim \sum_{s,t}\left[c_1^\rho(j,l,q,s,t)\text{e}^{-\eta(q+1-i\rho+2t)}+c_2^\rho(j,l,q,s,t)\text{e}^{-\eta(-q+1+i\rho+2j-2s)}\right],
\end{align}
where we remind that the sum over $s$ and $t$ is constrained by the condition that the factorials exist. In particular, from the denominator of our initial expression we see that we have the following conditions for $q$, $t$ and $j-s$:
\begin{equation}
j-s\ge 0\,,\qquad j-s\ge q\,,\qquad t\ge 0\,,\qquad q+t\ge 0\,.
\end{equation}
Now, let us consider separately the cases in which $q\ge 0$ and $q<0$:
\begin{description}
\item[$q\ge 0$] When $q\ge 0$ the two relevant constraints from those above are $j-s\ge q$ and $t\ge 0$. They determine the most dominating contribution in the sum in $s$ and $t$ in the limit of large $r$, which therefore are $2j-2s=2q$ and $t=0$. Substituting these equations into the previous one we obtain
\begin{equation}
d^{(\rho,0)}_{jlq}(\eta)\sim c_1^\rho(j,l,q)\text{e}^{-\eta(q+1-i\rho)}+c_2^\rho(j,l,q)\text{e}^{-\eta(q+1+i\rho)}\,.
\end{equation}
Since we know that the function must be symmetric in $\rho$ we can deduce that $c_2^\rho(j,l,q)=c_1^{-\rho}(j,l,q)$, and thus we conclude that
\begin{equation}
d^{(\rho,0)}_{jlq}(\eta)\sim \text{e}^{-\eta(q+1)}\left[c^\rho(j,l,q)\text{e}^{i\eta\rho}+c^{-\rho}(j,l,q)\text{e}^{-i\eta\rho}\right]\,,
\end{equation}
where we have dropped the subscript $1$ for the sake of notation.
\item[$q<0$] When $q<0$, instead, the two relevant constraints for $s$ and $t$ are $j-s\ge 0$ and $t\ge \vert q\vert\ge 0$. Substituting the extremal values $j-s=0$ and $t=\vert q\vert$ in the asymptotic form of $d^{(\rho,0)}_{jlq}(\eta)$, we find 
\begin{equation}
d^{(\rho,0)}_{jlq}(\eta)\sim \text{e}^{-\eta(\vert q\vert+1)}\left[c^\rho(j,l,q)\text{e}^{i\eta\rho}+c^{-\rho}(j,l,q)\text{e}^{-i\eta\rho}\right]\,,
\end{equation}
where as before we have used the symmetry properties of the function.
\end{description}
We thus conclude that the general behaviour
\begin{equation}\label{eqn:asymptoticrepfunctions}
d^{(\rho,0)}_{jlq}(\eta)\sim \text{e}^{-\eta(\vert q\vert+1)}\left[c^\rho(j,l,q)\text{e}^{i\eta\rho}+c^{-\rho}(j,l,q)\text{e}^{-i\eta\rho}\right]\,,
\end{equation}
holds for both positive and negative $q$.

\section{Basic group structure and representation theory of $\mathrm{Spin}(4)$}\label{app:Spin4}

In this appendix, we briefly review essential features of $\mathrm{Spin}(4)$ and its representation theory tailored to the application in this article. To this aim, we follow the notation of~\cite{doi:10.1063/1.1703728,Perez:2012wv} and in particular that of~\cite{Dona:2021ldn}. Based on this, we give the Fourier decomposition of a rank-$4$ group field subject to right-covariance and simplicity constraints.

The Lie group $\mathrm{Spin}(4)=\mathrm{SU}(2)\times\mathrm{SU}(2)$ has the Lie algebra $\mathrm{spin}(4)=\mathrm{su}(2)\oplus\mathrm{su}(2)$ with the two commuting Lie algebras of $\mathrm{SU}(2)$ with generators $J^{i}_{L/R}$ such that
\begin{equation}
    [J^{i}_{L/R},J^{j}_{L/R}]=i \epsilon^{ij}_k J^{k}_{L/R}
\end{equation}
holds. Its center is generated by the two Casimir operators $J^2_{L/R}$. The unitary irreducible representations of this group are given in terms of the tensor product of unitary irreducible representations of $\mathrm{SU}(2)$, are labelled by the tuple of half-integers $(j_L,j_R)$ and thus have dimension $d_{j_L}d_{j_R}=(2j_L+1)(2j_R+1)$. In this basis, the coefficients of the Wigner matrices of $g=(g_L,g_R)\in\text{Spin}(4)$ are given by
\begin{equation}
    D^{(j_L,j_R)}_{m_L m_R n_L n_R}(g)=\langle j_L, m_L; j_R, m_R|g|j_L, n_L; j_R, n_R\rangle=D^{j_L}_{m_L n_L}(g_L)D^{j_R}_{m_R n_R}(g_R).
\end{equation}
Obviously, it is advantageous to work within the standard basis since its Wigner matrices have a simple form. 
However, the mapping to unitary irreducible representations of $\text{SL}(2,\mathbb{C})$ in the principal series is most transparent 
%in the same basis  %via compactification and Wick rotation, see Section~\ref{section:wickrotation},
employing the canonical basis of $\text{Spin}(4)$ to be constructed hereafter.

To this aim, we first notice that due to the local isomorphism between $\text{Spin}(4)$ and $\text{SO}(4)$ their Lie algebras are in one-to-one correspondence with each other. Its generators are given in terms of those of the spatial rotation subgroup $L^i:=J_L^i+J_R^i$ together with those of the Euclidean boosts $K_{E}^{i}:=J_L^i-J_R^i$. Together they satisfy the algebra
\begin{equation}\label{eq:Euclideanizedalgebra}
[K_{E,i},K_{E,j}] = i\varepsilon_{ijk}L_k,\quad [L_i,K_{E,j}] = i\varepsilon_{ijk}K_{E,k},\quad [L_i,L_j] = i \varepsilon_{ijk}L_k.
\end{equation}
This allows to construct two other invariant operators
\begin{equation}\label{eq:spin4casimirs}
    \cas_1:= \vec{K}_E^2+\vec{L}^2=2\left(\vec{J_L}^2+\vec{J_R}^2\right)~~~\text{and}~~~
    \cas_2:=\vec{L}\cdot\vec{K}_E= \left(\vec{J_L}^2-\vec{J_R}^2\right)
\end{equation}
with which we relabel the representations $(j_L,j_R)$ in terms of the half-integers $p=j_L+j_R+1$ and $k=j_L-j_R$.
Without loss of generality one can assume that $j_L\geq j_R$ implying $p>k\geq 0$. 
It follows that these representations are of dimension $(p^2-k^2)$. With this, one has
\begin{align}
    \cas_1|p,k\rangle=(p^2+k^2-1)|p,k\rangle
    ~~~\text{and}~~~
    \cas_2|p,k\rangle&=pk|p,k\rangle.
\end{align}

In terms of irreducible representations of $\mathrm{SU}(2)$ one defines the canonical basis of $\text{Spin}(4)$ with basis vectors $|p,k;j,m\rangle$ where $p-1\geq j\geq k$ and $m=-j,...,j$. 
It simultaneously diagonalizes the $\mathrm{SU}(2)$-Casimir $\vec{L}^2$ and $L_3$, i.e.
\begin{equation}
    L^2|p,k;j,m\rangle 
    =j(j+1)|p,k;j,m\rangle~~~\text{and}~~~L_3|p,k;j,m\rangle
    =m|p,k;j,m\rangle.
\end{equation}
Hence, the unitary representations of $\text{Spin}(4)$ can be compactly given as the direct sum of those of $\text{SU}(2)$ according to
\begin{equation}
    \mathcal{H}^{(j_L,j_R)}=\mathcal{H}^{(p,k)}=\bigoplus_{j=k}^{p-1} \mathcal{H}^j.
\end{equation}
The matrix coefficients of a Wigner matrix of an element $g$ of $\text{Spin}(4)$ are then given by
\begin{equation}
    D^{(p,k)}_{jm ln}(g)=\langle p, k ; j,m|g|p, k ; l,n\rangle
\end{equation}
and the orthogonality relation
\begin{align}
    \int \text{d}g D^{(j_L,j_R)}_{m_L m_R n_L n_R}(g) D^{(j_L',j_R')}_{m_L' m_R' n_L' n_R'}(g)
    &=\frac{1}{d_{j_L}d_{j_R}}\delta_{j_L j_L'}\delta_{j_R j_R'}\delta_{m_L m_L'}\delta_{m_R m_R'}\delta_{n_L n_L'}\delta_{n_R n_R'}
\end{align}
becomes
\[
\int \text{d}g D^{(p,k)}_{j m ln}(g) D^{(p',k')}_{j' m' l' n'}(g)
=\frac{1}{p^2-k^2}\delta_{pp'}\delta_{kk'}\delta_{jj'}\delta_{mm'}\delta_{ll'}\delta_{nn'}
\]
with $\text{d}g$ denoting the respective Haar measure. Clearly, for $u\in \text{SU}(2)$ one finds $D^{(p,k)}_{j m ln}(u)=\delta_{jl}D^{j}_{mn}(u)$ such that
\begin{equation}
    \int \text{d}u D^{(p,k)}_{j m ln}(u)=\delta_{jl}\delta_{j0}\delta_{m0}\delta_{n0}\delta_{k0}.
\end{equation}
As above, the Haar measure $\text{d}u$ of $\text{SU}(2)$ is considered to be normalized to $1$ hereafter. With these details set up, the Fourier decomposition of a function $f\in L^2(\text{Spin}(4))$ is
\begin{equation}
    f(g)=\sum_{p,k}\sum_{j,l}\sum_{m,n}(p^2-k^2)f^{pk}_{jm ln}D^{(p,k)}_{j m ln}(g).
\end{equation}

For the purpose of our work it is also relevant to give the Cartan decomposition of $\text{Spin}(4)$, as discussed in~\cite{Dona:2021ldn}, i.e.
\begin{align}\label{eq:cartandecompositionspin4}
    \text{SU}(2)\times \text{T}^{+}\times \text{SU}(2)&\to \text{Spin}(4)\nonumber\\
    (u, \text{e}^{-i\frac{t}{2}\sigma_3},v)&\mapsto (u\text{e}^{-i\frac{t}{2}\sigma_3}v^{-1}, u\text{e}^{i\frac{t}{2}\sigma_3}v^{-1})
\end{align}
with
\begin{equation}
    \text{T}^{+}=\{\text{e}^{-i\frac{t}{2}\sigma_3}|t\in[0,2\pi)\}
\end{equation} 
which is derived from the torus subgroup of $\text{SU}(2)$, defined by
\begin{equation}
    \text{T}=\{\text{e}^{-i\frac{t}{2}\sigma_3}|t\in[0,4\pi)\}.
\end{equation}
Equipped with the Cartan decomposition of $\text{Spin}(4)$, we can decompose the respective Wigner matrices as
\begin{equation}\label{eq:Cartan decomposition of Spin4-Wigner matrix}
D^{(p,k)}_{jmln}(g) 
=
\sum_{q = -\min(j,l)}^{\min(j,l)} D^j_{mq}(u) \, d^{(p,k)}_{jlq}(t) \, D^l_{qn}(v^{-1}).
\end{equation}
In this decomposition the \textit{reduced} $\text{Spin}(4)$-Wigner matrices
\begin{align}
\label{eq:definition of reduced Spin 4 Wigner matrix}
d^{(p,k)}_{jlq}(t) 
&\equiv D^{(p,k)}_{jqlq}\left(\e^{i\frac{t}{2}\sigma_3},\e^{-i\frac{t}{2}\sigma_3}\right) \\
&=\sum_{m_L,m_R}\langle j_L,m_L;j_R,m_R|j,q\rangle \, \text{e}^{it(m_L-m_R)}\langle j_L, m_L; j_R,m_R|l,q\rangle
\end{align}
are given in terms of $\text{SU}(2)$-Clebsch-Gordan coefficients which intertwine the $(j_L,m_L;j_R,m_R)$-representation with the $(j,q)$- or $(l,q)$-$\text{SU}(2)$-representations in~\eqref{eq:definition of reduced Spin 4 Wigner matrix}. In addition, the Cartan decomposition induces a decomposition of the Haar measure as
\begin{equation}\label{eq:HaarmeasureSpin4}
    \d g=\sin^2(t)\d t \, \d u \, \d v.
\end{equation}

As an application with direct relevance for the main part of this work, we seek to apply this to rank-$4$ group fields in the extended formulation, i.e.
\begin{equation}\label{eq:gftfieldriemannian}
   \Phi(\boldsymbol{\phi},\boldsymbol{g},X)= \Phi(\phi_1,...,\phi_{\ld},g_1,...,g_4,X): \mathbb{R}^{\ld}\times\text{Spin}(4)^4\times S^3\to \mathbb{R}~ \text{or}~\mathbb{C}
\end{equation}
where we also included the local degrees of freedom to the domain. We assume these fields to be square-integrable functions $\Phi,\Phi'\in L^2(\mathbb{R}^{\ld}\times\text{Spin}(4)^4\times S^3)$ with the scalar product
\begin{equation}
    \left( \Phi , \Phi'\right) = \int_{\mathbb{R}^{\ld}} \text{d}\boldsymbol{\phi}\int_{\text{Spin}(4)^4}\text{d}\boldsymbol{g}\int_{S^3}\text{d}X~\Phi(\boldsymbol{\phi},\boldsymbol{g},X)  \Phi'(\boldsymbol{\phi},\boldsymbol{g},X).
\end{equation}
The domain of the field is extended by the non-dynamical variable $X$ denoting a vector normal to the tetrahedra described by the group fields. As such it is an element of the homogeneous space $S^3\cong \text{Spin}(4)/\text{SU}(2)_{X}$. The extension of the group field domain guarantees that the geometricity constraints are covariantly imposed and commute with each other. Notice that $\text{SU}(2)_{X}$ is the stabilizer subgroup $\text{SU}(2)_X\subset \text{Spin}(4)$ of $X$ defined by $\text{SU}(2)_X:=\{h\in \text{Spin}(4) | h \cdot X = X\}$. Using that the round metric on $S^3$ is
\begin{equation}
\text{d}S^2=\text{d}t^2+\sin^2(t)\text{d}\Omega_2,
\end{equation}
where $\text{d}\Omega_2$ denotes the metric on the $2$-sphere, the integration measure on $X$ can be simply obtained from the Jacobian or is just yielded from the Haar measure on $\text{Spin}(4)$, see~\eqref{eq:HaarmeasureSpin4}, that is
\begin{equation}
    \text{d}X=\sin^2(t)\d t \,\d \Omega_2.
\end{equation}
Now, we require the field to be subject to the following symmetries
\begin{align}
    &\Phi(\boldsymbol{\phi}, g_1,g_2,g_3, g_4, X)=\Phi(\boldsymbol{\phi}, g_1 u_1,g_2 u_2,g_3 u_3, g_4 u_4, X),~~~\forall u_i \in \text{SU}(2)_X,\label{eq:simplicityriemannian}\\
    &\Phi(\boldsymbol{\phi}, g_1,g_2,g_3, g_4, X)=\Phi(\boldsymbol{\phi}, g_1 h^{-1},g_2 h^{-1},g_3 h^{-1}, g_4 h^{-1}, h\cdot X),~~~\forall h \in \text{Spin}(4),\label{eq:gaugeinvarianceriemannian}
\end{align}
known as simplicity and right-covariance. Their geometric interpretation, in spite of working here in the Riemannian case, is analogous to that given for Lorentzian signature in Sec.~\ref{sec:modelsinlorentziansetting}. The expansion of a group field obeying these symmetries in terms of representation labels reads,
\begin{equation}\label{eq:fieldfourier1riemannian}
    \Phi(\boldsymbol{\phi}, \boldsymbol{g},X)=\prod_{i=1}^4\left(\sum_{p_i} p_i^2\sum_{j_i,m_i} D^{(p_i, 0)}_{j_i m_i 0 0}(g_i X)\right) \Phi_{j_1 m_1 j_2 m_2 j_3 m_3 j_4 m_4}^{p_1 p_2 p_3 p_4}(\boldsymbol{\phi})
\end{equation}
with $\Phi_{j_1 m_1 j_2 m_2 j_3 m_3 j_4 m_4}^{p_1 p_2 p_3 p_4}(\boldsymbol{\phi})\equiv\Phi_{j_1 m_1 0 0 j_2 m_2 0 0 j_3 m_3 0 0 j_4 m_4 0 0}^{p_1 0 p_2 0 p_3 0 p_4 0}(\boldsymbol{\phi})$ where we note that a part of the discrete representation labels vanishes because of the imposition of simplicity. Due to the fact that the normals are non-dynamical, we can simply integrate them out which together with~\eqref{eq:gaugeinvarianceriemannian} leads to the closure of the (Euclidean Barrett-Crane) tetrahedron, see also~\cite{Baratin:2011tx}. The Fourier expansion of the fields is then given by
\begin{align}\label{eq:fieldfourier1riemanniannormalintegrated}
    \Phi(\boldsymbol{\phi},\boldsymbol{g})&=\int_{S^3}\text{d}X\Phi(\boldsymbol{\phi},\boldsymbol{g},X)\nonumber\\&=\prod_{i=1}^4\left(\sum_{p_i} p_i^2\sum_{\substack{j_i,m_i;\\l_i,n_i}}D^{(p_i,0)}_{j_i m_i l_i n_i}(g_i)\right) B^{E,p_1 p_2 p_3 p_4}_{l_1 n_1 l_2 n_2 l_3 n_3 l_4 n_4} \Phi_{j_1 m_1 j_2 m_2 j_3 m_3 j_4 m_4}^{p_1 p_2 p_3 p_4}(\boldsymbol{\phi}),
\end{align}
with the BC intertwiner (in Euclidean signature)~\cite{Barrett:1999qw,Oriti:2003wf,Baratin:2011tx}
\begin{equation}
    B^{E,p_1 p_2 p_3 p_4}_{j_1 m_1 j_2 m_2 j_3 m_3 j_4 m_4}\equiv\int_{S^3}\text{d}X\prod_{i=1}^4 D^{(p_i,0)}_{j_i m_i 0 0}(X).
\end{equation}
Finally, the Fourier transform with respect to the local degrees of freedom on $\mathbb{R}^{\ld}$ is done in the standard way.

\section{Useful integrals for the computation of correlation functions}\label{app:integrals}
This Appendix provides useful expressions for the explicit computation of correlation functions in the main text. 

Generally, one needs the following typical integral to yield the correlation function in coordinate space, i.e.
\begin{equation}
    I_D(\mu)\equiv \int\diff^Dp\frac{\text{e}^{i\mathbf{p}\cdot\mathbf{x}}}{\vert\mathbf{p}\vert^2+\mu}=\int\diff\Omega_D\int_0^\infty\diff p\, p^{D-1}\frac{\text{e}^{i\mathbf{p}\cdot\mathbf{x}}}{p^2+\mu}\,,
\end{equation}
where $p^2\equiv \vert\mathbf{p}\vert^2\equiv \sum_{i=1}^Dp_i^2$, $D$ is a non-zero positive integer, $D\in \mathbb{N}^+$, and $\Omega_D$ denotes the angular measure on the Euclidean $D$-dimensional space. If we define $r^2\equiv \vert\mathbf{x}\vert^2\equiv \sum_{i=1}^Dx_i^2$ and $\mathbf{p}\cdot\mathbf{x}\equiv pr\cos\theta$, we can rewrite this integral as~\cite{Hong_Hao_2010}
\begin{equation}
    I_D(\mu)=\frac{2\pi^{\frac{D-1}{2}}}{\Gamma(\frac{D-1}{2})}\int_0^\pi\sin^{D-2}\theta\diff\theta\int_0^\infty\diff p\, p^{D-1}\frac{\text{e}^{ipr\cos\theta}}{p^2+\mu}\,.
\end{equation}
The angular contribution can be easily integrated, yielding
\begin{equation}
    \int_0^\pi\sin^{D-2}\theta\diff\theta\,\text{e}^{ipr\cos\theta}=\sqrt{\pi}\left(\frac{2}{pr}\right)^{\frac{D-2}{2}}\Gamma\left(\frac{D-1}{2}\right)J_{\frac{D-2}{2}}(pr)\,,
\end{equation}
where $J_\alpha(z)$ corresponds to the Bessel function of the first kind~\cite{Hong_Hao_2010}. If one combines these last two equations and changes the variable to $q\equiv pr$, on finds
\begin{equation}
    I_D(\mu)=\frac{2^D\pi^{D/2}}{r^{D-2}}\int_0^\infty\diff q\frac{q^{D/2}}{q^2+\mu r^2}J_{\frac{D-2}{2}}(q)\,.
\end{equation}
The explicit value of this integral now depends decisively on the sign of $\mu$. We discuss these two cases one after the other:
\begin{description}
\item[\underline{$\mu>0$}:] When $\mu>0$ the integrand is regular over the whole domain of integration and by virtue of the residue theorem, we obtain that
\begin{equation}
    I_D(\mu)=\frac{2^D\pi^{D/2}}{r^{D-2}}(\mu r^2)^{\frac{D-2}{4}}K_{\frac{D-2}{2}}(\sqrt{\mu} r)=2^D\pi^{D/2}(\mu r^{-2})^{\frac{D-2}{4}}K_{\frac{D-2}{2}}(\sqrt{\mu} r)\,,
\end{equation}
where $K_\alpha(z)$ denotes the modified Bessel function of the second kind, the asymptotic behaviour of which for large $z$ and for $\vert\text{arg}(z)\vert <\frac{3}{2}\pi$ reads
\begin{equation}\label{eqn:asymptoticexpmodbessel}
    K_\alpha(z)\sim \sqrt{\frac{\pi}{2z}}\text{e}^{-z}\left(1+\mathcal{O}(z^{-1})\right).
\end{equation}
Consequently, the correlation function is exponentially suppressed for large values of $r$. (More precisely, one has some power-law decay behaviour together with the exponential suppression.) One extracts that the scale $\sqrt{\mu}$ sets the scale of exponential suppression and is interpreted as the correlation length of the system.
\item[\underline{$\mu<0$}:] When $\mu<0$, the integrand is singular on the domain of integration. The integration can still be performed by employing the Feynman prescription, so that $q^2+\mu r^2\to q^2+(\mu -i\epsilon)r^2$, with $\epsilon\in \mathbb{R}$ being set to zero after integration. This results in
\begin{equation}
    I_D(\mu)=2^D\pi^{D/2}(-\vert\mu \vert r^{-2})^{\frac{D-2}{4}}K_{\frac{D-2}{2}}(i\sqrt{\vert \mu\vert}r)\,.
\end{equation}
One concludes from the asymptotic expansion~\eqref{eqn:asymptoticexpmodbessel} that apart from an unimportant proportionality factor, the behaviour of this integral for large $r$ is
\begin{equation}
    I_D(\mu)\sim 2^{D-1/2}\pi^{D/2+1}(-\vert\mu \vert r^{-2})^{\frac{D-2}{4}} \frac{\text{e}^{-i\sqrt{\vert \mu \vert}r}}{(\sqrt{\vert \mu\vert} r)^{1/2}}\,.
\end{equation}
Hence, the function oscillates and is suppressed by $r^{-(D-3)/2}$ at large $r$.
\end{description}

\bibliographystyle{jhep}
\bibliography{references.bib} 

\providecommand{\href}[2]{#2}\begingroup\raggedright\begin{thebibliography}{100}

\bibitem{Goldenfeld:1992qy}
N.~Goldenfeld, \emph{{Lectures on phase transitions and the renormalization
  group}} (1992).

\bibitem{Konopka:2008hp}
T.~Konopka, F.~Markopoulou and S.~Severini, \emph{{Quantum Graphity: A Model of
  emergent locality}},
  \href{https://doi.org/10.1103/PhysRevD.77.104029}{\emph{Phys. Rev. D}
  {\bfseries 77} (2008) 104029}
  [\href{https://arxiv.org/abs/0801.0861}{{\ttfamily 0801.0861}}].

\bibitem{Koslowski:2011vn}
T.~Koslowski and H.~Sahlmann, \emph{{Loop quantum gravity vacuum with
  nondegenerate geometry}},
  \href{https://doi.org/10.3842/SIGMA.2012.026}{\emph{SIGMA} {\bfseries 8}
  (2012) 026} [\href{https://arxiv.org/abs/1109.4688}{{\ttfamily 1109.4688}}].

\bibitem{Oriti:2013jga}
D.~Oriti, \emph{{Disappearance and emergence of space and time in quantum
  gravity}}, \href{https://doi.org/10.1016/j.shpsb.2013.10.006}{\emph{Stud.
  Hist. Phil. Sci. B} {\bfseries 46} (2014) 186}
  [\href{https://arxiv.org/abs/1302.2849}{{\ttfamily 1302.2849}}].

\bibitem{Oriti:2007qd}
D.~Oriti, \emph{{Group field theory as the microscopic description of the
  quantum spacetime fluid: A New perspective on the continuum in quantum
  gravity}}, {\emph{PoS} {\bfseries QG-PH} (2007) 030}
  [\href{https://arxiv.org/abs/0710.3276}{{\ttfamily 0710.3276}}].

\bibitem{Gurau:2016cjo}
R.~Gurau, \emph{{Invitation to Random Tensors}},
  \href{https://doi.org/10.3842/SIGMA.2016.094}{\emph{SIGMA} {\bfseries 12}
  (2016) 094} [\href{https://arxiv.org/abs/1609.06439}{{\ttfamily
  1609.06439}}].

\bibitem{Eichhorn:2018phj}
A.~Eichhorn, T.~Koslowski and A.D.~Pereira, \emph{{Status of
  background-independent coarse-graining in tensor models for quantum
  gravity}}, \href{https://doi.org/10.3390/universe5020053}{\emph{Universe}
  {\bfseries 5} (2019) 53} [\href{https://arxiv.org/abs/1811.12909}{{\ttfamily
  1811.12909}}].

\bibitem{Loll:2019rdj}
R.~Loll, \emph{{Quantum Gravity from Causal Dynamical Triangulations: A
  Review}}, \href{https://doi.org/10.1088/1361-6382/ab57c7}{\emph{Class. Quant.
  Grav.} {\bfseries 37} (2020) 013002}
  [\href{https://arxiv.org/abs/1905.08669}{{\ttfamily 1905.08669}}].

\bibitem{Steinhaus:2020lgb}
S.~Steinhaus, \emph{{Coarse Graining Spin Foam Quantum Gravity\textemdash{}A
  Review}}, \href{https://doi.org/10.3389/fphy.2020.00295}{\emph{Front. in
  Phys.} {\bfseries 8} (2020) 295}
  [\href{https://arxiv.org/abs/2007.01315}{{\ttfamily 2007.01315}}].

\bibitem{Surya:2019ndm}
S.~Surya, \emph{{The causal set approach to quantum gravity}},
  \href{https://doi.org/10.1007/s41114-019-0023-1}{\emph{Living Rev. Rel.}
  {\bfseries 22} (2019) 5} [\href{https://arxiv.org/abs/1903.11544}{{\ttfamily
  1903.11544}}].

\bibitem{Ashtekar:2004eh}
A.~Ashtekar and J.~Lewandowski, \emph{{Background independent quantum gravity:
  A Status report}},
  \href{https://doi.org/10.1088/0264-9381/21/15/R01}{\emph{Class. Quant. Grav.}
  {\bfseries 21} (2004) R53}
  [\href{https://arxiv.org/abs/gr-qc/0404018}{{\ttfamily gr-qc/0404018}}].

\bibitem{Perez:2003vx}
A.~Perez, \emph{{Spin foam models for quantum gravity}},
  \href{https://doi.org/10.1088/0264-9381/20/6/202}{\emph{Class. Quant. Grav.}
  {\bfseries 20} (2003) R43}
  [\href{https://arxiv.org/abs/gr-qc/0301113}{{\ttfamily gr-qc/0301113}}].

\bibitem{Perez:2012wv}
A.~Perez, \emph{{The Spin Foam Approach to Quantum Gravity}},
  \href{https://doi.org/10.12942/lrr-2013-3}{\emph{Living Rev. Rel.} {\bfseries
  16} (2013) 3} [\href{https://arxiv.org/abs/1205.2019}{{\ttfamily
  1205.2019}}].

\bibitem{Rovelli:2011eq}
C.~Rovelli, \emph{{Zakopane lectures on loop gravity}},
  \href{https://doi.org/10.22323/1.140.0003}{\emph{PoS} {\bfseries QGQGS2011}
  (2011) 003} [\href{https://arxiv.org/abs/1102.3660}{{\ttfamily 1102.3660}}].

\bibitem{Conrady:2010kc}
F.~Conrady and J.~Hnybida, \emph{{A spin foam model for general Lorentzian
  4-geometries}},
  \href{https://doi.org/10.1088/0264-9381/27/18/185011}{\emph{Class. Quant.
  Grav.} {\bfseries 27} (2010) 185011}
  [\href{https://arxiv.org/abs/1002.1959}{{\ttfamily 1002.1959}}].

\bibitem{Conrady:2010vx}
F.~Conrady, \emph{{Spin foams with timelike surfaces}},
  \href{https://doi.org/10.1088/0264-9381/27/15/155014}{\emph{Class. Quant.
  Grav.} {\bfseries 27} (2010) 155014}
  [\href{https://arxiv.org/abs/1003.5652}{{\ttfamily 1003.5652}}].

\bibitem{williams2009quantum}
R.~Williams, \emph{Quantum regge calculus}, {\emph{Approaches to Quantum
  Gravity, Editor D. Oriti, Cambridge University Press, Cambridge} (2009) 360}.

\bibitem{Ambjorn:2012jv}
J.~Ambjorn, A.~Goerlich, J.~Jurkiewicz and R.~Loll, \emph{{Nonperturbative
  Quantum Gravity}},
  \href{https://doi.org/10.1016/j.physrep.2012.03.007}{\emph{Phys. Rept.}
  {\bfseries 519} (2012) 127}
  [\href{https://arxiv.org/abs/1203.3591}{{\ttfamily 1203.3591}}].

\bibitem{gurau2017random}
R.G.~Gur{\u{a}}u, \emph{Random tensors}, Oxford University Press (2017).

\bibitem{Gurau:2019qag}
R.G.~Gurau, \emph{{Notes on tensor models and tensor field theories}},
  \href{https://doi.org/10.4171/aihpd/117}{\emph{Ann. Inst. H. Poincare D Comb.
  Phys. Interact.} {\bfseries 9} (2022) 159}
  [\href{https://arxiv.org/abs/1907.03531}{{\ttfamily 1907.03531}}].

\bibitem{Freidel:2005qe}
L.~Freidel, \emph{{Group field theory: An Overview}},
  \href{https://doi.org/10.1007/s10773-005-8894-1}{\emph{Int. J. Theor. Phys.}
  {\bfseries 44} (2005) 1769}
  [\href{https://arxiv.org/abs/hep-th/0505016}{{\ttfamily hep-th/0505016}}].

\bibitem{Oriti:2011jm}
D.~Oriti, \emph{{The microscopic dynamics of quantum space as a group field
  theory}},  in \emph{{Foundations of Space and Time: Reflections on Quantum
  Gravity}}, pp.~257--320, 10, 2011
  [\href{https://arxiv.org/abs/1110.5606}{{\ttfamily 1110.5606}}].

\bibitem{Krajewski:2011zzu}
T.~Krajewski, \emph{{Group field theories}},
  \href{https://doi.org/10.22323/1.140.0005}{\emph{PoS} {\bfseries QGQGS2011}
  (2011) 005} [\href{https://arxiv.org/abs/1210.6257}{{\ttfamily 1210.6257}}].

\bibitem{Carrozza:2013oiy}
S.~Carrozza, \emph{{Tensorial methods and renormalization in Group Field
  Theories}}, Ph.D. thesis, Orsay, LPT, 2013.
\newblock \href{https://arxiv.org/abs/1310.3736}{{\ttfamily 1310.3736}}.
\newblock 10.1007/978-3-319-05867-2.

\bibitem{Oriti:2014uga}
D.~Oriti, \emph{{Group Field Theory and Loop Quantum Gravity}},  in \emph{{Loop
  Quantum Gravity}: {The First 30 Years}}, A.~Ashtekar and J.~Pullin, eds.,
  pp.~125--151, WSP (2017),
  \href{https://doi.org/10.1142/9789813220003_0005}{DOI}
  [\href{https://arxiv.org/abs/1408.7112}{{\ttfamily 1408.7112}}].

\bibitem{DiFrancesco:1993cyw}
P.~Di~Francesco, P.H.~Ginsparg and J.~Zinn-Justin, \emph{{2-D Gravity and
  random matrices}},
  \href{https://doi.org/10.1016/0370-1573(94)00084-G}{\emph{Phys. Rept.}
  {\bfseries 254} (1995) 1}
  [\href{https://arxiv.org/abs/hep-th/9306153}{{\ttfamily hep-th/9306153}}].

\bibitem{sachs2006elements}
I.~Sachs, S.~Sen and J.~Sexton, \emph{Elements of statistical mechanics: with
  an introduction to quantum field theory and numerical simulation}, Cambridge
  University Press (2006).

\bibitem{Kopietz:2010zz}
P.~Kopietz, L.~Bartosch and F.~Sch\"utz, \emph{{Introduction to the functional
  renormalization group}}, vol.~798 (2010),
  \href{https://doi.org/10.1007/978-3-642-05094-7}{10.1007/978-3-642-05094-7}.

\bibitem{zinn2021quantum}
J.~Zinn-Justin, \emph{Quantum field theory and critical phenomena}, vol.~171,
  Oxford university press (2021).

\bibitem{Pereira:2019dbn}
A.D.~Pereira, \emph{{Quantum spacetime and the renormalization group: Progress
  and visions}},  in \emph{{Progress and Visions in Quantum Theory in View of
  Gravity}: {Bridging foundations of physics and mathematics}}, 4, 2019
  [\href{https://arxiv.org/abs/1904.07042}{{\ttfamily 1904.07042}}].

\bibitem{Eichhorn:2021vid}
A.~Eichhorn, B.~Bahr and A.D.~Pereira, \emph{{Editorial: Coarse graining in
  quantum gravity -- Bridging the gap between microscopic models and spacetime
  physics}},  \href{https://arxiv.org/abs/2103.14605}{{\ttfamily 2103.14605}}.

\bibitem{Eichhorn:2013isa}
A.~Eichhorn and T.~Koslowski, \emph{{Continuum limit in matrix models for
  quantum gravity from the Functional Renormalization Group}},
  \href{https://doi.org/10.1103/PhysRevD.88.084016}{\emph{Phys. Rev. D}
  {\bfseries 88} (2013) 084016}
  [\href{https://arxiv.org/abs/1309.1690}{{\ttfamily 1309.1690}}].

\bibitem{Eichhorn:2014xaa}
A.~Eichhorn and T.~Koslowski, \emph{{Towards phase transitions between discrete
  and continuum quantum spacetime from the Renormalization Group}},
  \href{https://doi.org/10.1103/PhysRevD.90.104039}{\emph{Phys. Rev. D}
  {\bfseries 90} (2014) 104039}
  [\href{https://arxiv.org/abs/1408.4127}{{\ttfamily 1408.4127}}].

\bibitem{Eichhorn:2017xhy}
A.~Eichhorn and T.~Koslowski, \emph{{Flowing to the continuum limit in tensor
  models for quantum gravity}},
  \href{https://doi.org/10.4171/aihpd/52}{\emph{Ann. Inst. H. Poincare D Comb.
  Phys. Interact.} {\bfseries 5} (2018) 173}
  [\href{https://arxiv.org/abs/1701.03029}{{\ttfamily 1701.03029}}].

\bibitem{Eichhorn:2018ylk}
A.~Eichhorn, T.~Koslowski, J.~Lumma and A.D.~Pereira, \emph{{Towards background
  independent quantum gravity with tensor models}},
  \href{https://doi.org/10.1088/1361-6382/ab2545}{\emph{Class. Quant. Grav.}
  {\bfseries 36} (2019) 155007}
  [\href{https://arxiv.org/abs/1811.00814}{{\ttfamily 1811.00814}}].

\bibitem{Eichhorn:2019hsa}
A.~Eichhorn, J.~Lumma, A.D.~Pereira and A.~Sikandar, \emph{{Universal critical
  behavior in tensor models for four-dimensional quantum gravity}},
  \href{https://doi.org/10.1007/JHEP02(2020)110}{\emph{JHEP} {\bfseries 02}
  (2020) 110} [\href{https://arxiv.org/abs/1912.05314}{{\ttfamily
  1912.05314}}].

\bibitem{Castro:2020dzt}
A.~Castro and T.~Koslowski, \emph{{Renormalization Group Approach to the
  Continuum Limit of Matrix Models of Quantum Gravity with Preferred
  Foliation}}, \href{https://doi.org/10.3389/fphy.2021.531766}{\emph{Front. in
  Phys.} {\bfseries 9} (2021) 114}
  [\href{https://arxiv.org/abs/2008.10090}{{\ttfamily 2008.10090}}].

\bibitem{Eichhorn:2020sla}
A.~Eichhorn, A.D.~Pereira and A.G.A.~Pithis, \emph{{The phase diagram of the
  multi-matrix model with ABAB-interaction from functional renormalization}},
  \href{https://doi.org/10.1007/JHEP12(2020)131}{\emph{JHEP} {\bfseries 12}
  (2020) 131} [\href{https://arxiv.org/abs/2009.05111}{{\ttfamily
  2009.05111}}].

\bibitem{Benedetti:2015et}
D.~Benedetti, J.~Ben~Geloun and D.~Oriti, \emph{{Functional Renormalisation
  Group Approach for Tensorial Group Field Theory: a Rank-3 Model}},
  {\emph{JHEP} {\bfseries 03} (2015) 084}
  [\href{https://arxiv.org/abs/1411.3180}{{\ttfamily 1411.3180}}].

\bibitem{BenGeloun:2015ej}
J.~Ben~Geloun, R.~Martini and D.~Oriti, \emph{{Functional Renormalization Group
  analysis of a Tensorial Group Field Theory on ${R}^3$}}, {\emph{EPL}
  {\bfseries 112} (2015) 31001}
  [\href{https://arxiv.org/abs/1508.01855}{{\ttfamily 1508.01855}}].

\bibitem{BenGeloun:2016kw}
J.~Ben~Geloun, R.~Martini and D.~Oriti, \emph{{Functional renormalization group
  analysis of tensorial group field theories on $R^d$}}, {\emph{Phys. Rev. D}
  {\bfseries 94} (2016) 024017}
  [\href{https://arxiv.org/abs/1601.08211}{{\ttfamily 1601.08211}}].

\bibitem{Benedetti:2016db}
D.~Benedetti and V.~Lahoche, \emph{{Functional renormalization group approach
  for tensorial group field theory: a rank-6 model with closure constraint}},
  {\emph{Classical And Quantum Gravity} {\bfseries 33} (2016) }
  [\href{https://arxiv.org/abs/1508.06384}{{\ttfamily 1508.06384}}].

\bibitem{Carrozza:2016vsq}
S.~Carrozza, \emph{{Flowing in Group Field Theory Space: a Review}},
  \href{https://doi.org/10.3842/SIGMA.2016.070}{\emph{SIGMA} {\bfseries 12}
  (2016) 070} [\href{https://arxiv.org/abs/1603.01902}{{\ttfamily
  1603.01902}}].

\bibitem{Carrozza:2016tih}
S.~Carrozza and V.~Lahoche, \emph{{Asymptotic safety in three-dimensional SU(2)
  Group Field Theory: evidence in the local potential approximation}},
  \href{https://doi.org/10.1088/1361-6382/aa6d90}{\emph{Class. Quant. Grav.}
  {\bfseries 34} (2017) 115004}
  [\href{https://arxiv.org/abs/1612.02452}{{\ttfamily 1612.02452}}].

\bibitem{Carrozza:2017vkz}
S.~Carrozza, V.~Lahoche and D.~Oriti, \emph{{Renormalizable Group Field Theory
  beyond melonic diagrams: an example in rank four}},
  \href{https://doi.org/10.1103/PhysRevD.96.066007}{\emph{Phys. Rev. D}
  {\bfseries 96} (2017) 066007}
  [\href{https://arxiv.org/abs/1703.06729}{{\ttfamily 1703.06729}}].

\bibitem{BenGeloun:2018ekd}
J.~Ben~Geloun, T.A.~Koslowski, D.~Oriti and A.D.~Pereira, \emph{{Functional
  Renormalization Group analysis of rank 3 tensorial group field theory: The
  full quartic invariant truncation}},
  \href{https://doi.org/10.1103/PhysRevD.97.126018}{\emph{Phys. Rev. D}
  {\bfseries 97} (2018) 126018}
  [\href{https://arxiv.org/abs/1805.01619}{{\ttfamily 1805.01619}}].

\bibitem{Pithis:2018bw}
A.G.A.~Pithis and J.~Th{\"u}rigen, \emph{{Phase transitions in group field
  theory: The Landau perspective}},
  \href{https://doi.org/10.1103/PhysRevD.98.126006}{\emph{Phys. Rev. D}
  {\bfseries 98} (2018) 126006}
  [\href{https://arxiv.org/abs/1808.09765}{{\ttfamily 1808.09765}}].

\bibitem{Pithis:2020sxm}
A.G.A.~Pithis and J.~Th\"urigen, \emph{{(No) phase transition in tensorial
  group field theory}},
  \href{https://doi.org/10.1016/j.physletb.2021.136215}{\emph{Phys. Lett. B}
  {\bfseries 816} (2021) 136215}
  [\href{https://arxiv.org/abs/2007.08982}{{\ttfamily 2007.08982}}].

\bibitem{Pithis:2020kio}
A.G.A.~Pithis and J.~Th\"urigen, \emph{{Phase transitions in TGFT: functional
  renormalization group in the cyclic-melonic potential approximation and
  equivalence to O$(N)$ models}},
  \href{https://doi.org/10.1007/JHEP12(2020)159}{\emph{JHEP} {\bfseries 12}
  (2020) 159} [\href{https://arxiv.org/abs/2009.13588}{{\ttfamily
  2009.13588}}].

\bibitem{Marchetti:2021xvf}
L.~Marchetti, D.~Oriti, A.G.A.~Pithis and J.~Th\"urigen, \emph{{Phase
  transitions in tensorial group field theories: Landau-Ginzburg analysis of
  models with both local and non-local degrees of freedom}},
  \href{https://doi.org/10.1007/JHEP12(2021)201}{\emph{JHEP} {\bfseries 21}
  (2021) 201} [\href{https://arxiv.org/abs/2110.15336}{{\ttfamily
  2110.15336}}].

\bibitem{Baloitcha:2020lha}
E.~Baloitcha, V.~Lahoche and D.~Ousmane~Samary, \emph{{Flowing in discrete
  gravity models and Ward identities: a review}},
  \href{https://doi.org/10.1140/epjp/s13360-021-01823-z}{\emph{Eur. Phys. J.
  Plus} {\bfseries 136} (2021) 982}
  [\href{https://arxiv.org/abs/2001.02631}{{\ttfamily 2001.02631}}].

\bibitem{Lahoche:2022gkz}
V.~Lahoche and D.O.~Samary, \emph{{Stochastic dynamics for group field
  theories}},  \href{https://arxiv.org/abs/2209.02321}{{\ttfamily 2209.02321}}.

\bibitem{Gielen:2016dss}
S.~Gielen and L.~Sindoni, \emph{{Quantum Cosmology from Group Field Theory
  Condensates: a Review}},
  \href{https://doi.org/10.3842/SIGMA.2016.082}{\emph{SIGMA} {\bfseries 12}
  (2016) 082} [\href{https://arxiv.org/abs/1602.08104}{{\ttfamily
  1602.08104}}].

\bibitem{Oriti:2016acw}
D.~Oriti, \emph{{The universe as a quantum gravity condensate}},
  \href{https://doi.org/10.1016/j.crhy.2017.02.003}{\emph{Comptes Rendus
  Physique} {\bfseries 18} (2017) 235}
  [\href{https://arxiv.org/abs/1612.09521}{{\ttfamily 1612.09521}}].

\bibitem{Pithis:2019tvp}
A.G.A.~Pithis and M.~Sakellariadou, \emph{{Group field theory condensate
  cosmology: An appetizer}},
  \href{https://doi.org/10.3390/universe5060147}{\emph{Universe} {\bfseries 5}
  (2019) 147} [\href{https://arxiv.org/abs/1904.00598}{{\ttfamily
  1904.00598}}].

\bibitem{Oriti:2021oux}
D.~Oriti, \emph{{Tensorial Group Field Theory condensate cosmology as an
  example of spacetime emergence in quantum gravity}},  12, 2021
  [\href{https://arxiv.org/abs/2112.02585}{{\ttfamily 2112.02585}}].

\bibitem{Gielen:2013kla}
S.~Gielen, D.~Oriti and L.~Sindoni, \emph{{Cosmology from Group Field Theory
  Formalism for Quantum Gravity}},
  \href{https://doi.org/10.1103/PhysRevLett.111.031301}{\emph{Phys. Rev. Lett.}
  {\bfseries 111} (2013) 031301}
  [\href{https://arxiv.org/abs/1303.3576}{{\ttfamily 1303.3576}}].

\bibitem{Gielen:2013naa}
S.~Gielen, D.~Oriti and L.~Sindoni, \emph{{Homogeneous cosmologies as group
  field theory condensates}},
  \href{https://doi.org/10.1007/JHEP06(2014)013}{\emph{JHEP} {\bfseries 06}
  (2014) 013} [\href{https://arxiv.org/abs/1311.1238}{{\ttfamily 1311.1238}}].

\bibitem{Oriti:2016qtz}
D.~Oriti, L.~Sindoni and E.~Wilson-Ewing, \emph{{Emergent Friedmann dynamics
  with a quantum bounce from quantum gravity condensates}},
  \href{https://doi.org/10.1088/0264-9381/33/22/224001}{\emph{Class. Quant.
  Grav.} {\bfseries 33} (2016) 224001}
  [\href{https://arxiv.org/abs/1602.05881}{{\ttfamily 1602.05881}}].

\bibitem{deCesare:2016rsf}
M.~de~Cesare, A.G.A.~Pithis and M.~Sakellariadou, \emph{{Cosmological
  implications of interacting Group Field Theory models: cyclic Universe and
  accelerated expansion}},
  \href{https://doi.org/10.1103/PhysRevD.94.064051}{\emph{Phys. Rev. D}
  {\bfseries 94} (2016) 064051}
  [\href{https://arxiv.org/abs/1606.00352}{{\ttfamily 1606.00352}}].

\bibitem{Marchetti:2020umh}
L.~Marchetti and D.~Oriti, \emph{{Effective relational cosmological dynamics
  from Quantum Gravity}},
  \href{https://doi.org/10.1007/JHEP05(2021)025}{\emph{JHEP} {\bfseries 05}
  (2021) 025} [\href{https://arxiv.org/abs/2008.02774}{{\ttfamily
  2008.02774}}].

\bibitem{Marchetti:2021gcv}
L.~Marchetti and D.~Oriti, \emph{{Effective dynamics of scalar cosmological
  perturbations from quantum gravity}},
  \href{https://doi.org/10.1088/1475-7516/2022/07/004}{\emph{JCAP} {\bfseries
  07} (2022) 004} [\href{https://arxiv.org/abs/2112.12677}{{\ttfamily
  2112.12677}}].

\bibitem{Oriti:2016ueo}
D.~Oriti, L.~Sindoni and E.~Wilson-Ewing, \emph{{Bouncing cosmologies from
  quantum gravity condensates}},
  \href{https://doi.org/10.1088/1361-6382/aa549a}{\emph{Class. Quant. Grav.}
  {\bfseries 34} (2017) 04LT01}
  [\href{https://arxiv.org/abs/1602.08271}{{\ttfamily 1602.08271}}].

\bibitem{Jercher:2021bie}
A.F.~Jercher, D.~Oriti and A.G.A.~Pithis, \emph{{Emergent cosmology from
  quantum gravity in the Lorentzian Barrett-Crane tensorial group field theory
  model}}, \href{https://doi.org/10.1088/1475-7516/2022/01/050}{\emph{JCAP}
  {\bfseries 01} (2022) 050}
  [\href{https://arxiv.org/abs/2112.00091}{{\ttfamily 2112.00091}}].

\bibitem{strocchi2013introduction}
F.~Strocchi, \emph{An introduction to non-perturbative foundations of quantum
  field theory}, vol.~158, Oxford University Press (2013).

\bibitem{strocchi2005symmetry}
F.~Strocchi, \emph{Symmetry breaking}, vol.~643, Springer (2005).

\bibitem{BenGeloun:2022xyz}
J.~Ben~Geloun, A.G.A.~Pithis and J.~Thurigen, \emph{{Phase transitions in TGFT:
  adding local degrees of freedom to the functional RG flow}}, {\emph{to
  appear} } [\href{https://arxiv.org/abs/22ab.cdefg}{{\ttfamily 22ab.cdefg}}].

\bibitem{Li:2017uao}
Y.~Li, D.~Oriti and M.~Zhang, \emph{{Group field theory for quantum gravity
  minimally coupled to a scalar field}},
  \href{https://doi.org/10.1088/1361-6382/aa85d2}{\emph{Class. Quant. Grav.}
  {\bfseries 34} (2017) 195001}
  [\href{https://arxiv.org/abs/1701.08719}{{\ttfamily 1701.08719}}].

\bibitem{Gielen:2018fqv}
S.~Gielen, \emph{{Group field theory and its cosmology in a matter reference
  frame}}, \href{https://doi.org/10.3390/universe4100103}{\emph{Universe}
  {\bfseries 4} (2018) 103} [\href{https://arxiv.org/abs/1808.10469}{{\ttfamily
  1808.10469}}].

\bibitem{Oriti:2006jk}
D.~Oriti and J.~Ryan, \emph{{Group field theory formulation of 3-D quantum
  gravity coupled to matter fields}},
  \href{https://doi.org/10.1088/0264-9381/23/22/027}{\emph{Class. Quant. Grav.}
  {\bfseries 23} (2006) 6543}
  [\href{https://arxiv.org/abs/gr-qc/0602010}{{\ttfamily gr-qc/0602010}}].

\bibitem{Brown:1994py}
J.D.~Brown and K.V.~Kuchar, \emph{{Dust as a standard of space and time in
  canonical quantum gravity}},
  \href{https://doi.org/10.1103/PhysRevD.51.5600}{\emph{Phys. Rev. D}
  {\bfseries 51} (1995) 5600}
  [\href{https://arxiv.org/abs/gr-qc/9409001}{{\ttfamily gr-qc/9409001}}].

\bibitem{Rovelli:2001bz}
C.~Rovelli, \emph{{Partial observables}},
  \href{https://doi.org/10.1103/PhysRevD.65.124013}{\emph{Phys. Rev. D}
  {\bfseries 65} (2002) 124013}
  [\href{https://arxiv.org/abs/gr-qc/0110035}{{\ttfamily gr-qc/0110035}}].

\bibitem{Dittrich:2005kc}
B.~Dittrich, \emph{{Partial and complete observables for canonical general
  relativity}}, \href{https://doi.org/10.1088/0264-9381/23/22/006}{\emph{Class.
  Quant. Grav.} {\bfseries 23} (2006) 6155}
  [\href{https://arxiv.org/abs/gr-qc/0507106}{{\ttfamily gr-qc/0507106}}].

\bibitem{Ashtekar:2011ni}
A.~Ashtekar and P.~Singh, \emph{{Loop Quantum Cosmology: A Status Report}},
  \href{https://doi.org/10.1088/0264-9381/28/21/213001}{\emph{Class. Quant.
  Grav.} {\bfseries 28} (2011) 213001}
  [\href{https://arxiv.org/abs/1108.0893}{{\ttfamily 1108.0893}}].

\bibitem{Giesel:2012rb}
K.~Giesel and T.~Thiemann, \emph{{Scalar Material Reference Systems and Loop
  Quantum Gravity}},
  \href{https://doi.org/10.1088/0264-9381/32/13/135015}{\emph{Class. Quant.
  Grav.} {\bfseries 32} (2012) 135015}
  [\href{https://arxiv.org/abs/1206.3807}{{\ttfamily 1206.3807}}].

\bibitem{Carrozza:2022xut}
S.~Carrozza, S.~Eccles and P.A.~Hoehn, \emph{{Edge modes as dynamical frames:
  charges from post-selection in generally covariant theories}},
  \href{https://arxiv.org/abs/2205.00913}{{\ttfamily 2205.00913}}.

\bibitem{Goeller:2022rsx}
C.~Goeller, P.A.~Hoehn and J.~Kirklin, \emph{{Diffeomorphism-invariant
  observables and dynamical frames in gravity: reconciling bulk locality with
  general covariance}},  \href{https://arxiv.org/abs/2206.01193}{{\ttfamily
  2206.01193}}.

\bibitem{Pithis:2019mlv}
A.G.A.~Pithis, \emph{{Aspects of quantum gravity}}, Ph.D. thesis, King's Coll.
  London, 2019.
\newblock \href{https://arxiv.org/abs/1903.07735}{{\ttfamily 1903.07735}}.

\bibitem{Jercher:2022mky}
A.F.~Jercher, D.~Oriti and A.G.A.~Pithis, \emph{{The Complete Barrett-Crane
  Model and its Causal Structure}},
  \href{https://arxiv.org/abs/2206.15442}{{\ttfamily 2206.15442}}.

\bibitem{Baratin:2011tx}
A.~Baratin and D.~Oriti, \emph{{Quantum simplicial geometry in the group field
  theory formalism: reconsidering the Barrett-Crane model}},
  \href{https://doi.org/10.1088/1367-2630/13/12/125011}{\emph{New J. Phys.}
  {\bfseries 13} (2011) 125011}
  [\href{https://arxiv.org/abs/1108.1178}{{\ttfamily 1108.1178}}].

\bibitem{Wipf:2021mns}
A.~Wipf, \emph{{Statistical Approach to Quantum Field Theory: An
  Introduction}}, vol.~992 of \emph{Lecture Notes in Physics} (10, 2021),
  \href{https://doi.org/10.1007/978-3-030-83263-6}{10.1007/978-3-030-83263-6}.

\bibitem{hohenberg2015introduction}
P.~Hohenberg and A.~Krekhov, \emph{An introduction to the ginzburg--landau
  theory of phase transitions and nonequilibrium patterns}, {\emph{Physics
  Reports} {\bfseries 572} (2015) 1}.

\bibitem{levanyuk1959contribution}
A.~Levanyuk, \emph{Contribution to the theory of light scattering near the
  second-order phase-transition points}, {\emph{Sov. Phys. JETP} {\bfseries 9}
  (1959) 571}.

\bibitem{ginzburg1961some}
V.~Ginzburg, \emph{Some remarks on phase transitions of the second kind and the
  microscopic theory of ferroelectric materials}, {\emph{Soviet Phys. Solid
  State} {\bfseries 2} (1961) 1824}.

\bibitem{wilson1983renormalization}
K.G.~Wilson, \emph{The renormalization group and critical phenomena},
  {\emph{Reviews of Modern Physics} {\bfseries 55} (1983) 583}.

\bibitem{Dupuis:2020fhh}
N.~Dupuis, L.~Canet, A.~Eichhorn, W.~Metzner, J.M.~Pawlowski, M.~Tissier
  et~al., \emph{{The nonperturbative functional renormalization group and its
  applications}},
  \href{https://doi.org/10.1016/j.physrep.2021.01.001}{\emph{Phys. Rept.}
  {\bfseries 910} (2021) 1} [\href{https://arxiv.org/abs/2006.04853}{{\ttfamily
  2006.04853}}].

\bibitem{Benedetti1403}
D.~Benedetti, \emph{{Critical behavior in spherical and hyperbolic spaces}},
  \href{https://doi.org/10.1088/1742-5468/2015/01/P01002}{\emph{Journal of
  Statistical Mechanics: Theory and Experiment} {\bfseries 1501} (2015) P01002}
  [\href{https://arxiv.org/abs/1403.6712}{{\ttfamily 1403.6712}}].

\bibitem{Barrett:1999qw}
J.W.~Barrett and L.~Crane, \emph{{A Lorentzian signature model for quantum
  general relativity}},
  \href{https://doi.org/10.1088/0264-9381/17/16/302}{\emph{Class. Quant. Grav.}
  {\bfseries 17} (2000) 3101}
  [\href{https://arxiv.org/abs/gr-qc/9904025}{{\ttfamily gr-qc/9904025}}].

\bibitem{Perez:2000ec}
A.~Perez and C.~Rovelli, \emph{{Spin foam model for Lorentzian general
  relativity}}, \href{https://doi.org/10.1103/PhysRevD.63.041501}{\emph{Phys.
  Rev. D} {\bfseries 63} (2001) 041501}
  [\href{https://arxiv.org/abs/gr-qc/0009021}{{\ttfamily gr-qc/0009021}}].

\bibitem{Perez:2000ep}
A.~Perez and C.~Rovelli, \emph{{3+1 spinfoam model of quantum gravity with
  space - like and time - like components}},
  \href{https://doi.org/10.1103/PhysRevD.64.064002}{\emph{Phys. Rev. D}
  {\bfseries 64} (2001) 064002}
  [\href{https://arxiv.org/abs/gr-qc/0011037}{{\ttfamily gr-qc/0011037}}].

\bibitem{DePietri:1999bx}
R.~De~Pietri, L.~Freidel, K.~Krasnov and C.~Rovelli, \emph{{Barrett-Crane model
  from a Boulatov-Ooguri field theory over a homogeneous space}},
  \href{https://doi.org/10.1016/S0550-3213(00)00005-5}{\emph{Nucl. Phys. B}
  {\bfseries 574} (2000) 785}
  [\href{https://arxiv.org/abs/hep-th/9907154}{{\ttfamily hep-th/9907154}}].

\bibitem{Ruehl1970}
W.~Ruehl, \emph{Lorentz group and harmonic analysis}, W A Benjamin, Inc, United
  States (1970).

\bibitem{Ooguri:1992eb}
H.~Ooguri, \emph{{Topological lattice models in four-dimensions}},
  \href{https://doi.org/10.1142/S0217732392004171}{\emph{Mod. Phys. Lett. A}
  {\bfseries 7} (1992) 2799}
  [\href{https://arxiv.org/abs/hep-th/9205090}{{\ttfamily hep-th/9205090}}].

\bibitem{Baratin:2010wi}
A.~Baratin and D.~Oriti, \emph{{Group field theory with non-commutative metric
  variables}},
  \href{https://doi.org/10.1103/PhysRevLett.105.221302}{\emph{Phys. Rev. Lett.}
  {\bfseries 105} (2010) 221302}
  [\href{https://arxiv.org/abs/1002.4723}{{\ttfamily 1002.4723}}].

\bibitem{Guedes:2013vi}
C.~Guedes, D.~Oriti and M.~Raasakka, \emph{{Quantization maps, algebra
  representation and non-commutative Fourier transform for Lie groups}},
  \href{https://doi.org/10.1063/1.4818638}{\emph{J. Math. Phys.} {\bfseries 54}
  (2013) 083508} [\href{https://arxiv.org/abs/1301.7750}{{\ttfamily
  1301.7750}}].

\bibitem{Oriti:2018bwr}
D.~Oriti and G.~Rosati, \emph{{Noncommutative Fourier transform for the Lorentz
  group via the Duflo map}},
  \href{https://doi.org/10.1103/PhysRevD.99.106005}{\emph{Phys. Rev. D}
  {\bfseries 99} (2019) 106005}
  [\href{https://arxiv.org/abs/1812.08616}{{\ttfamily 1812.08616}}].

\bibitem{Finocchiaro:2018hks}
M.~Finocchiaro and D.~Oriti, \emph{{Spin foam models and the Duflo map}},
  \href{https://doi.org/10.1088/1361-6382/ab58da}{\emph{Class. Quant. Grav.}
  {\bfseries 37} (2020) 015010}
  [\href{https://arxiv.org/abs/1812.03550}{{\ttfamily 1812.03550}}].

\bibitem{Oriti:2003wf}
D.~Oriti, \emph{{Spin foam models of quantum space-time}},  other thesis, 11,
  2003, [\href{https://arxiv.org/abs/gr-qc/0311066}{{\ttfamily
  gr-qc/0311066}}].

\bibitem{BenGeloun:2011jnm}
J.~Ben~Geloun and V.~Bonzom, \emph{{Radiative corrections in the
  Boulatov-Ooguri tensor model: The 2-point function}},
  \href{https://doi.org/10.1007/s10773-011-0782-2}{\emph{Int. J. Theor. Phys.}
  {\bfseries 50} (2011) 2819}
  [\href{https://arxiv.org/abs/1101.4294}{{\ttfamily 1101.4294}}].

\bibitem{BenGeloun:2011rc}
J.~Ben~Geloun and V.~Rivasseau, \emph{{A Renormalizable 4-Dimensional Tensor
  Field Theory}},
  \href{https://doi.org/10.1007/s00220-012-1549-1}{\emph{Commun. Math. Phys.}
  {\bfseries 318} (2013) 69} [\href{https://arxiv.org/abs/1111.4997}{{\ttfamily
  1111.4997}}].

\bibitem{BenGeloun:2013mgx}
J.~Ben~Geloun, \emph{{On the finite amplitudes for open graphs in Abelian
  dynamical colored Boulatov-Ooguri models}},
  \href{https://doi.org/10.1088/1751-8113/46/40/402002}{\emph{J. Phys. A}
  {\bfseries 46} (2013) 402002}
  [\href{https://arxiv.org/abs/1307.8299}{{\ttfamily 1307.8299}}].

\bibitem{Rivasseau:2016zco}
V.~Rivasseau, \emph{{Random Tensors and Quantum Gravity}},
  \href{https://doi.org/10.3842/SIGMA.2016.069}{\emph{SIGMA} {\bfseries 12}
  (2016) 069} [\href{https://arxiv.org/abs/1603.07278}{{\ttfamily
  1603.07278}}].

\bibitem{DePietri:2000ii}
R.~De~Pietri and C.~Petronio, \emph{{Feynman diagrams of generalized matrix
  models and the associated manifolds in dimension 4}},
  \href{https://doi.org/10.1063/1.1290053}{\emph{J. Math. Phys.} {\bfseries 41}
  (2000) 6671} [\href{https://arxiv.org/abs/gr-qc/0004045}{{\ttfamily
  gr-qc/0004045}}].

\bibitem{Gurau:2010nd}
R.~Gurau, \emph{{Lost in Translation: Topological Singularities in Group Field
  Theory}}, \href{https://doi.org/10.1088/0264-9381/27/23/235023}{\emph{Class.
  Quant. Grav.} {\bfseries 27} (2010) 235023}
  [\href{https://arxiv.org/abs/1006.0714}{{\ttfamily 1006.0714}}].

\bibitem{Gurau:2010mhz}
R.~Gurau, \emph{{Topological Graph Polynomials in Colored Group Field Theory}},
  \href{https://doi.org/10.1007/s00023-010-0035-6}{\emph{Annales Henri
  Poincare} {\bfseries 11} (2010) 565}
  [\href{https://arxiv.org/abs/0911.1945}{{\ttfamily 0911.1945}}].

\bibitem{Gurau:2009tw}
R.~Gurau, \emph{{Colored Group Field Theory}},
  \href{https://doi.org/10.1007/s00220-011-1226-9}{\emph{Commun. Math. Phys.}
  {\bfseries 304} (2011) 69} [\href{https://arxiv.org/abs/0907.2582}{{\ttfamily
  0907.2582}}].

\bibitem{Gurau:2011xp}
R.~Gurau and J.P.~Ryan, \emph{{Colored Tensor Models - a review}},
  \href{https://doi.org/10.3842/SIGMA.2012.020}{\emph{SIGMA} {\bfseries 8}
  (2012) 020} [\href{https://arxiv.org/abs/1109.4812}{{\ttfamily 1109.4812}}].

\bibitem{Bonzom:2012hw}
V.~Bonzom, R.~Gurau and V.~Rivasseau, \emph{{Random tensor models in the large
  N limit: Uncoloring the colored tensor models}},
  \href{https://doi.org/10.1103/PhysRevD.85.084037}{\emph{Phys. Rev. D}
  {\bfseries 85} (2012) 084037}
  [\href{https://arxiv.org/abs/1202.3637}{{\ttfamily 1202.3637}}].

\bibitem{Riello:2013bzw}
A.~Riello, \emph{{Self-energy of the Lorentzian Engle-Pereira-Rovelli-Livine
  and Freidel-Krasnov model of quantum gravity}},
  \href{https://doi.org/10.1103/PhysRevD.88.024011}{\emph{Phys. Rev. D}
  {\bfseries 88} (2013) 024011}
  [\href{https://arxiv.org/abs/1302.1781}{{\ttfamily 1302.1781}}].

\bibitem{landau2013statistical}
L.~Landau and E.~Lifshitz, \emph{Statistical Physics: Volume 5}, no.~v. 5,
  Elsevier Science (2013).

\bibitem{dmitriev1996reconstructive}
V.~Dmitriev and P.~Toledano, \emph{Reconstructive Phase Transitions: In
  Crystals And Quasicrystals}, World Scientific Publishing Company (1996).

\bibitem{Catterall:1994pg}
S.~Catterall, J.B.~Kogut and R.~Renken, \emph{{Phase structure of
  four-dimensional simplicial quantum gravity}},
  \href{https://doi.org/10.1016/0370-2693(94)91480-X}{\emph{Phys. Lett. B}
  {\bfseries 328} (1994) 277}
  [\href{https://arxiv.org/abs/hep-lat/9401026}{{\ttfamily hep-lat/9401026}}].

\bibitem{Bialas:1996wu}
P.~Bialas, Z.~Burda, A.~Krzywicki and B.~Petersson, \emph{{Focusing on the
  fixed point of 4-D simplicial gravity}},
  \href{https://doi.org/10.1016/0550-3213(96)00214-3}{\emph{Nucl. Phys. B}
  {\bfseries 472} (1996) 293}
  [\href{https://arxiv.org/abs/hep-lat/9601024}{{\ttfamily hep-lat/9601024}}].

\bibitem{Coumbe:2014nea}
D.~Coumbe and J.~Laiho, \emph{{Exploring Euclidean Dynamical Triangulations
  with a Non-trivial Measure Term}},
  \href{https://doi.org/10.1007/JHEP04(2015)028}{\emph{JHEP} {\bfseries 04}
  (2015) 028} [\href{https://arxiv.org/abs/1401.3299}{{\ttfamily 1401.3299}}].

\bibitem{Laiho:2016nlp}
J.~Laiho, S.~Bassler, D.~Coumbe, D.~Du and J.T.~Neelakanta, \emph{{Lattice
  Quantum Gravity and Asymptotic Safety}},
  \href{https://doi.org/10.1103/PhysRevD.96.064015}{\emph{Phys. Rev. D}
  {\bfseries 96} (2017) 064015}
  [\href{https://arxiv.org/abs/1604.02745}{{\ttfamily 1604.02745}}].

\bibitem{Ambjorn:2022dvx}
J.~Ambjorn, J.~Gizbert-Studnicki, A.~G\"orlich and D.~N\'emeth, \emph{{Topology
  induced first-order phase transitions in lattice quantum gravity}},
  \href{https://doi.org/10.1007/JHEP04(2022)103}{\emph{JHEP} {\bfseries 04}
  (2022) 103} [\href{https://arxiv.org/abs/2202.07392}{{\ttfamily
  2202.07392}}].

\bibitem{Dona:2021ldn}
P.~Don\`a, F.~Gozzini and A.~Nicotra, \emph{{Wick rotation for spin foam
  quantum gravity}},
  \href{https://doi.org/10.1103/PhysRevD.104.126008}{\emph{Phys. Rev. D}
  {\bfseries 104} (2021) 126008}
  [\href{https://arxiv.org/abs/2106.14672}{{\ttfamily 2106.14672}}].

\bibitem{kostorz2001phase}
G.~Kostorz, \emph{Phase transformations in materials}, Wiley Online Library
  (2001).

\bibitem{doi:10.1002/9783527603978.mst0387}
K.~Binder, \emph{Statistical theories of phase transitions},  in
  \emph{Materials Science and Technology}, American Cancer Society (2013),
  \href{https://doi.org/10.1002/9783527603978.mst0387}{DOI}
  [\href{https://arxiv.org/abs/https://onlinelibrary.wiley.com/doi/pdf/10.1002/9783527603978.mst0387}{{\ttfamily
  https://onlinelibrary.wiley.com/doi/pdf/10.1002/9783527603978.mst0387}}].

\bibitem{grouptheoryphysicists}
Z.Q.~Ma, \emph{Real orthogonal groups},  in \emph{Group Theory for Physicists},
  pp.~399--460 \href{https://doi.org/10.1142/9789812771438_0009}{DOI}.

\bibitem{gallier2020differential}
J.~Gallier and J.~Quaintance, \emph{Differential geometry and Lie groups: a
  computational perspective}, vol.~12, Springer Nature (2020).

\bibitem{alexandrino2009introduction}
M.M.~Alexandrino and R.G.~Bettiol, \emph{Introduction to lie groups, isometric
  and adjoint actions and some generalizations}, {\emph{arXiv preprint
  arXiv:0901.2374} (2009) }.

\bibitem{einsiedler2013ergodic}
M.~Einsiedler and T.~Ward, \emph{Ergodic theory}, {\emph{Springer} {\bfseries
  4} (2013) 4}.

\bibitem{Rivasseau:2015ova}
V.~Rivasseau, \emph{{Why are tensor field theories asymptotically free?}},
  \href{https://doi.org/10.1209/0295-5075/111/60011}{\emph{EPL} {\bfseries 111}
  (2015) 60011} [\href{https://arxiv.org/abs/1507.04190}{{\ttfamily
  1507.04190}}].

\bibitem{Banerjee:2016laq}
R.~Banerjee and P.~Mukherjee, \emph{{Torsional Newton\textendash{}Cartan
  geometry from Galilean gauge theory}},
  \href{https://doi.org/10.1088/0264-9381/33/22/225013}{\emph{Class. Quant.
  Grav.} {\bfseries 33} (2016) 225013}
  [\href{https://arxiv.org/abs/1604.06893}{{\ttfamily 1604.06893}}].

\bibitem{Bergshoeff:2017dqq}
E.~Bergshoeff, A.~Chatzistavrakidis, L.~Romano and J.~Rosseel,
  \emph{{Newton-Cartan Gravity and Torsion}},
  \href{https://doi.org/10.1007/JHEP10(2017)194}{\emph{JHEP} {\bfseries 10}
  (2017) 194} [\href{https://arxiv.org/abs/1708.05414}{{\ttfamily
  1708.05414}}].

\bibitem{Guerrieri:2020vhp}
A.~Guerrieri and R.F.~Sobreiro, \emph{{Non-relativistic limit of gravity
  theories in the first order formalism}},
  \href{https://doi.org/10.1007/JHEP03(2021)104}{\emph{JHEP} {\bfseries 03}
  (2021) 104} [\href{https://arxiv.org/abs/2010.14918}{{\ttfamily
  2010.14918}}].

\bibitem{rovelli2004quantum}
C.~Rovelli, \emph{Quantum gravity}, Cambridge university press (2004).

\bibitem{BenGeloun:2014kpa}
J.~Ben~Geloun, \emph{{Renormalizable Models in Rank $d\geq 2$ Tensorial Group
  Field Theory}}, {\emph{Comm. Math. Phys.} {\bfseries 332} (2014) 117}
  [\href{https://arxiv.org/abs/1306.1201}{{\ttfamily 1306.1201}}].

\bibitem{Marchetti:2022nrf}
L.~Marchetti, D.~Oriti, A.G.A.~Pithis and J.~Th\"urigen, \emph{{Mean-field
  phase transitions in TGFT quantum gravity}},
  \href{https://arxiv.org/abs/2211.12768}{{\ttfamily 2211.12768}}.

\bibitem{Pithis:2016wzf}
A.G.A.~Pithis, M.~Sakellariadou and P.~Tomov, \emph{{Impact of nonlinear
  effective interactions on group field theory quantum gravity condensates}},
  \href{https://doi.org/10.1103/PhysRevD.94.064056}{\emph{Phys. Rev. D}
  {\bfseries 94} (2016) 064056}
  [\href{https://arxiv.org/abs/1607.06662}{{\ttfamily 1607.06662}}].

\bibitem{Kegeles:2017ems}
A.~Kegeles, D.~Oriti and C.~Tomlin, \emph{{Inequivalent coherent state
  representations in group field theory}},
  \href{https://doi.org/10.1088/1361-6382/aac39f}{\emph{Class. Quant. Grav.}
  {\bfseries 35} (2018) 125011}
  [\href{https://arxiv.org/abs/1709.00161}{{\ttfamily 1709.00161}}].

\bibitem{Kegeles:2018tyo}
A.~Kegeles, \emph{{Algebraic foundation of Group Field Theory}}, Ph.D. thesis,
  University of Potsdam, 2018.

\bibitem{honegger2015photons}
R.~Honegger and A.~Rieckers, \emph{Photons in Fock space and beyond}, World
  Scientific (2015).

\bibitem{Bonzom:2011zz}
V.~Bonzom, R.~Gurau, A.~Riello and V.~Rivasseau, \emph{{Critical behavior of
  colored tensor models in the large N limit}},
  \href{https://doi.org/10.1016/j.nuclphysb.2011.07.022}{\emph{Nucl. Phys. B}
  {\bfseries 853} (2011) 174}
  [\href{https://arxiv.org/abs/1105.3122}{{\ttfamily 1105.3122}}].

\bibitem{Gurau:2011tj}
R.~Gurau, \emph{{A generalization of the Virasoro algebra to arbitrary
  dimensions}},
  \href{https://doi.org/10.1016/j.nuclphysb.2011.07.009}{\emph{Nucl. Phys. B}
  {\bfseries 852} (2011) 592}
  [\href{https://arxiv.org/abs/1105.6072}{{\ttfamily 1105.6072}}].

\bibitem{Gurau:2013cbh}
R.~Gurau and J.P.~Ryan, \emph{{Melons are branched polymers}},
  \href{https://doi.org/10.1007/s00023-013-0291-3}{\emph{Annales Henri
  Poincare} {\bfseries 15} (2014) 2085}
  [\href{https://arxiv.org/abs/1302.4386}{{\ttfamily 1302.4386}}].

\bibitem{Jordan:2013awa}
S.~Jordan and R.~Loll, \emph{{Causal Dynamical Triangulations without Preferred
  Foliation}},
  \href{https://doi.org/10.1016/j.physletb.2013.06.007}{\emph{Phys. Lett. B}
  {\bfseries 724} (2013) 155}
  [\href{https://arxiv.org/abs/1305.4582}{{\ttfamily 1305.4582}}].

\bibitem{Loll:2015yaa}
R.~Loll and B.~Ruijl, \emph{{Locally Causal Dynamical Triangulations in Two
  Dimensions}}, \href{https://doi.org/10.1103/PhysRevD.92.084002}{\emph{Phys.
  Rev. D} {\bfseries 92} (2015) 084002}
  [\href{https://arxiv.org/abs/1507.04566}{{\ttfamily 1507.04566}}].

\bibitem{Asante:2021phx}
S.K.~Asante, B.~Dittrich and J.~Padua-Arg\"uelles, \emph{{Complex actions and
  causality violations: Applications to Lorentzian quantum cosmology}},
  \href{https://arxiv.org/abs/2112.15387}{{\ttfamily 2112.15387}}.

\bibitem{Dittrich:2021kzs}
B.~Dittrich, \emph{{Modified Graviton Dynamics From Spin Foams: The Area Regge
  Action}},  \href{https://arxiv.org/abs/2105.10808}{{\ttfamily 2105.10808}}.

\bibitem{Berges:2002ga}
J.~Berges, N.~Tetradis and C.~Wetterich, \emph{{Non-perturbative
  renormalization flow in quantum field theory and statistical physics}},
  \href{https://doi.org/10.1016/S0370-1573(01)00098-9}{\emph{Phys. Rept.}
  {\bfseries 363} (2002) 223}
  [\href{https://arxiv.org/abs/hep-ph/0005122}{{\ttfamily hep-ph/0005122}}].

\bibitem{Fairbairn:2007sv}
W.J.~Fairbairn and E.R.~Livine, \emph{{3d Spinfoam Quantum Gravity: Matter as a
  Phase of the Group Field Theory}},
  \href{https://doi.org/10.1088/0264-9381/24/20/021}{\emph{Class. Quant. Grav.}
  {\bfseries 24} (2007) 5277}
  [\href{https://arxiv.org/abs/gr-qc/0702125}{{\ttfamily gr-qc/0702125}}].

\bibitem{Livine:2011yb}
E.R.~Livine, D.~Oriti and J.P.~Ryan, \emph{{Effective Hamiltonian Constraint
  from Group Field Theory}},
  \href{https://doi.org/10.1088/0264-9381/28/24/245010}{\emph{Class. Quant.
  Grav.} {\bfseries 28} (2011) 245010}
  [\href{https://arxiv.org/abs/1104.5509}{{\ttfamily 1104.5509}}].

\bibitem{BenGeloun:2018eoe}
J.~Ben~Geloun, A.~Kegeles and A.G.A.~Pithis, \emph{{Minimizers of the dynamical
  Boulatov model}},
  \href{https://doi.org/10.1140/epjc/s10052-018-6483-8}{\emph{Eur. Phys. J. C}
  {\bfseries 78} (2018) 996}
  [\href{https://arxiv.org/abs/1806.09961}{{\ttfamily 1806.09961}}].

\bibitem{gel1950unitary}
I.M.~Gel'fand and M.A.~Naimark, \emph{Unitary representations of the classical
  groups}, {\emph{Trudy Matematicheskogo Instituta imeni VA Steklova}
  {\bfseries 36} (1950) 3}.

\bibitem{gel1963representations}
I.~Gelfand, I.~Gelfand, H.~Farahat, Z.~Shapiro, R.~Minlos, Z.~Shapiro et~al.,
  \emph{Representations of the Rotation and Lorentz Groups and Their
  Applications}, (Pergamon Press book), Pergamon Press (1963).

\bibitem{naimark1964linear}
M.~Na{\u\i}mark and H.~Farahat, \emph{Linear Representations of the Lorentz
  Group}, International series of monographs in pure and applied mathematics,
  Pergamon Press (1964).

\bibitem{duc1967theory}
D.V.~Duc and N.~Van~Hieu, \emph{On the theory of unitary representations of the
  $ sl (2, c) $ group},  in \emph{Annales de l'IHP Physique th{\'e}orique},
  vol.~6, pp.~17--37, 1967.

\bibitem{gelfand1968representation}
I.~Gelfand, M.~Graev and Piatetski-Shapiro, \emph{Representation Theory and
  Automorphic Functions}, Generalized functions, Saunders (1968).

\bibitem{knapp2016representation}
A.W.~Knapp, \emph{Representation theory of semisimple groups}, Princeton
  university press (2016).

\bibitem{gel2016generalized1}
I.M.~Gel'fand and G.E.~Shilov, \emph{Generalized functions, Volume 2: Spaces of
  fundamental and generalized functions}, vol.~261, American Mathematical Soc.
  (2016).

\bibitem{gel2016generalized2}
I.~Gelfand, M.~Graev and N.~Vilenkin, \emph{Generalized Functions, Volume 5},
  AMS Chelsea Publishing, American Mathematical Society (2016).

\bibitem{turski1998harmonic}
J.~Turski, \emph{Harmonic analysis on sl (2, c) and projectively adapted
  pattern representation}, {\emph{Journal of Fourier Analysis and Applications}
  {\bfseries 4} (1998) 67}.

\bibitem{Martin-Dussaud:2019ypf}
P.~Martin-Dussaud, \emph{{A Primer of Group Theory for Loop Quantum Gravity and
  Spin-foams}}, \href{https://doi.org/10.1007/s10714-019-2583-5}{\emph{Gen.
  Rel. Grav.} {\bfseries 51} (2019) 110}
  [\href{https://arxiv.org/abs/1902.08439}{{\ttfamily 1902.08439}}].

\bibitem{hormander2015analysis}
L.~H{\"o}rmander, \emph{The analysis of linear partial differential operators
  I: Distribution theory and Fourier analysis}, Springer (2015).

\bibitem{Speziale:2016axj}
S.~Speziale, \emph{{Boosting Wigner\textquoteright{}s nj-symbols}},
  \href{https://doi.org/10.1063/1.4977752}{\emph{J. Math. Phys.} {\bfseries 58}
  (2017) 032501} [\href{https://arxiv.org/abs/1609.01632}{{\ttfamily
  1609.01632}}].

\bibitem{rashid}
M.A.~Rashid, \emph{Boost matrix elements of the homogeneous lorentz group},
  \href{https://doi.org/10.1063/1.524211}{\emph{Journal of Mathematical
  Physics} {\bfseries 20} (1979) 1514}
  [\href{https://arxiv.org/abs/https://doi.org/10.1063/1.524211}{{\ttfamily
  https://doi.org/10.1063/1.524211}}].

\bibitem{hyperentire}
N.M.~Temme, \emph{Hypergeometric functions},  in \emph{Special Functions},
  pp.~107--132, John Wiley \& Sons, Ltd (1996),
  \href{https://doi.org/https://doi.org/10.1002/9781118032572.ch5}{DOI}
  [\href{https://arxiv.org/abs/https://onlinelibrary.wiley.com/doi/pdf/10.1002/9781118032572.ch5}{{\ttfamily
  https://onlinelibrary.wiley.com/doi/pdf/10.1002/9781118032572.ch5}}].

\bibitem{hyperentire2}
F.~Olver, \emph{Differential equations with irregular singularities; bessel and
  confluent hypergeometric functions},  in \emph{Asymptotics and Special
  Functions}, pp.~229--278, Academic Press (1974),
  \href{https://doi.org/10.1016/B978-0-12-525850-0.50012-2}{DOI}.

\bibitem{hypergeo}
A.~Gil, J.~Segura and N.M.~Temme, \emph{Convergent and divergent series},  in
  \emph{Numerical Methods for Special Functions}, pp.~15--50 (2007),
  \href{https://doi.org/10.1137/1.9780898717822.ch2}{DOI}
  [\href{https://arxiv.org/abs/https://epubs.siam.org/doi/pdf/10.1137/1.9780898717822.ch2}{{\ttfamily
  https://epubs.siam.org/doi/pdf/10.1137/1.9780898717822.ch2}}].

\bibitem{doi:10.1063/1.1703728}
L.C.~Biedenharn, \emph{Wigner coefficients for the r4 group and some
  applications}, \href{https://doi.org/10.1063/1.1703728}{\emph{Journal of
  Mathematical Physics} {\bfseries 2} (1961) 433}
  [\href{https://arxiv.org/abs/https://doi.org/10.1063/1.1703728}{{\ttfamily
  https://doi.org/10.1063/1.1703728}}].

\bibitem{Hong_Hao_2010}
Z.~Hong-Hao, F.~Kai-Xi, Q.~Si-Wei, Z.~An and L.~Xue-Song, \emph{On analytic
  formulas of feynman propagators in position space},
  \href{https://doi.org/10.1088/1674-1137/34/10/005}{\emph{Chinese Physics C}
  {\bfseries 34} (2010) 1576–1582}.

\end{thebibliography}\endgroup



\providecommand{\href}[2]{#2}\begingroup\raggedright\endgroup

\end{document}